\documentclass[aip,superscriptaddress]{revtex4-1}
\usepackage{float}
\usepackage{amsmath,amssymb}
\usepackage{amsfonts}
\usepackage{psfrag}
\usepackage{subfigure}
\usepackage{epsfig}
\usepackage{multirow}
\usepackage{algorithm}
\usepackage{algorithmic}

\DeclareMathOperator{\pa}{+\!\!=}

\usepackage[normalem]{ulem}
\usepackage[usenames,dvipsnames]{color}

\begin{document}
\title{Accelerating Image Reconstruction in Three-Dimensional 
Optoacoustic Tomography on Graphics Processing Units
}
\author{Kun~Wang$^*$ 
}
\altaffiliation{
Contributed equally to this work
}
\affiliation{
Department of Biomedical Engineering, Washington University in St. Louis, 
St. Louis, MO 63130
}
\author{Chao~Huang$^*$}
\altaffiliation{
Contributed equally to this work
}
\affiliation{
Department of Biomedical Engineering, Washington University in St. Louis, 
St. Louis, MO 63130
}
\author{Yu-Jiun Kao}
\affiliation{
Department of Bio-Industrial Mechatronics Engineering,
National Taiwan University, Taipei 106, Taiwan
}

\author{Cheng-Ying~Chou}
\affiliation{
Department of Bio-Industrial Mechatronics Engineering, 
National Taiwan University, Taipei 106, Taiwan
}
\author{Alexander~A.~Oraevsky}
\affiliation{
TomoWave Laboratories, 
675 Bering drive, Suite 575, Houston,
Texas, Houston, TX 77057
}
\author{Mark~A.~Anastasio}
\email{anastasio@seas.wustl.edu}
\affiliation{
Department of Biomedical Engineering, Washington University in St. Louis, St. Louis, MO 63130
}

\date{\today}

\begin{abstract}
\noindent{\bf Purpose:}
% Optoacoustic tomography (OAT), also known as photoacoustic computed tomography, is an emerging 
% hybrid imaging modality that combines the advantages of optical imaging and ultrasound 
% imaging.
Optoacoustic tomography (OAT) is inherently a three-dimensional 
(3D) inverse problem.
However, most studies of OAT image reconstruction still employ 
two-dimensional (2D) imaging models. 
One important reason is because 3D image reconstruction is computationally
burdensome. 
The aim of this work is to 
accelerate existing image reconstruction algorithms 
for 3D OAT by use of parallel programming techniques. 

\noindent{\bf Methods:}
Parallelization strategies are proposed to accelerate 
a filtered backprojection (FBP) algorithm and two different pairs 
of projection/backprojection operations that correspond to
two different numerical imaging models.
The algorithms are designed to fully exploit the parallel computing 
power of graphic processing units (GPUs).
In order to evaluate the parallelization strategies for the 
projection/backprojection pairs, 
an iterative image reconstruction algorithm is implemented. 
Computer-simulation and experimental studies are conducted to 
investigate the computational efficiency and numerical accuracy
of the developed algorithms.

\noindent{\bf Results:}
The GPU implementations improve the computational efficiency by 
factors of $1,000$, $125$, and $250$ for the FBP algorithm 
and the two pairs of projection/backprojection
operators, respectively.
Accurate images are reconstructed by use of the FBP and 
iterative image reconstruction algorithms from both 
computer-simulated and experimental data.

\noindent{\bf Conclusions:}
Parallelization strategies for 3D OAT image reconstruction 
are proposed for the first time.  
These GPU-based implementations significantly reduce the computational time
for 3D image reconstruction, complementing our earlier work on 3D OAT iterative 
image reconstruction. 
% \rd{[considering mentioning SIMD]}
\end{abstract}

% make the title area
\maketitle

\section{Introduction}
\label{sect:intro}
Optoacoustic tomography (OAT), also known as photoacoustic computed tomography, 
is an emerging imaging modality that has great potential for a wide range of biomedical imaging applications 
\cite{OraBook,WangTutorial,KrugerMP1999,CoxOSA2006}. 
In OAT, a short laser pulse is employed to irradiate biological tissues. 
When the biological tissues absorb the optical energy, acoustic wave fields can be generated 
via the thermoacoustic effect. 
The acoustic wave fields propagate outward in three-dimensional (3D) space and are measured by use of ultrasonic transducers that are distributed outside the object. 
% Because the distribution of absorbed optical energy is closely related to the physiological and pathological conditions of the tissue
The goal of OAT is to obtain an estimate of the absorbed energy density map within the object from the measured acoustic signals. 
To accomplish this, an image reconstruction algorithm is required. 

A variety of analytic image reconstruction algorithms have been proposed \cite{Kunyansky:07,Finch:02,Xu:2005bp,Xu:planar}. 
These algorithms generally assume an idealized transducer model and an acoustically homogeneous medium. 
Also, since they are based on discretization of continuous reconstruction formulae, 
these algorithms require the acoustic pressure to be densely sampled over a surface that encloses the object to obtain an accurate reconstruction. 
To overcome these limitations, iterative image reconstruction algorithms have been proposed 
\cite{GPaltauf:2002,HJiang:2007,JCarson:2008,Jin:2009,OATTV09:Provost,TMI:transmodel,GuoCS:2010,Huang:SPIEBoundEnhance,xu:water,xu:edema,Ntziachristos:2011,Bu:2012,KunSPIETV:2012,Kun:PMB,HuangchaoJBObrain,German:3D12}. 
Although the optoacoustic wave intrinsically propagates in 3D space, 
when applying to experimental data, most studies 
have employed two-dimensional (2D) imaging models 
by making certain assumptions on the transducer responses and/or the object structures \cite{HJiang:2007,OATTV09:Provost,GuoCS:2010,Huang:SPIEBoundEnhance,Ntziachristos:2011,HuangchaoJBOatten}. 
An important reason is because the computation required for 3D OAT image reconstruction is excessively burdensome. 
Therefore, acceleration of 3D image reconstruction will facilitate 
algorithm development and many applications including real-time 3D PACT
% there remains a great need to develop fast implementations of 
% 3D reconstruction algorithms. 
\cite{Wang:12,Buehler:12}.

A graphics processing unit (GPU) card is a specialized device specifically designed for 
parallel computations \cite{NVIDIA:09arch}. 
Compute unified device architecture (CUDA) is an extension of the C/Fortran language 
that provides a convenient programming platform to exploit the parallel computational 
power of GPUs \cite{CUDAGuide2.0}. 
The CUDA-based parallel programming technique has been successfully applied 
to accelerate image reconstruction in mature imaging modalities such as
X-ray computed tomography (CT) \cite{Zhao09:GPU,Okitsu10:GPU,ChouMP:GPU} 
and magnetic resonance imaging (MRI) \cite{MRIGPU:2008}. 
In OAT, however, only a few works on utilization of GPUs to accelerate image  
reconstruction have been reported \cite{Bu:2012,kwave:toolbox}. 
For example, the k-wave toolbox employs the NVIDIA CUDA Fast Fourier Transform library 
(cuFFT) to accelerate the computation of 3D FFT \cite{kwave:toolbox}. 
Also a GPU-based sparse matrix-vector multiplication strategy has been applied to 3D OAT image reconstruction for the case that the system matrix is sparse and can be stored in memory \cite{Bu:2012}. 
However, there remains an important need to develop efficient implementations of 
OAT reconstruction algorithms for general applications in which the system matrix 
is too large to be stored. 

In this work, we propose parallelization strategies, for use with GPUs, 
to accelerate 3D image reconstruction in OAT. 
Both filtered backprojection (FBP) and iterative image reconstruction algorithms are 
investigated. 
For use with iterative image reconstruction algorithms, we focus on the parallelization of 
projection and backprojection operators. 
Specifically, we develop two pairs of projection/backprojection operators that correspond to 
two distinct discrete-to-discrete (D-D) imaging models employed in OAT, 
namely the interpolation-based and the spherical-voxel-based D-D imaging models. 
Note that our implementations of the backprojection operators compute the exact adjoint of the 
forward operators, and therefore the projector pairs are `matched' \cite{BarrettBook}. 
%Note that our implementation of each projection and backprojection pair are 
%exactly adjoint, ensuring the convergence when applying iterative image reconstruction 
%algorithms \cite{BarrettBook}. 

The remainder of the article is organized as follows.
In Section \ref{sect:background}, we briefly review OAT imaging models in 
their continuous and discrete forms.   
We propose GPU-based parallelization strategies in Section \ref{sect:GPUmethods}. 
Numerical studies and results are described in Section \ref{sect:Nummethods} 
and Section \ref{sect:results} respectively. 
Finally, a brief discussion and summary of the proposed algorithms are provided in 
Section \ref{sect:summary}. 

\section{Background}
\label{sect:background}
\subsection{Continuous-to-continuous imaging models and analytic image reconstruction 
algorithms
}
A continuous-to-continuous (C-C) OAT imaging model neglects sampling effects and 
provides a mapping from the absorbed energy density function $A(\mathbf r)$ 
to the induced acoustic pressure function $p(\mathbf r^s, t)$. 
Here, $t$ is the temporal coordinate, 
$\mathbf r \in V$ and $\mathbf r^s \in S $ denote the locations within the object support $V$ 
and on the measurement surface $S$, respectively. 
A canonical OAT C-C imaging model can be expressed as 
\cite{LVbook:optics,OraBook,TMI:transmodel}:
\begin{equation}\label{eqn:fwdpcc}
 p(\mathbf r^s,t) = {\frac{\beta}{4\pi C_p} } 
   \int_V\!\! d \mathbf{r}\, A(\mathbf{r}) 
    {\frac{d}{dt}} 
    \frac{\delta\left(t-{
         \frac{\vert \mathbf r^s-\mathbf{r}\vert}{ c_0}
       }\right)}
  {\vert \mathbf r^s-\mathbf r\vert}
  \equiv \mathcal H_{\rm CC}A
,
\end{equation}
where $\delta (t)$ is the Dirac delta function, 
$\beta$, $c_0$, and $C_p$ denote the thermal coefficient
of volume expansion, (constant) speed-of-sound,
and the specific heat capacity of the medium at constant pressure, respectively.
We introduce an operator notation $\mathcal H_{\rm CC}$ to denote this C-C mapping.

Alternatively, Eqn.\ \eqref{eqn:fwdpcc} can be reformulated as the well-known 
spherical Radon transform (SRT) \cite{XuReview,Jin:2009}:
\begin{equation}\label{eqn:srt}
  g(\mathbf r^s,t)=\int_V\!\! d \mathbf r\, 
     A(\mathbf r) \delta(c_0t-|\mathbf r^s-\mathbf r |),
\end{equation}
where the function $g(\mathbf r^s, t)$ is related to $p(\mathbf r^s, t)$ as 
\begin{equation}\label{eqn:g2p}
  p(\mathbf r^s, t)=\frac{\beta}{4\pi C_p}
     \frac{\partial}{\partial t}\Big(\frac{g(\mathbf r^s, t)}{t}\Big). 
\end{equation}
The SRT model provides an intuitive interpretation of each value of $g(\mathbf r^s, t)$ 
as a surface integral of $A(\mathbf r)$ over a sphere centered 
at $\mathbf r^s$ with radius $tc_0$.

Based on C-C imaging models, a variety of analytic image 
reconstruction algorithms have been developed 
\cite{Kunyansky:07, Finch:02, Xu:2005bp, Xu:planar}.
For the case of a spherical measurement geometry, an FBP algorithm 
in its continuous form is given by \cite{Finch:02}:
\begin{equation}\label{eqn:fbp}
  A(\mathbf r) = -\frac{C_p}{2\pi \beta c_0^2 R^s}
   \int_{S}\!\! d \mathbf r^s\, 
   \Big[
       \frac{2p(\mathbf r^s, t)}
            {|\mathbf r-\mathbf r^s|}
        + \frac{1}{c_0}
           \frac{\partial p(\mathbf r^s, t)}
           {\partial t} 
        \Big]_{t=\frac{|\mathbf r-\mathbf r^s|}{c_0}},
\end{equation}
where $R^s$ denotes the radius of the measurement surface $S$. 

\subsection{Discrete-to-discrete (D-D) imaging models and iterative image reconstruction algorithms}

When sampling effects are considered, an OAT system is properly described 
as a continuous-to-discrete (C-D) imaging model
\cite{BarrettBook,MarkBookChapter,TMI:transmodel,Ntziachristos:CaliTrans11,Kun:PMB}: 
\begin{equation}\label{eqn:CDmodel}
  \big[\mathbf u\big]_{qK+k} = 
      h^e(t) *_t \frac{1}{S_q} \int_{S_q}\!\! d \mathbf r^s\,
     p(\mathbf r^s, t)\Big|_{t=k\Delta_t},\quad 
  \substack{ q=0,1,\cdots, Q-1\\ 
   k=0,1,\cdots, K-1} , 
\end{equation}
where
$Q$ and $K$ denote the total numbers of transducers (indexed by $q$) and 
the time samples (indexed by $k$) respectively. 
$S_q$ is the surface area of the $q$-th transducer, which is assumed to be a subset of 
$S$;
$h^e(t)$ denotes the acousto-electric impulse response (EIR) of each transducer
that, without loss of generality, is assumed to be identical 
for all transducers; 
`$*_t$' denotes a linear convolution with respect to time coordinate; 
and 
$\Delta_t$ is the temporal sampling interval.  
The vector $\mathbf u$ represents the lexicographically ordered 
measured voltage signals 
whose $(qK+k)$-th element is denoted by $[\mathbf u]_{qK+k}$. 

In order to apply iterative image reconstruction algorithms
a D-D imaging model 
is required, which necessitates the discretization of $A(\mathbf r)$. 
The following $N$-dimensional representation of the object 
function can be employed \cite{BarrettBook, MarkBookChapter}: 
\begin{equation}\label{eqn:dis_obj}
  A(\mathbf r) \approx \sum_{n=0}^{N-1} [\boldsymbol \alpha]_n 
    \psi_n(\mathbf r), 
\end{equation}
where 
$\boldsymbol\alpha$ is a coefficient vector whose $n$-th element is denoted by 
$[\boldsymbol\alpha]_n$ and $\psi_n(\mathbf r)$ is the expansion function.
On substitution from Eqn.\ \eqref{eqn:dis_obj} into Eqn.\ \eqref{eqn:CDmodel},
where $p(\mathbf r^s, t)$ is defined by Eqn.\ \eqref{eqn:fwdpcc},
one obtains 
a D-D mapping from $\boldsymbol\alpha$ to $\mathbf u$, expressed as 
\begin{equation}\label{eqn:DDmodel}
  \mathbf u \approx \mathbf H \boldsymbol \alpha, 
\end{equation}
where each element of the matrix $\mathbf H$ is defined as 
\begin{equation}\label{eqn:imagoptor}
 \big[\mathbf H\big]_{qK+k,n} = \big[
          h^e *_t   \frac{1}{S_q} 
       \int_{S_q}\!\!d \mathbf r^s 
       \mathcal H_{\rm CC} \psi_n
       \big]_{t=k\Delta_t}.
\end{equation}
Here, $\mathbf H$ is the D-D imaging operator also known as system matrix
or projection operator. 
Note that the `$\approx$' in Eqn.\ \eqref{eqn:DDmodel} is due to the use of   
the finite-dimensional representation of the object function 
(i.e., Eqn.\ \eqref{eqn:dis_obj}). 
No additional approximations have been introduced. 

Below we describe two types of D-D imaging models that have been employed in OAT 
\cite{Ntziachristos:CaliTrans11,TMI:transmodel,Kun:PMB}: 
the interpolation-based imaging model and the spherical-voxel-based imaging model. 
The quantities $\mathbf u$, $\mathbf H$, and $\boldsymbol \alpha$ (or $\psi_n$) in the two models will be distinguished by the subscripts (or superscripts) `int' and `sph', respectively.

% \noindent {\it B.1 Interpolation-based D-D imaging model}
\subsubsection{Interpolation-based D-D imaging model}

The interpolation-based D-D imaging model defines the coefficient vector as samples  
of the object function on the nodes of a uniform Cartesian grid: 
\begin{equation}\label{eqn:coeffint}
  \big[\boldsymbol\alpha_{\rm int}\big]_n = \int_V\!\!d
           \mathbf r\, \delta(\mathbf r-\mathbf r_n)
           A(\mathbf r),\quad n=0,1,\cdots,N-1, 
\end{equation}
where, $\mathbf r_n=(x_n,y_n,z_n)^T$ specifies the location of the $n$-th node of the uniform 
Cartesian grid. 
The definition of the expansion function depends on the choice of interpolation method
\cite{Ntziachristos:2011}.
If a trilinear interpolation method is employed, the expansion function can be 
expressed as \cite{KakBook}: 
\begin{equation}\label{eqn:expfunI}
  \psi^{\rm int}_n(\mathbf r)=\left\{\begin{array}{ll}
                    (1-\frac{|x-x_n|}{\Delta_s})
                    (1-\frac{|y-y_n|}{\Delta_s})
                    (1-\frac{|z-z_n|}{\Delta_s}), & \text{if}\,
  |x-x_n|, |y-y_n|, |z-z_n| \leq  \Delta_s\\
                0, & \text{otherwise}
                  \end{array}\right.,
\end{equation}
where $\Delta_s$ is the distance between two neighboring grid points.

In principle, the interpolation-based D-D imaging model can be constructed by 
substitution from Eqns.\ \eqref{eqn:coeffint} and \eqref{eqn:expfunI} 
to Eqn.\ \eqref{eqn:imagoptor}. 
In practice, however, implementation of the surface integral over $S_q$ 
is difficult 
for the choice of expansion functions in Eqn.\ \eqref{eqn:expfunI}. 
Also, implementations of the temporal convolution and 
$\mathcal H_{\rm CC}\psi_n^{\rm int}$ usually require extra discretization
procedures. 
Therefore, utilization of the interpolation-based D-D model commonly 
assumes the transducers to be point-like.
In this case, 
the implementation of $\mathbf H_{\rm int}$ is decomposed as a three-step operation: 
\begin{equation}
  \mathbf u_{\rm int} = \mathbf H_{\rm int}\boldsymbol \alpha_{\rm int} 
     \equiv \mathbf H^e \mathbf D \mathbf G \boldsymbol \alpha_{\rm int}, 
\end{equation}
where $\mathbf G$, $\mathbf D$, and  $\mathbf H^e$
are discrete approximations of the SRT (Eqn.\ \eqref{eqn:srt}), 
the differential operator (Eqn.\eqref{eqn:g2p}), 
and 
the operator that implements a temporal convolution with EIR, respectively. 
We implemented $\mathbf G$ 
in a way 
\cite{AnastasioTATHT,AnastasioTATHTfeasible,Jin:2009}
that is similar to the `ray-driven' implementation of 
Radon transform in X-ray CT \cite{KakBook}, 
i.e, for each data sample, we accumulated the contributions from the voxels that resided on the 
spherical shell specified by the data sample. 
By use of Eqns. \eqref{eqn:srt}, \eqref{eqn:dis_obj}, \eqref{eqn:coeffint}, and 
\eqref{eqn:expfunI}, one obtains:   
\begin{equation}\label{eqn:dis_srt}
\big[\mathbf G \boldsymbol\alpha_{\rm int}\big]_{qK+k}
      = \Delta_s^2 
      \sum_{n=0}^{N-1}
      \big[\boldsymbol \alpha_{\rm int}\big]_n
      \sum_{i=0}^{N_i-1} 
      \sum_{j=0}^{N_j-1}
      \psi_n^{\rm int}(\mathbf r_{k,i,j})
      \equiv 
  \big[\mathbf g\big]_{qK+k} ,
\end{equation}
where $[\mathbf g]_{qK+k} \approx g(\mathbf r^s_q, t)|_{t=k\Delta_t}$ 
with $\mathbf r_q^s$ specifying the location of the $q$-th point-like transducer,
and 
$N_i$ and $N_j$ denote the numbers of divisions over the two angular coordinates of a local 
spherical coordinate system shown in Fig.\ \ref{fig:geo}-(b).  
A derivation of Eqn.\ \eqref{eqn:dis_srt} is provided in Appendix. 
The differential operator in Eqn.\ \eqref{eqn:g2p} is approximated 
as  
\begin{equation}\label{eqn:dis_g2p}
\big[\mathbf D \mathbf g\big]_{qK+k}
= \frac{\beta}{8\pi C_p\Delta_t^2}
 \Big(\frac{[\mathbf g]_{qK+k+1}}{k+1}
     -\frac{[\mathbf g]_{qK+k-1}}{k-1}
 \Big)
    \equiv 
\big[\mathbf p_{\rm int}\big]_{qK+k}, 
\end{equation}
where $[\mathbf p_{\rm int}]_{qK+k} \approx p(\mathbf r^s_q, t)|_{t=k\Delta_t}$. 
Finally, the continuous temporal convolution is approximated by a discrete linear convolution as 
\cite{ClaerboutBook}
\begin{equation}\label{eqn:EIR}
  \big[\mathbf H^e \mathbf p_{\rm int}\big]_{qK+k} = 
  \sum_{\kappa = 0}^{K-1}  [ \mathbf h^e ]_{k-1-\kappa}  [\mathbf p_{\rm int}]_{qK+\kappa}
  \equiv [\mathbf u_{\rm int}]_{qK+k}, 
\end{equation}
where 
$[\mathbf h^e]_{k} = \Delta_t h^e(t)|_{t=k\Delta_t}$. 

% \vskip 0.3cm
% \noindent{\it B.2 Spherical-voxel-based D-D imaging model}
% \vskip 0.3cm

\subsubsection{Spherical-voxel-based D-D imaging model}

The spherical-voxel-based imaging model is also widely employed in OAT 
\cite{Khokhlova:07,GPaltauf:2002,JCarson:2009,TMI:transmodel,Kun:PMB}. 
It employs the expansion functions 
\begin{equation}\label{eqn:basissph}
  \psi^{\rm sph}_n(\mathbf r) = \left\{\begin{array}{ll}
                1, & \text{if}\quad|\mathbf r - \mathbf r_n| \leq \Delta_s/2 \\
                0, & \text{otherwise} 
                \end{array}\right. ,
\end{equation}
where $\mathbf r_n$ is defined as in Eqn.\ \eqref{eqn:coeffint}. 
The $n$-th expansion function $\psi^{\rm sph}_n (\mathbf r)$ is a uniform sphere that is 
inscribed by the $n$-th cuboid of a Cartesian grid. 
The $n$-th component of the coefficient vector 
$\boldsymbol \alpha_{\rm sph}$ is defined as:
\begin{equation}\label{eqn:coeffsph}
  \big[\boldsymbol \alpha_{\rm sph}\big]_n = \frac{V_{\text{cube}}}{V_{\text{sph}}}
              \int_{V}\!\! d\mathbf{r} \;
               \psi^{\rm sph}_n (\mathbf{r})A(\mathbf{r}), 
\end{equation}
where $V_{\text{cube}}$ and $V_{\text{sph}}$ are the volumes of a cubic voxel of dimension $\Delta_s$ and of a spherical voxel of radius $\Delta_s/2$ respectively.

Unlike the interpolation-based imaging model, 
by use of the expansion functions defined in Eqn.\ \eqref{eqn:basissph}, 
the surface integral over $S_q$ and $\mathcal H_{\rm CC}\psi_n^{\rm sph}$  
in Eqn.\ \eqref{eqn:imagoptor} can be converted to a temporal convolution
and calculated analytically \cite{TMI:transmodel,Kun:PMB}.
To avoid utilizing excessively high sampling rate to mitigate aliasing, 
the spherical-voxel-based imaging model can be conveniently implemented in the 
temporal frequency domain as \cite{Kun:PMB}: 
\begin{equation}\label{eqn:dis_sph}
 \big[\tilde{\mathbf u}_{\rm sph}\big]_{qL+l}=\tilde p_0(f) 
  \sum_{n=0}^{N-1}\big[\boldsymbol\alpha_{\rm sph}\big]_n
  \frac{1}{S_q}
  \tilde h_q^s(\mathbf r_n, f)
  \Big|_{f=l\Delta_f}, \quad
  {\rm for}\,  l=0, 1, \cdots, L-1, 
\end{equation}
where $\Delta_f$ is the frequency sampling interval,
and $L$ denotes the total number of temporal-frequency samples indexed by $l$.
A derivation of Eqn.\ \eqref{eqn:dis_sph} can be found in Ref.\ \onlinecite{Kun:PMB}.
The function  $\tilde h_q^s(\mathbf r_n, f)$ represents the temporal Fourier transform
of the spatial impulse response (SIR) of the $q$-th transducer for the source 
located at $\mathbf r_n$, 
expressed as: 
\begin{equation}
  \tilde h_q(\mathbf r_n, f) = 
    \int_{S_q}\!\! d \mathbf r^s\,
    \frac{ \exp(-\hat{\jmath}\,2\pi f\frac{|\mathbf r^s-\mathbf r_n|}{c_0})}
      {2\pi |\mathbf r^s-\mathbf r_n|}. 
\end{equation}
Also, $\tilde p_0(f)$ is defined as 
\begin{equation}\label{eqn:p0f}
  \tilde p_0(f) = -\hat\jmath \frac{\beta c_0^3}{C_pf}
    \Bigg[
    \frac{\Delta_s}{2c_0}\cos\big(\frac{\pi f\Delta_s}{c_0}\big)
   -\frac{1}{2\pi f}\sin\big(\frac{\pi f\Delta_s}{c_0}\big)
    \Bigg]
     \tilde{h}^e(f), 
\end{equation}
where $\tilde{h}^e(f)$ is the EIR in temporal-frequency domain. 
% However, this advantage is accompanied by a significant increase in computational complexity. 
In summary, the imaging model can be expressed in matrix form as: 
\begin{equation}\label{eqn:ddmodelsp}
  \tilde {\mathbf u}_{\rm sph} = \mathbf H_\text{sph} 
      \boldsymbol \alpha_{\rm sph}. 
\end{equation}

% \vskip 0.3cm
% \noindent{\it B.3 Iterative image reconstruction algorithms and adjoint operators}
% \vskip 0.3cm
\subsubsection{Adjoints of the system matrices}

Iterative image reconstruction algorithms employ numerical 
implementations of the 
projection operator, i.e., the system matrix $\mathbf H$, as well as its adjoint, denoted 
by $\mathbf H^\dagger$\ \ \cite{Wernickbookchap21}. 
The adjoint is also referred to as the backprojection operator. 
Note that for most practical applications, $\mathbf H$ and $\mathbf H^\dagger$ are too large 
to be stored in the random access memory of currently available computers.
Therefore, in practice, the actions of 
$\mathbf H$ and $\mathbf H^\dagger$ are almost always 
calculated on the fly.
The same strategy was adopted in this work. 

According to the definition of the adjoint operator \cite{BarrettBook,ClaerboutBook}, 
$\mathbf H_{\rm int}^\dagger = \mathbf G^\dagger \mathbf D^\dagger \mathbf {H^e}^\dagger$, 
where
\begin{equation}
  \big[ \mathbf {H^e}^\dagger \mathbf u_{\rm int}  \big]_{qK+k} = 
  \sum_{\kappa=0}^{K-1} 
     [\mathbf h^e]_{\kappa-1-k} 
     [\mathbf u]_{qK+\kappa}
  \equiv [\mathbf p'_{\rm int}]_{qK+k}, 
\end{equation}
\begin{equation}
  \bigg[\mathbf D^\dagger \mathbf p'_{\rm int}\bigg]_{qK+k}=\frac{\beta}{8\pi C_p\Delta_t^2k}
      \Big( \big[\mathbf p'_{\rm int}\big]_{qK+k-1}
            -\big[\mathbf p'_{\rm int}\big]_{qK+k+1}\Big)
    \equiv \big[\mathbf g'\big]_{qK+k}, 
\end{equation}
and 
\begin{equation}
  \bigg[\mathbf G^\dagger \mathbf g'\bigg]_{n} = \Delta_s^2
     \sum_{q=0}^{Q-1} \sum_{k=0}^{K-1} \big[\mathbf g'\big]_{qK+k}
        \sum_{i=0}^{N_i-1}\sum_{j=0}^{N_j-1}\psi_n^{\rm int} (\mathbf r_{k,i,j})
    \equiv [\boldsymbol\alpha'_{\rm int}]_n.
\end{equation}
It can also be verified that the adjoint operator $\mathbf H_{\rm sph}^\dagger$ is given by: 
\begin{equation}\label{eqn:sphbwd}
  \bigg[\mathbf H^{\dagger}_{\rm sph} \tilde{\mathbf u}_{\rm sph}\bigg]_{n}
     = \sum_{q=0}^{Q-1}\sum_{l=0}^{L-1} 
      \big[\tilde{\mathbf u}_{\rm sph}\big]_{qL+l} 
       \tilde{p}^*_0(f)
       \tilde{h}^*_q(\mathbf r_n, f)\Big|_{f=l\Delta_f},
\end{equation}
where the superscript `*' denotes the complex conjugate. 
Unlike the unmatched backprojection opertors 
\cite{Zeng:unmatch}
that are obtained by discretization of 
the continuous adjoint operator, utilization of the exact adjoint operator facilitates 
the convergence of iterative image reconstruction algorithms. 

\subsection{GPU architecture and CUDA programming}

The key features of GPU architecture and the 
basics of CUDA programming are briefly summarized in this section. 
We refer the readers to Refs. \onlinecite{CUDAGuide2.0,NVIDIA:09arch} for 
additional details.
% a comprehensive description on the topic. 

A GPU card contains multiple streaming multiprocessors. 
Each streaming multiprocessor is configured with multiple processor cores. 
For example,
the Tesla C1060  possesses $30$ streaming multiprocessors with $8$ processor cores on each;
and the Tesla C2050 possesses $14$ streaming multiprocessors with $32$ processor cores on each
\cite{NVIDIA:09arch}. 
The processor cores in each multiprocessor execute the same instruction on different 
pieces of data, which is referred to as ``single instruction, multiple data'' (SIMD) model
of parallel programming. 
In order to fully exploit the computing power of GPUs, one of the major challenges is to 
design a parallelization strategy fitting in the SIMD framework such that the largest number of 
processor cores can execute the computation simultaneously \cite{CUDAGuide2.0}. 

A GPU card has six types of memory that have varying capacities and different 
access rules and efficiencies: 
(1) Registers are assigned for each thread and have the fastest access.   
(2) Shared memory is assigned for each block and can be efficiently accessed by all threads in the block 
if designed appropriately. 
(3) Constant memory is read-only and can be accessed by all threads efficiently.  
(4) Texture memory is also read-only and is optimized for interpolation operations. 
(5) Global memory has the slowest access that takes hundreds times more clock cycles than does the 
computation of basic arithmetic operations.
(6) Local memory is assigned for each thread but has a slow access as 
does the global memory.
Therefore, an efficient GPU-based implementation in general requires a limited number of global and local memory access.

CUDA is a platform and programming model
developed by NVIDIA that includes a collection of functions and keywords 
to exploit the parallel computing power of GPUs \cite{CUDAGuide2.0}. 
A CUDA parallel program is composed of a host program and kernels.
The host program is executed by CPUs and launches the kernels, which 
are custom-designed functions executed by GPUs. 
A general parallel programming strategy is to launch multiple instances of a kernel 
and to run the multiple instances concurrently on GPUs. 
In CUDA, each instance of the kernel is named as a thread and processes only a portion of 
the data. A hierarchy of threads is employed: 
Threads are grouped into blocks, and blocks are grouped into a grid. 
Therefore, each thread is specified by a multi-index containing a
block index and a thread index within the block.

\section{GPU-accelerated reconstruction algorithms}
\label{sect:GPUmethods}

In this section, we propose GPU-based parallelization strategies for
 the FBP algorithm  
and the projection/backprojection operations corresponding to
 the interpolation-based 
and the spherical-voxel-based D-D imaging models. 
% These strategies are described assuming an idealized transducer model, i.e., Eqn.\ \eqref{eqn:idealsample}.
% However, when transducer responses are considered, only minor modifications need to be introduced 
% as described in Appendix-B. 

\subsection{Measurement geometry}
\begin{figure}[ht]
\centering
\hskip 3cm
\subfigure[]{\includegraphics[width=10cm]{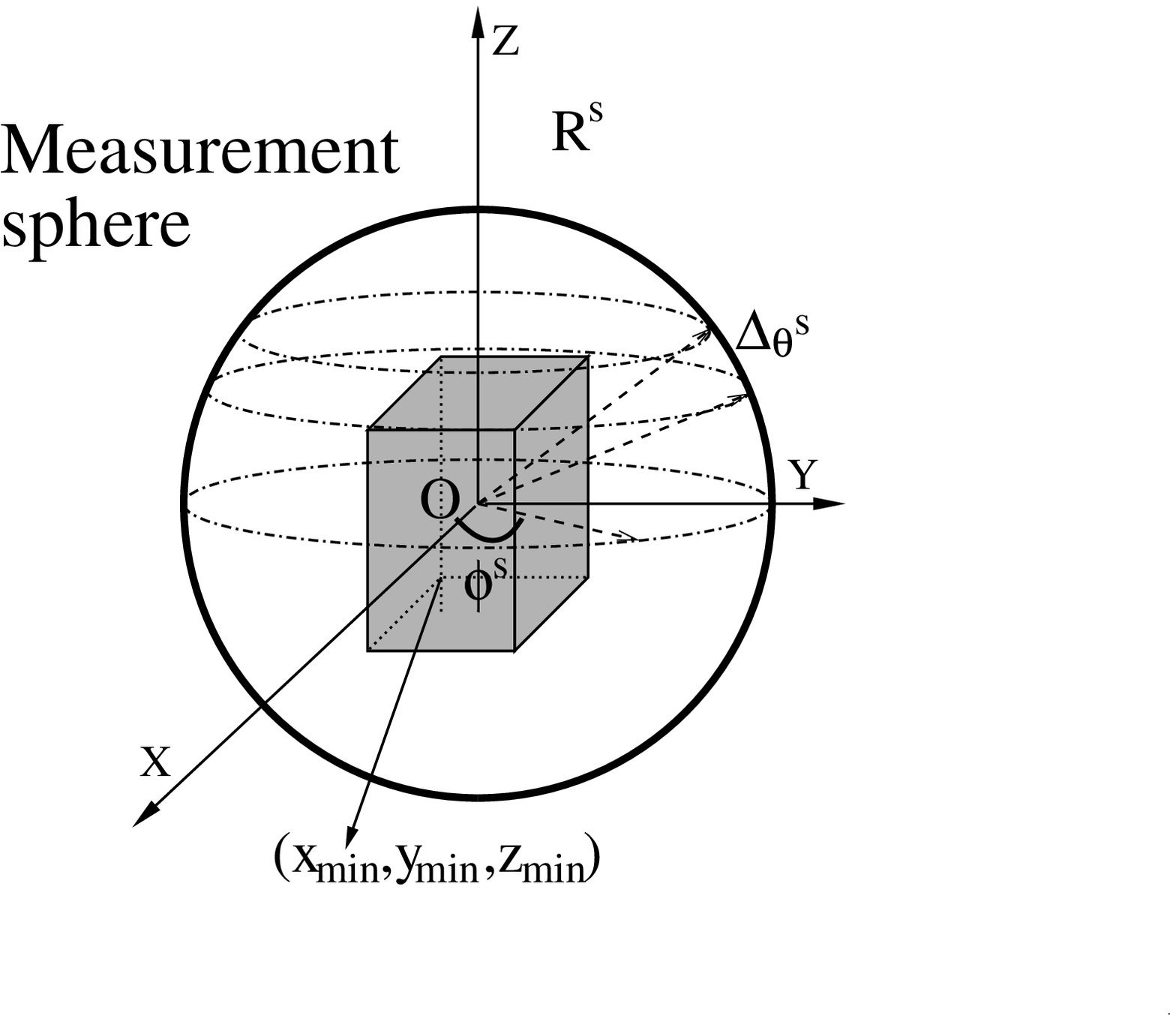}}\\
\subfigure[]{\includegraphics[width=7.5cm]{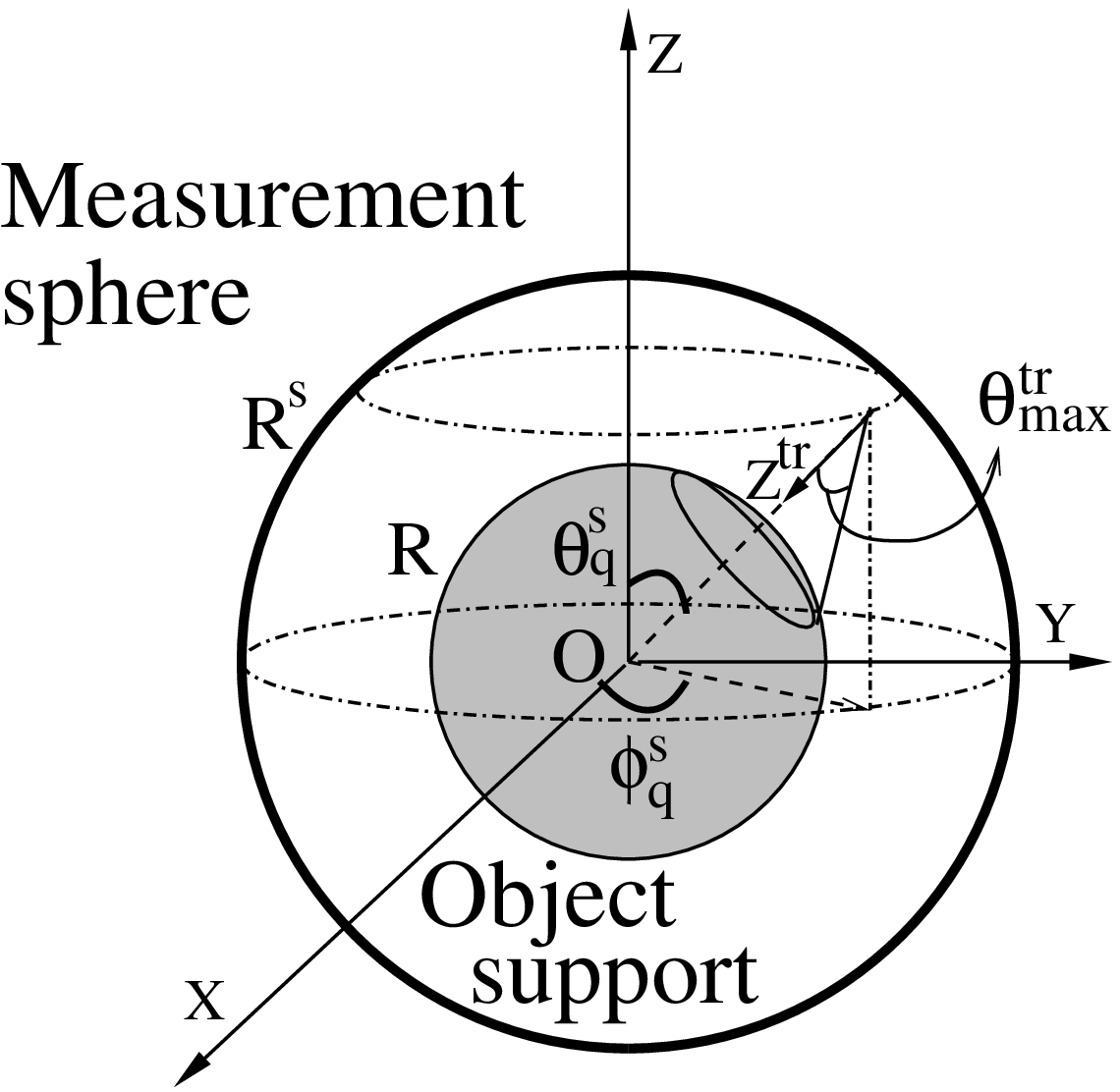}}
\caption{\label{fig:geo}
}
\end{figure}

We employed a spherical measurement geometry shown in Fig.\ \ref{fig:geo}-(a). 
The measurement sphere was of radius $R^s$ centered at the origin of the Cartesian coordinate system (or the equivalent spherical coordinate system). 
The polar angle $\theta^s\in[0,\pi]$ was equally divided with interval $\Delta_{\theta^s}=\pi/N_r$, starting from $\theta^s_{\rm min}$. 
At each polar angle, a ring on the sphere that was parallel to the plane $z=0$ can be specified, resulting $N_r$ rings. 
On each ring, $N_v$ ultrasonic transducers were assumed to be 
uniformly distributed with azimuth angle interval $\Delta_{\phi^s}=2\pi/N_v$. 
Hereafter, each azimuth angle will be referred to as a tomographic view.
At each view, we assumed that $N_t$ temporal samples were acquired and the first sample corresponded to time instance $t_{\rm min}$. 
For implementations in temporal-frequency domain, we assumed 
that $N_f$ temporal-frequency samples were available and the 
first sample corresponded to $f_{\rm min}$. 
The region to be reconstructed was a rectangular cuboid 
whose edges were parallel to the axes of the coordinate system 
and the left-bottom-back vertex was located at  
$(x_{\rm min}, y_{\rm min}, z_{\rm min})$. 
The numbers of voxels along the three coordinates will be denoted 
by $N_x$, $N_y$, and $N_z$, 
respectively, totally $N=N_xN_yN_z$ voxels. 
We also assumed the cuboid was contained in another sphere of radius $R$ 
that was concentric with the measurement sphere shown in Fig.\ \ref{fig:geo}-(b).

\subsection{Implementation of the FBP algorithm}

Central processing unit (CPU)-based implementations of continuous FBP formulae
%, i.e., Eqn.\ \eqref{eqn:fbp}, 
have been described in Refs. \onlinecite{Kunyansky:07,Finch:02,Xu:2005bp,Xu:planar}. 
Though the discretization methods vary, in general, three approximations have to be employed. 
Firstly, the first-order derivative term $\partial p(\mathbf r^s, t)/\partial t$ 
has to be approximated by a difference scheme up to certain order \cite{MortonBook:2005}. 
Secondly, the measurement sphere has to be divided into small patches, 
and the surface integral has to be approximated by a summation of the area of every patch weighted by the effective value of the integrand on the patch. 
Finally, the value of the integrand at an arbitrary time instance $t=|\mathbf r^s-\mathbf r|/c_0$ has 
to be approximated by certain interpolation method.
% since in practice only finite number of time samples of the pressure function are available. 

In this study, we approximated the surface integral by use of the trapezoidal rule. 
As described earlier, the spherical surface was divided into $N_r N_v$ patches.
For the transducer indexed by $q$ that was located at $\mathbf r^s_q=(R^s, \theta^s_q, \phi^s_q)$, 
the area of the patch was approximated by $(R^s)^2\Delta_{\theta^s}\Delta_{\phi^s}\sin\theta^s_q$. 
The value at time instance $t=|\mathbf r^s_q-\mathbf r_n|/c_0$ was approximated 
by the linear interpolation from its two neighboring samples as:
\begin{equation}\label{eqn:linearint}
  p(\mathbf r^s_q, t)\Big|_{t=\frac{|\mathbf r^s_q-\mathbf r_n|}{c_0}} \approx
   \big(k+1-\tilde k\big)\big[\mathbf p\big]_{qK+k} 
  +\big(\tilde k-k\big)\big[\mathbf p\big]_{qK+k+1},
\end{equation}
where $\tilde k = (|\mathbf r^s_q-\mathbf r_n|/c_0-t_{\rm min})/\Delta_t$, 
and $k$ is the integer part of $\tilde k$.  
Here $\mathbf p$ is a vector of lexicographically ordered samples of the 
pressure function $p(\mathbf r^s, t)$, which is estimated from the measured voltage 
data vector $\mathbf u$. 
Also, the first-order derivative term was approximated by: 
\begin{equation}
  \frac{\partial}{\partial t}p(\mathbf r^s_q, t) \Big|_{t=\frac{|\mathbf r^s_q-\mathbf r_n|}{c_0}}
\approx \frac{1}{\Delta_t}\Big(
     \big[\mathbf p\big]_{qK+k+1} - \big[\mathbf p\big]_{qK+k}
       \Big). 
\end{equation} 
By use of these three numerical approximations, the discretized FBP formula was expressed as: 
\begin{equation}\label{eqn:disfbp}
\begin{split}
  \big[\hat{\boldsymbol\alpha}_{\rm fbp}\big]_n = 
  -\frac{C_pR^s\Delta_{\theta^s}\Delta_{\phi^s}}{\pi\beta c_0^3\Delta_t}
   \sum_{n_r=0}^{N_r-1}\sin \theta^s_q
   &\sum_{n_v=0}^{N_v-1}
   \bigg\{
     \Big(1.5-\frac{ k+t_{\rm min}/\Delta_t }
                 {\tilde k +t_{\rm min}/\Delta_t}\Big)\big[\mathbf p\big]_{qK+k+1}\\
    &+\Big(\frac{ k+1+t_{\rm min}/\Delta_t }
                 {\tilde k +t_{\rm min}/\Delta_t}-1.5\Big)\big[\mathbf p\big]_{qK+k}  
   \bigg\}.
\end{split}
\end{equation}
Unlike the implementations of FBP formulas in X-ray cone beam CT \cite{Okitsu10:GPU,ChouMP:GPU}, 
we combined the filter and the linear interpolation. 
This reduced the number of visits to the global memory in the GPU implementation 
described below.
% with little increase on the computational complexity. 

We implemented the FBP formula in a way that is similar to the `pixel-driven' implementation 
in X-ray CT \cite{ChouMP:GPU}, i.e., we assigned each thread to execute the two accumulative summations
in Eqn.\ \eqref{eqn:disfbp} for each voxel. 
We bound the pressure data $\mathbf p$ to 
texture memory because it is cached and has a faster accessing rate. 
Therefore our implementation only requires access to texture memory twice and to global memory once. 
The pseudo-codes are provided in Algs.\ \ref{alg:fbp} and \ref{alg:K_fbp} 
for the host part and the device part respectively. 
Note that the pseudo-codes do not intend to be always optimal because the performance of the codes could 
depend on the dimensions of $\mathbf p$ and $\hat{\boldsymbol\alpha}_{\rm fbp}$. 
For example, we set the block size to be $(N_z, 1, 1)$ because for our applications,
$N_z$ was bigger than $N_x$ and $N_y$ and smaller than the limit number of threads that a block can 
support (i.e., 1024 for the NVIDIA Tesla C2050). 
If the values of $N_x$, $N_y$, and $N_z$ change, we may need to redesign the dimensions of the grid and blocks. 
However, the general SIMD parallelization strategy remains.

\subsection{Implementation of $\mathbf H_{\rm int}$ and $\mathbf H^\dagger_{\rm int}$}

The forward projection operation $\mathbf H_{\rm int}\boldsymbol\alpha_{\rm int}$ is composed of 
three consecutive operations $\mathbf g = \mathbf G\boldsymbol\alpha_{\rm int}$, 
$\mathbf p_{\rm int} = \mathbf D\mathbf g$,
and $\mathbf u_{\rm int} = \mathbf H^e \mathbf p_{\rm int}$ 
that are defined 
in Eqns.\ \eqref{eqn:dis_srt}, \eqref{eqn:dis_g2p}, and \eqref{eqn:EIR}, respectively.
Both the difference operator $\mathbf D$ and the one-dimensional (1D) convolution $\mathbf H^e$
have low computational complexities
while the SRT operator $\mathbf G$ is computationally burdensome. 
Hence, we developed the GPU-based implementation of $\mathbf G$ while leaving $\mathbf D$ 
and $\mathbf H^e$ to be implemented by CPUs.

The SRT in OAT shares many features with the Radon transform in X-ray CT. 
Thus, our GPU-based implementation is closely related to the implementations of Radon transform that have been optimized for X-ray CT\cite{Zhao09:GPU,Okitsu10:GPU,ChouMP:GPU}. 
The surface integral was approximated according to the trapezoidal rule. 
Firstly, the integral surface was divided into small patches, which is described in the Appendix. 
Secondly, each patch was assigned an effective value of the object function by trilinear interpolation. 
The trilinear interpolation was calculated by use of the texture memory of GPUs that 
is specifically designed for interpolation. 
Finally, GPU threads accumulated the areas of patches weighted by the effective values of the object function and wrote the final results to global memory. 
The pseudo-codes for implementation of $\mathbf G$ 
are provided in Algs. \ref{alg:intH} and \ref{alg:K_intH} 
for the host part and the device part, respectively.
Note that we employed the ``one-level''-strategy  \cite{ChouMP:GPU}, 
i.e., each thread calculates one data sample. 
Higher level strategies have been proposed to improve the performance by assigning each block 
to calculate multiple data samples \cite{ChouMP:GPU}, which, however, caused many thread idles in OAT 
mainly because the amount of computation required to calculate a data sample varies largely 
among samples for SRT. 
% the SRT is distinct from Radon transform in the way that the amount of computation varies among data samples at different time instances.  
% Assuming that the object is supported in a sphere, the integral surface was a spherical cap 
% determined by the intersection of the object's spherical support and the sphere of radius $tc_0$ that was centered at the transducer. 
% It is obvious that at the same time instance, the intersectional spherical caps have the same area for all transducers if the measurement sphere is concentric with the spherical support of the object, suggesting nearly the same amount of computation involved. 
% Therefore, we designed each block to calculate the data samples at the same time instance, effectively avoiding thread idle. 

Implementation of the backprojection operator $\mathbf H_{\rm int}^\dagger$ 
was very similar to the implementation of $\mathbf H_{\rm int}$. 
The operators $\mathbf D^\dagger$ and ${\mathbf H^e}^\dagger$ 
were calculated on CPUs while $\mathbf G^\dagger$ was 
calculated by use of GPUs.
The pseudo-codes are provided in Algs.\ \ref{alg:intHT} and \ref{alg:K_intHT}. 
We made use of the CUDA function `atomicAdd' to add weights to global memory from each thread. 

\subsection{Implementation of $\mathbf H_{\rm sph}$ and $\mathbf H^\dagger_{\rm sph}$ }

Implementation of the forward projection operation for the spherical-voxel-based imaging model is distinct from that of the interpolation-based model. 
The major difference is that calculation of each element of the data vector for 
the spherical-voxel-based imaging model requires the accumulation of the contributions 
from all voxels because the model is expressed in the temporal frequency domain. 
Because of this, the amount of computation required to calculate each data sample 
in the spherical-voxel-based imaging model is almost identical, 
simplifying the parallelization strategy. 

We proposed a parallelization strategy that was inspired by one  
applied in advanced MRI reconstruction \cite{MRIGPU:2008} and is summarized as follows.  
Discrete samples of $\tilde p_0(f)$ 
defined in Eqn.\ \eqref{eqn:p0f} were precalcualted and stored 
as a vector $\tilde{\mathbf p}_0$  
in constant memory. 
Because the size of the input vector $\boldsymbol\alpha_{\rm sph}$ is often too large to 
fit in the constant memory, we divided $\boldsymbol\alpha_{\rm sph}$ into sub-vectors
that matched the capacity of the constant memory. 
We employed a CPU loop to copy every sub-vector sequentially to the constant memory 
and call the GPU kernel function to accumulate a partial summation. 
The major advantage of this design is that the total number of global memory visits to 
calculate one data sample is reduced to the number of sub-vectors.

Implementation of the projection operator for the spherical-voxel-based imaging model
generally involves more arithmetic operations than does the interpolation-based imaging model. 
Moreover, the spherical-voxel-based imaging model has been employed to compensate for the 
finite aperture size effect 
of transducers \cite{TMI:transmodel,Kun:PMB}, which makes the computation even more burdensome. 
Because of this, we further developed an implementation that employed multiple GPUs. 
The pseudo-codes of the projection operation are provided in Algs. \ref{alg:sphH}, \ref{alg:fwd_pthread}, and \ref{alg:Kfwdsph}. 
We created $N_{\rm pth}$ pthreads on CPUs by use of the `pthread.h' library.
Here, we denote the threads on CPUs by `pthread' to distinguish from threads on GPUs.
We divided the input vector $\boldsymbol \alpha_{\rm sph}$ into $N_{\rm pth}$ sub-vectors 
(denoted by $\boldsymbol\alpha_{\rm pth}$'s)
of equal size
and declared an output vector $\tilde{\mathbf u}'_{\rm sph}$ of dimension $N_{\rm pth}N_rN_vN_f$. 
By calling the pthread function `fwd\_pthread', $N_{\rm sph}$ pthreads simultaneously calculated the projection. 
Each pthread projected an $\boldsymbol\alpha_{\rm pth}$ to a partial voltage data vector 
$\tilde{\mathbf u}'_{\rm pth}$ that filled in the larger vector $\tilde{\mathbf u}'_{\rm sph}$.
Once all pthreads finished filling their $\tilde{\mathbf u}'_{\rm pth}$ into $\tilde{\mathbf u}'_{\rm sph}$, 
the projection data $\tilde{\mathbf u}_{\rm sph}$ were obtained by 
a summation of the $N_{\rm pth}$ $\tilde{\mathbf u}'_{\rm pth}$'s. 

Implementation of the backprojection operator was similar except the 
dividing and looping were over the 
vector $\tilde{\mathbf u}_{\rm sph}$ instead of $\boldsymbol\alpha_{\rm sph}$. 
The pseudo-codes for the backprojection operation are provided in Algs. \ref{alg:sphHT}, \ref{alg:bwd_pthread}, and \ref{alg:Kbwdsph}.

\section{descriptions of computer simulation and experimental studies}
\label{sect:Nummethods}
The computational efficiency and accuracy of the proposed GPU-based
implementations of the FBP algorithm and projection/backprojection
operators for use with iterative image reconstruction algorithms
were quantified in computer simulation and experimental OAT imaging 
studies. 

% Both computer-simulation and experimental studies were conducted to investigate the computational 
% efficiency and accuracy of our GPU-based implementations. 
% We implemented a 3D iterative image reconstruction algorithm that utilized the proposed implementations of the projection/backprojection operators. 
% We applied the FBP and iterative image reconstruction algorithms to computer-simulated and experimental 
% 3D OAT measurement data. 

\subsection{Computer-simulation studies}

\noindent{\it Numerical phantom:} The numerical phantom consisted of $9$ uniform spheres 
that were blurred by a 3D Gaussian kernel possessing a full width at half maximum (FWHM) of 
$0.77$-mm. 
The phantom was contained within a cuboid of size $29.4\times29.4\times61.6$-mm$^3$.
A 2D image corresponding to the plane $y=0$ through the phantom is shown in Fig.\ \ref{fig:fbp}-(a). 
\vskip .5cm
\noindent{\it Simulated projection data:} 
The measurement surface was a sphere of radius $R^s=65$-mm.
corresponding to an exsiting OAT imaging system \cite{petermice:jbo,Kun:PMB}.
As described in Section \ref{sect:GPUmethods}, 
ideal point-like transducers were uniformly distributed over $128$ rings and $90$ tomographic views.
The $128$ rings covered the full $\pi$ polar angle, i.e., $\theta^s_{\rm min} = \pi/256$,
while the $90$ views covered the full $2\pi$ azimuth angle. 
The speed of sound was set at $c_0 = 1.54$-mm/$\mu$s. 
We selected the Gr\"uneisen coefficient as $\Gamma=\beta c_0^2/C_p = 2,000$ of arbitrary units (a.u.). 
For each transducer, we analytically calculated $1022$ temporal samples of the pressure function 
at the sampling rate of $f_{\rm sam}=20$-MHz by use of Eqn.\ \eqref{eqn:fwdpcc}.
Because we employed a smooth object function, the pressure data were calculated by 
the following two steps: 
Firstly, we calculated temporal samples of pressure function $p_{\rm us}(\mathbf r^s, t)$ 
that corresponds to the $9$ uniform spheres by \cite{OraBook,LVbook:optics}
\begin{equation}
  p_{\rm us} (\mathbf r^s, t)|_{t=k\Delta_t} =  
\sum_{i=0}^8
    \left\lbrace\begin{array}{ll}
       A_i\Big[- \frac{\beta c_0^3}{C_p |\mathbf r^s - \mathbf r_i|} t
        + \frac{\beta c_0^2}{2} \Big]_{t=k\Delta_t},
         & {\rm if} \; \big|c_0k\Delta_t-|\mathbf r^s -\mathbf r_i| \big| \leq R_i \\
        0, & {\rm otherwise} \end{array}\right.
\end{equation}
where $\mathbf r_i$, $R_i$ and $A_i$ denote the center location, 
the radius and the absorbed energy density of the $i$-th sphere, respectively.  
Subsequently, we convolved $p_{\rm us} (\mathbf r^s, t)$ with
a one-dimensional (1D) Gaussian kernel with  ${\rm FWHM}=0.5$-$\mu s$ \cite{FourierShell:07}
to produce the pressure data.
%  corresponding to the spatial blurring of the object function \cite{FourierShell:07}. 
% Note that in the simulation studies, we assumed point-like transducers with $h^e(t)=\delta(t)$. 
From the simulated pressure data, we calculated the temporal-frequency spectrum by use of 
fast Fourier transform (FFT), 
from which we created an alternative data vector that contained $511$ frequency components 
occupying $(0,5]$-MHz for each transducer. 
The simulated projection data in either the time domain or the temporal frequency domain 
will hereafter be referred to as ``$128\times 90$''-data. 
By undersampling the ``$128\times 90$''-data uniformly over rings and tomographic views, we created three subsets that contained varying number of transducers. 
These data sets will be referred to as ``$64\times 90$''-data, ``$64\times 45$''-data, and ``$32\times 45$''-data, where the two numbers specify the number of rings and the number of tomographic views, respectively.

\vskip .5cm
\noindent{\it Reconstruction algorithms:} 
The GPU accelerated FBP algorithm was employed to reconstruct the object function sampled on a 3D Cartesian 
grid with spacing $\Delta_s=0.14$-mm.  
The dimension of the reconstructed images $\hat{\boldsymbol\alpha}_{\rm fbp}$ was 
$210\times 210\times 440$. 

We employed an iterative image reconstruction algorithm that sought to minimize 
a penalized least-squares (PLS) objective
\cite{Fessler:94,Wernickbookchap21}. 
Two versions of the reconstruction algorithm were developed that utilized the interpolation-based 
imaging model and the spherical-voxel-based imaging model respectively. 
The two versions sought to solve the optimization problems
by use of the linear conjugate gradient (CG) method 
\cite{Shewchuk:1994,Fessler:1999}:
\begin{equation}\label{eqn:PLSObjFuncInt}
  \hat {\boldsymbol \alpha}_{\rm int} = \arg\min_{\boldsymbol \alpha_{\rm int}}
    \Vert \mathbf u - \mathbf H_{\rm int}\boldsymbol\alpha_{\rm int}\Vert^2
    +\mu R(\boldsymbol\alpha_{\rm int}),
\end{equation}
and 
\begin{equation}\label{eqn:PLSObjFuncSph}
  \hat {\boldsymbol \alpha}_{\rm sph} = \arg\min_{\boldsymbol \alpha_{\rm sph}}
    \Vert \tilde{\mathbf u} - \mathbf H_{\rm sph}\boldsymbol\alpha_{\rm sph}\Vert^2
    +\mu R(\boldsymbol\alpha_{\rm sph}),
\end{equation}
respectively, where
% $\mathbf u$ is the data vector in time-domain,  
% $\tilde{\mathbf u}$ is the data vector in temporal-frequency domain, 
$R(\boldsymbol\alpha)$ is a regularizing penalty term whose impact is controlled by 
the regularization parameter $\mu$.
The penalty term was employed only when processing the experimental data as 
described in Section \ref{sect:Nummethods}-B. 
% We set $\beta=0$ for all computer-simulations.
% The optimization problems Eqns.\ \eqref{eqn:PLSObjFuncInt} and \eqref{eqn:PLSObjFuncSph} 
% were solved iteratively by a linear conjugate gradient (CG) method \cite{Shewchuk:1994,Fessler:1999},
The reconstruction algorithms required computation of 
one projection and one backprojection operation at each iteration. 
Hereafter, the two reconstruction algorithms will be referred to as 
PLS-Int and PLS-Sph algorithms, respectively. 
We set $\Delta_s=0.14$-mm. 
Therefore, both the dimensions of $\hat{\boldsymbol\alpha}_{\rm int}$ 
and $\hat{\boldsymbol\alpha}_{\rm sph}$ were $210\times 210\times 440$.

\vskip .5cm
\noindent {\it Performance assessment:}
We compared the computational times of 3D image reconstruction corresponding to the GPU- and 
CPU-based implementations. 
The CPU-based implementations of the PLS-Int and PLS-Sph algorithms 
take several days to complete a single iteration even for the ``$32\times 45$''-data. 
Therefore, we only recorded the computational time for the CPU-based implementations to complete
a single iteration when the data vector contained a single transducer. 
We assumed that the computational times were linearly proportional to the number of transducers
in the data sets because the CPU-based implementations are sequential.  

The GPU-based implementations employed the single-precision floating-point format rather than the 
conventional double-precision utilized by CPU-based implementations. 
In order to quantify how the single-precision floating-point format would degrade the image accuracy,
we calculated the root mean square error (RMSE) between the reconstructed image and the phantom defined by:
\begin{equation}
{\rm RMSE} = \sqrt{\frac{1}{N}  
                   (\hat{\boldsymbol \alpha}-\boldsymbol \alpha)^{\rm T}
                                    (\hat{\boldsymbol\alpha}-\boldsymbol \alpha)},
\end{equation}
where $\boldsymbol\alpha$ and $\hat{\boldsymbol\alpha}$ are the samples of the phantom and the coefficients of the reconstructed images respectively.  

\vskip .5cm
\noindent {\it Hardware specifications:} 
All implementations were tested on the platform consisted of dual quad-core 
Intel(R) Xeon (R) CPUs with a clock speed $2.40$-GHz. 
The GPU-based implementations of the FBP and the PLS-Int algorithms were tested on a single 
Tesla C2050 GPU, 
while the PLS-Sph algorithm was tested on $8$ Tesla C1060 GPUs. 

\subsection{Experimental studies}

The FBP, PLS-Int and PLS-Sph algorithms were investigated by use of an existing 
data set corresponding to a live mouse \cite{petermice:jbo,Kun:PMB}. 
The scanning geometry and dimensions were the same as those employed in the computer-simulation 
studies except that only $64$ rings were uniformly distributed over the 
polar angle ranging from $14^\circ$ to $83^\circ$.   
The transducers were of size $2\times 2$-mm$^2$. 
The raw data were acquired at $180$ tomographic views, which are referred to as ``full data''. 
We undersampled the ``full data'' uniformly over the tomographic views, constructing a subset 
containing $45$ tomographic views. 
The subset will be referred to as ``quarter data''.

Unlike in the idealized computer-simulation studies, 
the transducer response has to be compensated for 
when processing the experimental data. 
% However, limited by each reconstruction algorithm, the transducer effects
% were compensated for in different ways \cite{Kun:PMB}. 
When implementing the FBP algorithm, the EIR was compensated for by a direct 
Fourier deconvolution, 
% we estimated the acoustic pressure $\mathbf p$ 
% from the measured voltage signals $\mathbf u$ by use of linear regularized Fourier 
% deconvolution \cite{KrugerMP1999}, 
expressed  in temporal frequency domain as \cite{KrugerMP1999}:
\begin{equation}
  \tilde p(\mathbf r^s, f) = \frac{\tilde u(\mathbf r^s, f)}{\tilde h^e(f)}
    \tilde W(f),
\end{equation}
where $\tilde W(f)$ is a window function for noise suppression.
In this study, we adopted the Hann window function defined as:
\begin{equation}
  \tilde W(f) = \frac{1}{2}\Big[1-\cos(\pi\frac{f_c-f}{f_c})\Big],
\end{equation}
where the cutoff frequency was chosen as $f_c = 5$-MHz. 
When applying iterative image reconstruction algorithms, 
the transducer effects were implicitly compensated for during iteration 
by employing imaging models that incorporates the transducer charactertics
\cite{TMI:transmodel,Kun:PMB}. 
We incorporated the EIR into the interpolation-based imaging model 
while incorporating both the EIR and the SIR into the 
spherical-voxel-based imaging model.

For both PLS-Int and PLS-Sph algorithms, 
we employed a quadratic smoothness penalty to mitigate measurement noise \cite{Fessler:94}: 
\begin{equation}
  R(\boldsymbol\alpha)=\sum_{n=0}^{N-1}
    \big([\boldsymbol\alpha]_n - [\boldsymbol\alpha]_{n_x}\big)^2 
   +\big([\boldsymbol\alpha]_n - [\boldsymbol\alpha]_{n_y}\big)^2
   +\big([\boldsymbol\alpha]_n - [\boldsymbol\alpha]_{n_z}\big)^2, 
\end{equation}
where $n_x$, $n_y$ and $n_z$ were the indices of the neighboring voxels before the $n$-th voxel along the three Cartesian axes, respectively.

\section{results}
\label{sect:results}
\subsection{Computational efficiency}
As shown in Table \ref{tab:time}, the GPU-based implementations took less than  
$0.1\%$, $0.8\%$ and $0.4\%$ of the computational times required by corresponding 
CPU-based implementations for the FBP, the PLS-Int, and the PLS-Sph algorithms, respectively. 
The relative computational times for the GPU-based implementations are nearly linearly proportional to 
the amount of data. 
Note that the ``$64\times90$''-data and the ``quarter data'' are of the same size. 
However, the computational times of the ``quarter data'' are more than $1.8$ times those of the 
``$64\times90$''-data. 
This is because the calculation of the SIR increases the computational complexity of the reconstruction
algorithm. 

\subsection{Computational accuracy}

\begin{figure}[ht]
\centering
\subfigure[]{\includegraphics[width=5cm]{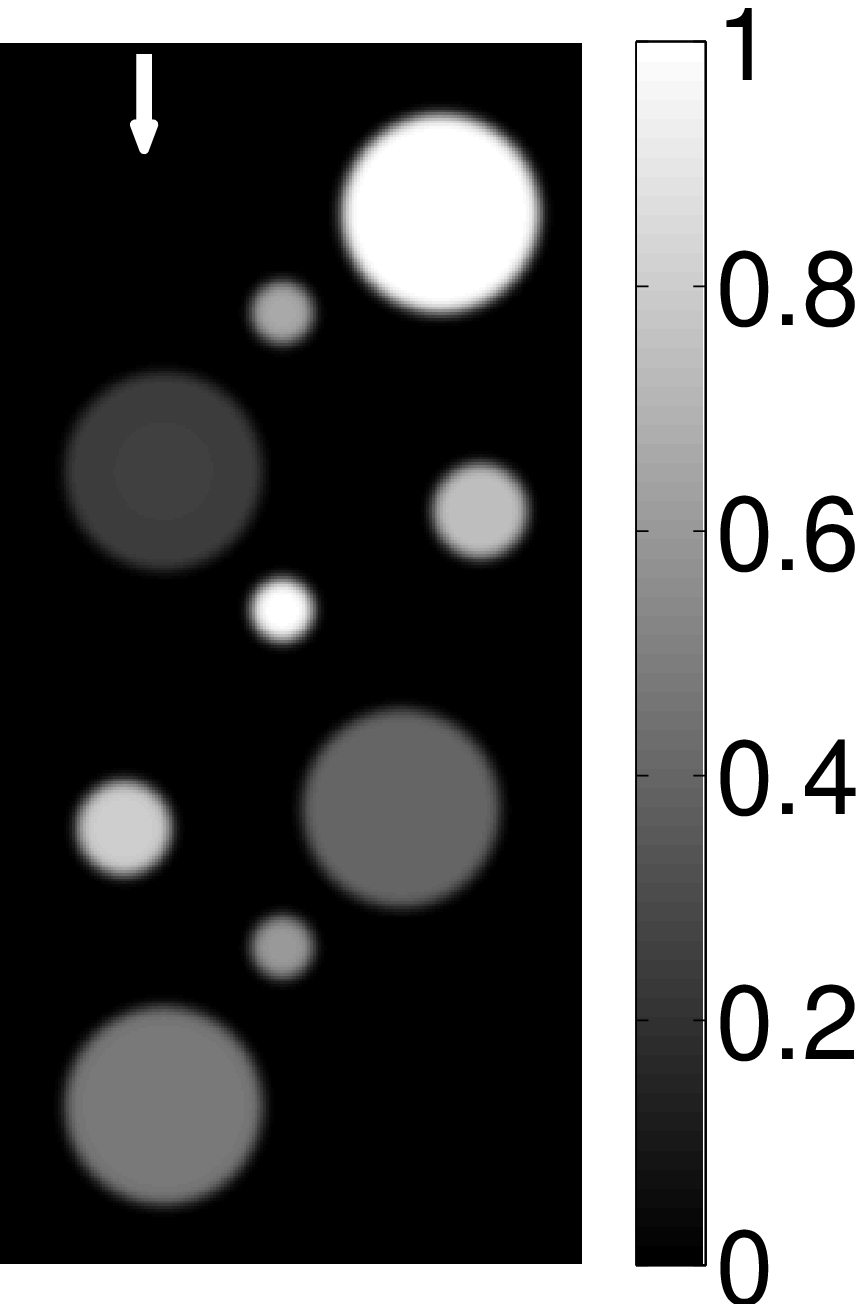}}
\subfigure[]{\includegraphics[width=5cm]{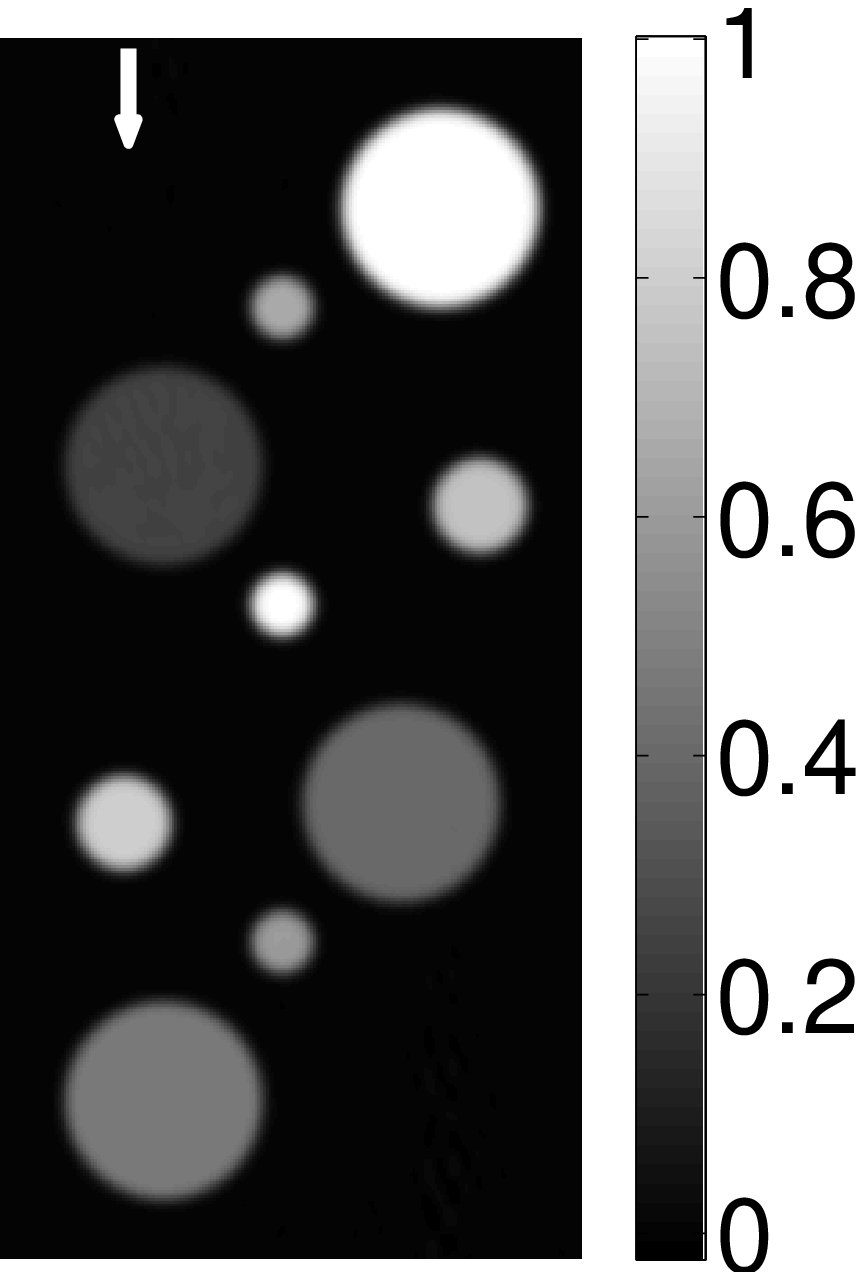}}
\subfigure[]{\includegraphics[width=5cm]{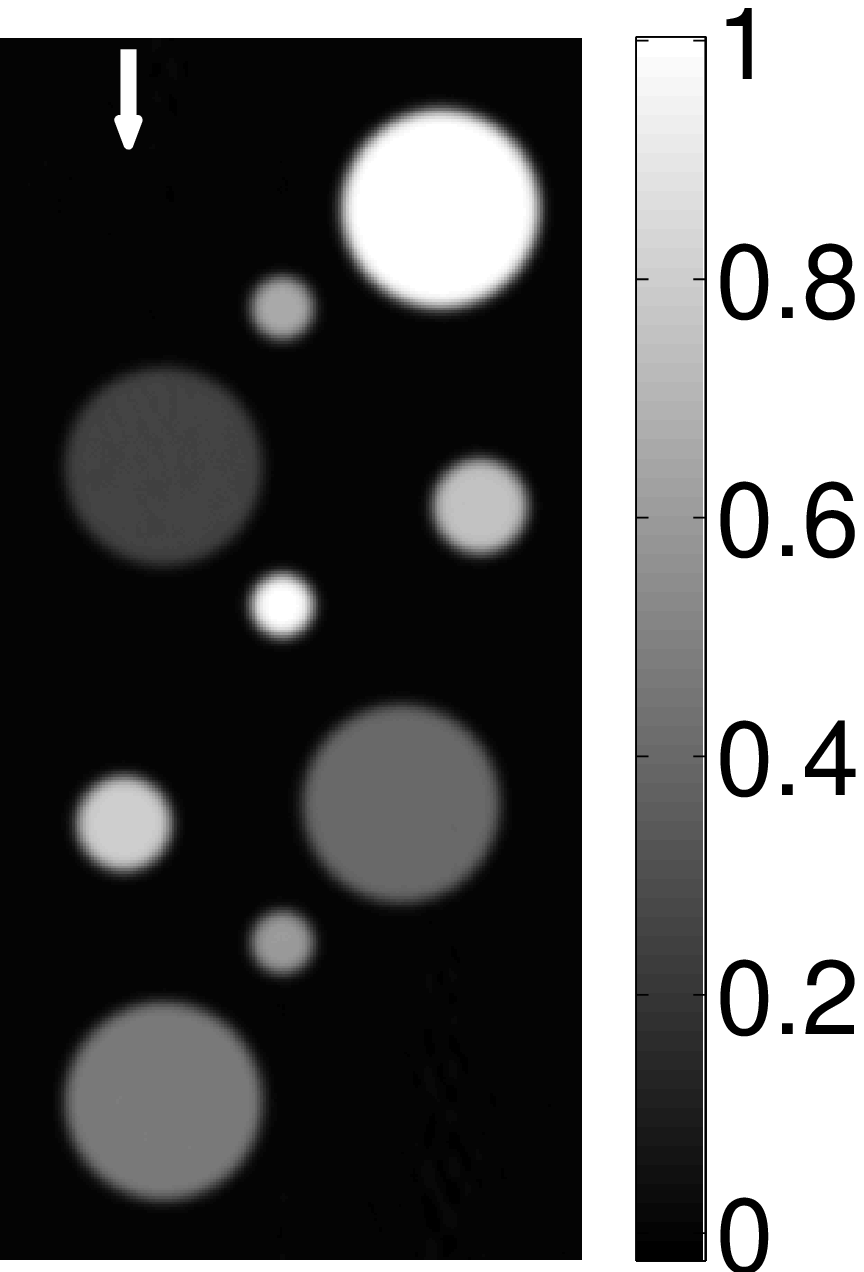}}
\caption{\label{fig:fbp}
}
\end{figure}

\begin{figure}[ht]
\centering
\subfigure[]{\includegraphics[width=5cm]{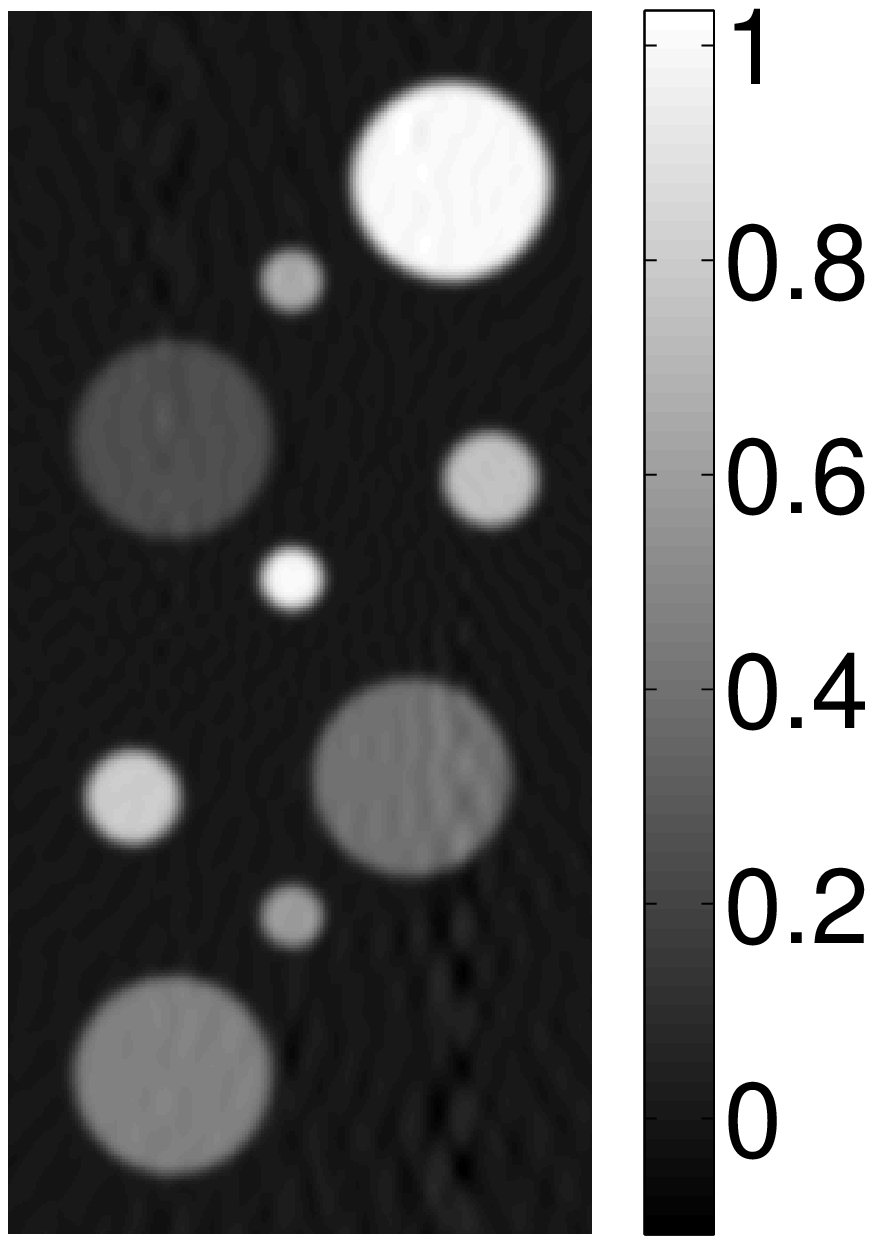}}
\subfigure[]{\includegraphics[width=5cm]{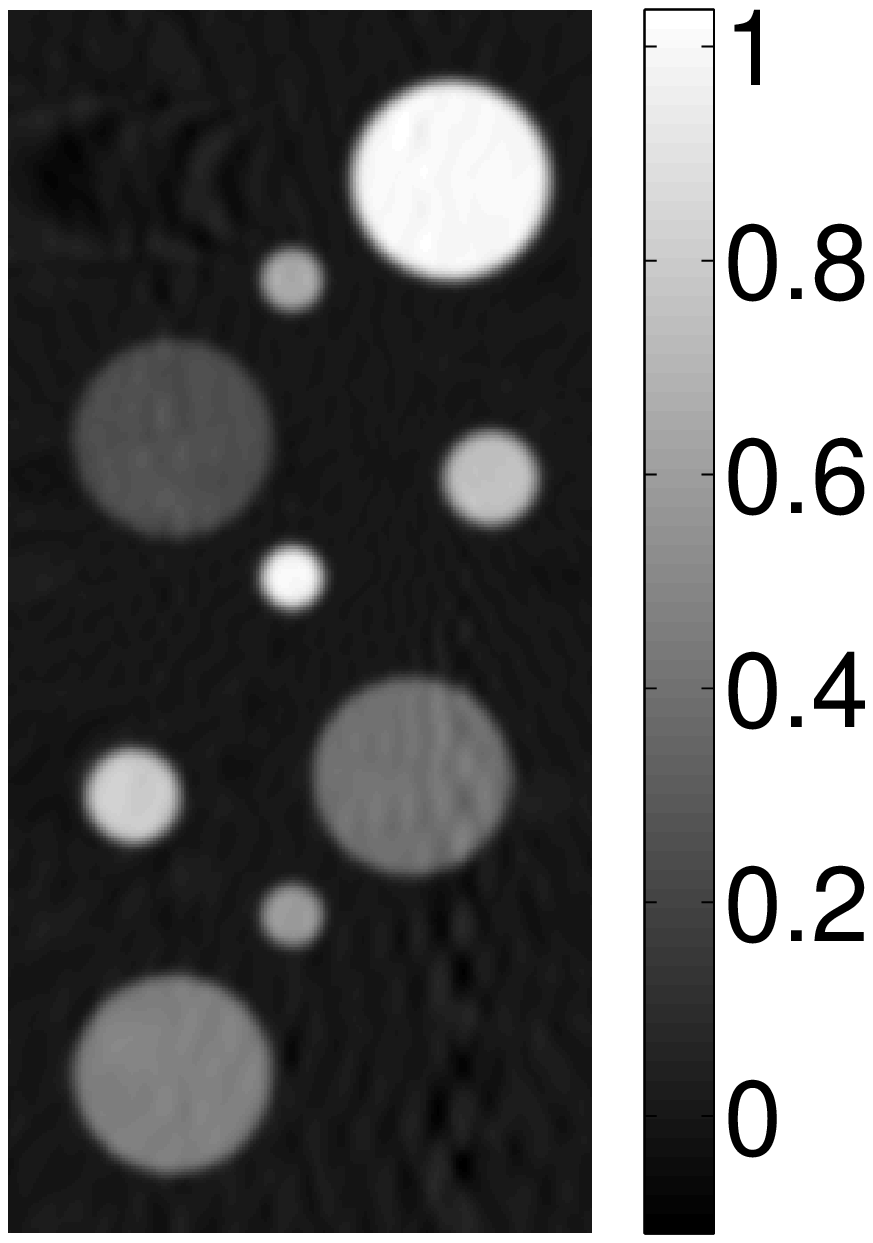}}
\subfigure[]{\includegraphics[width=5cm]{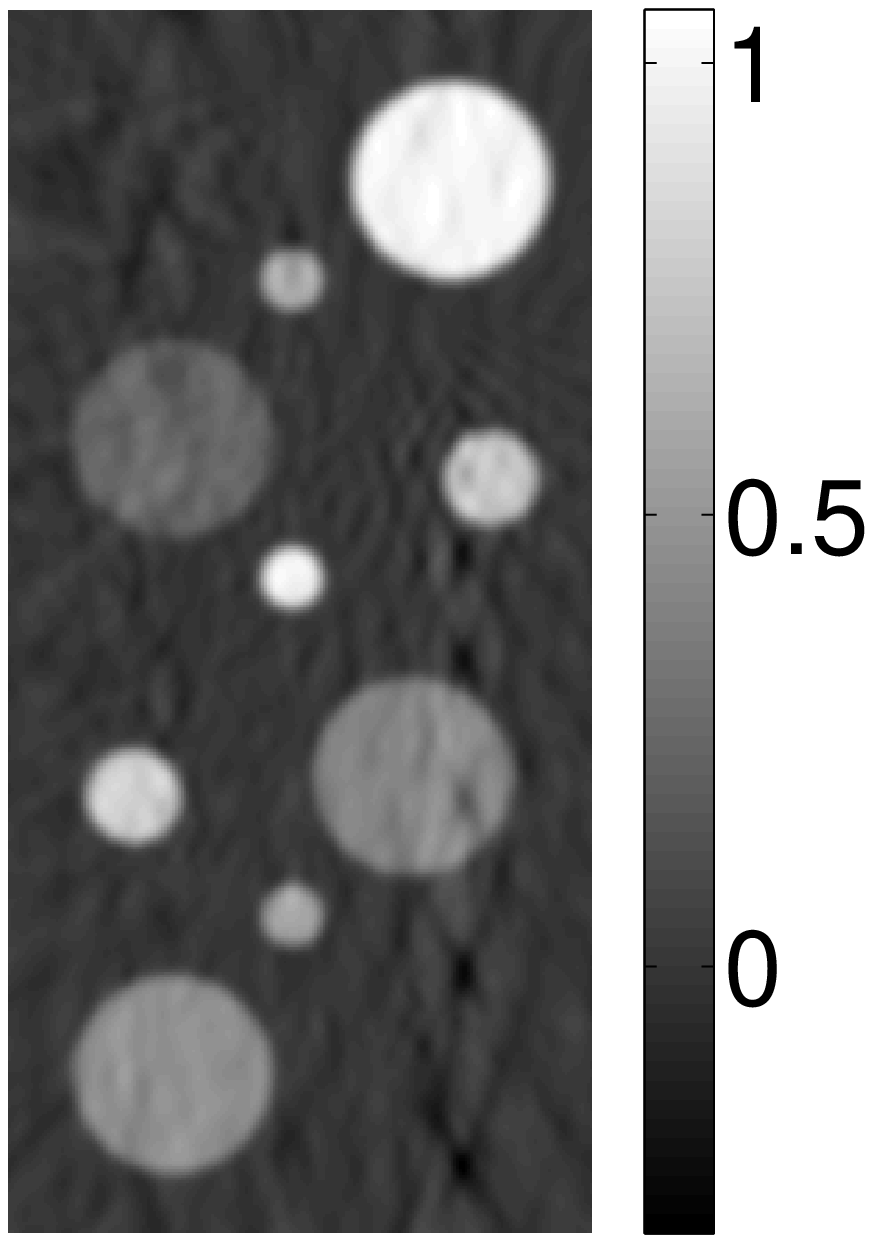}}\\
\subfigure[]{\includegraphics[width=5cm]{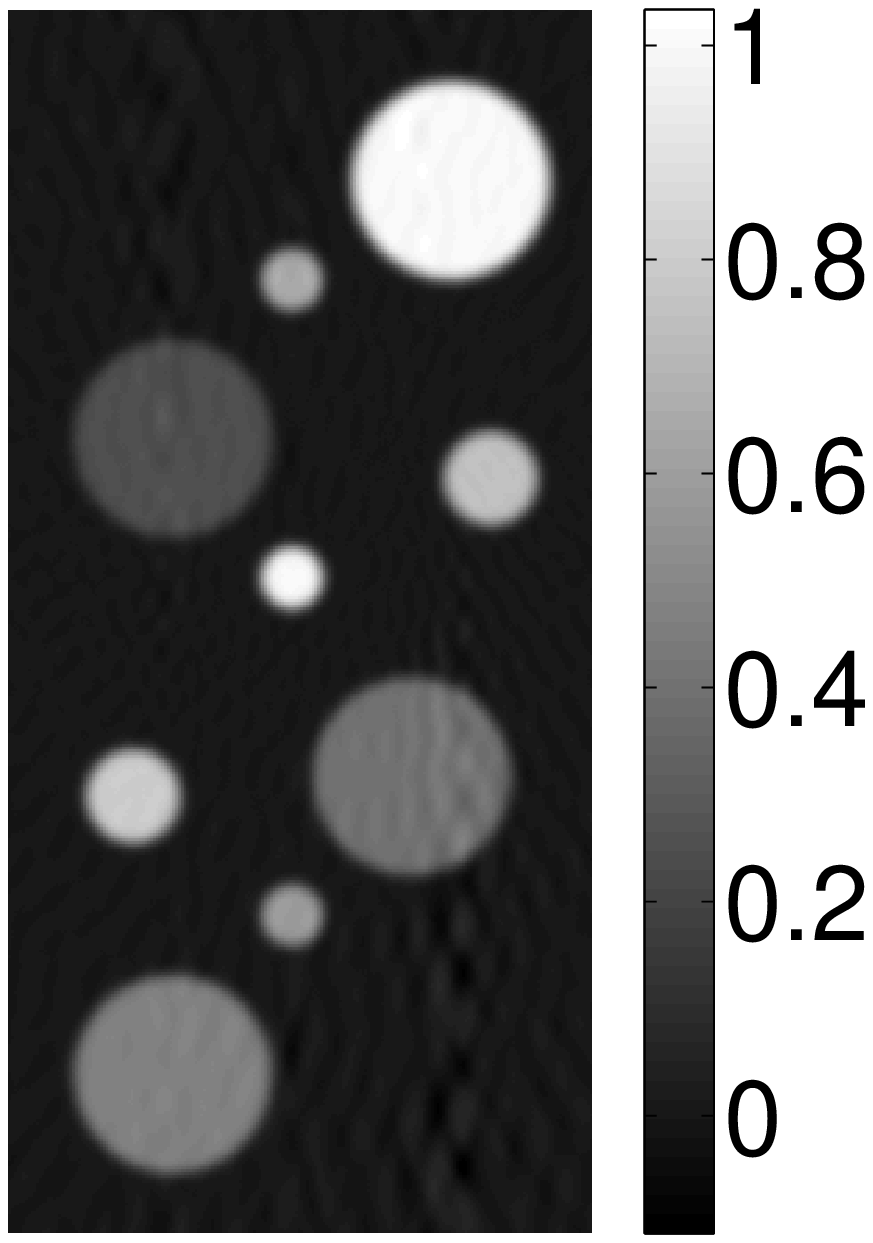}}
\subfigure[]{\includegraphics[width=5cm]{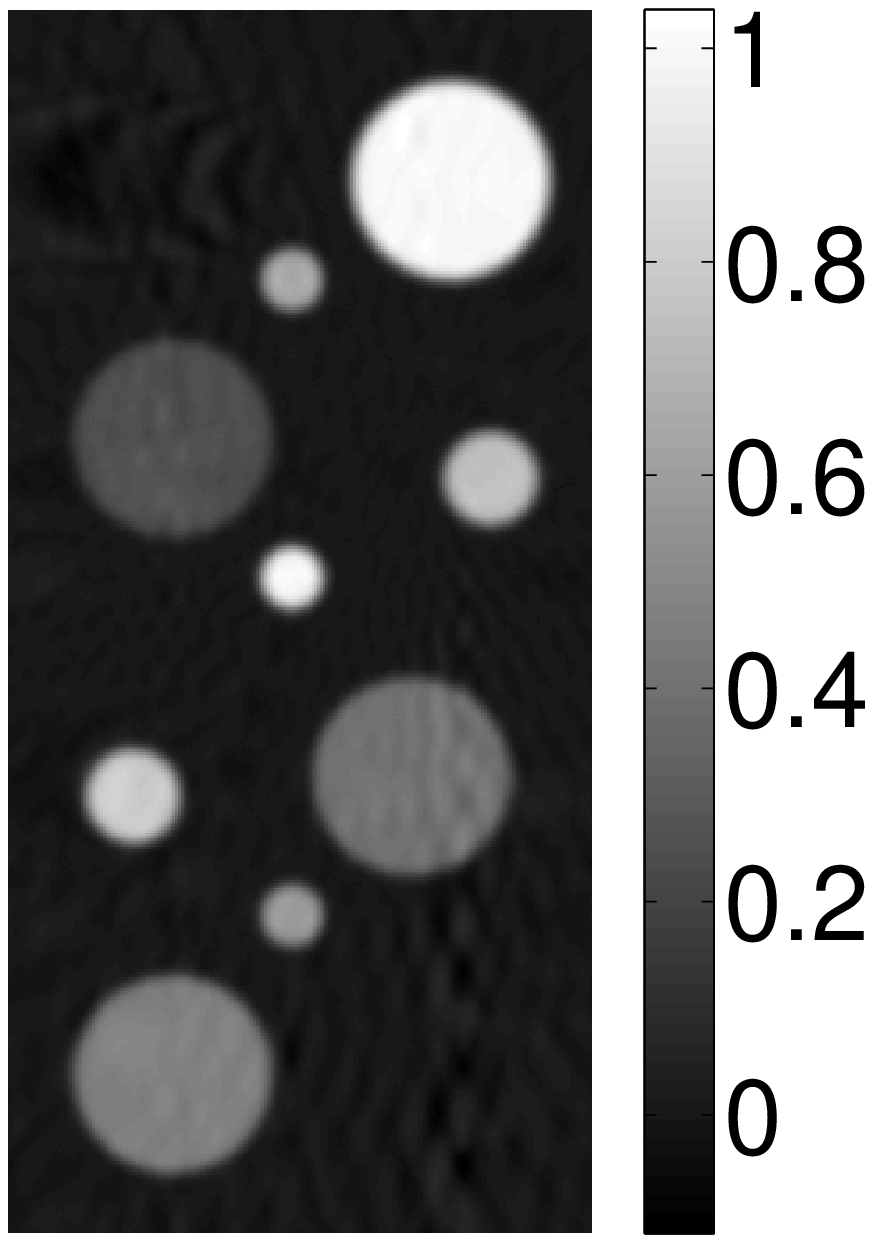}}
\subfigure[]{\includegraphics[width=5cm]{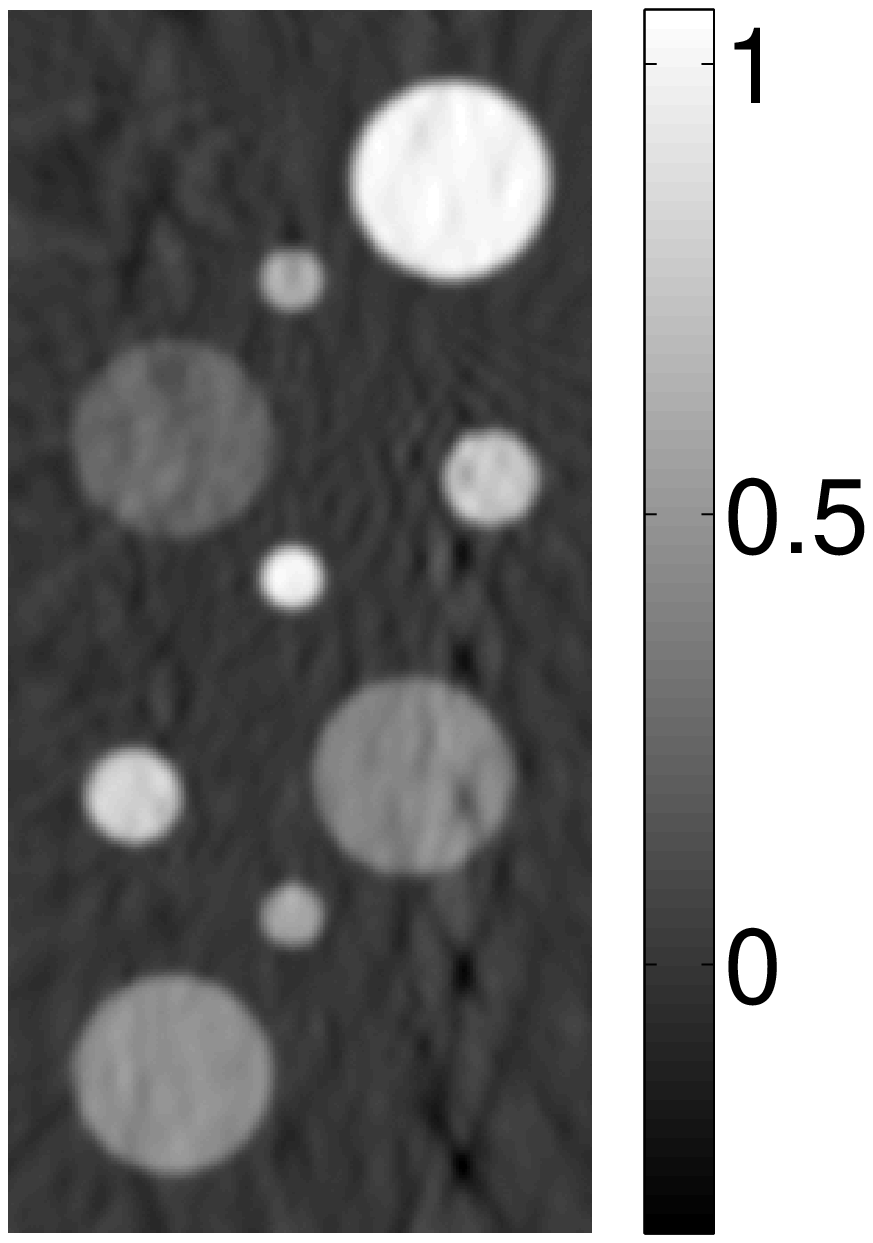}}
\caption{\label{fig:fbp2}
}
\end{figure}

\begin{figure}[ht]
\centering
\subfigure[]{\includegraphics[width=8cm]{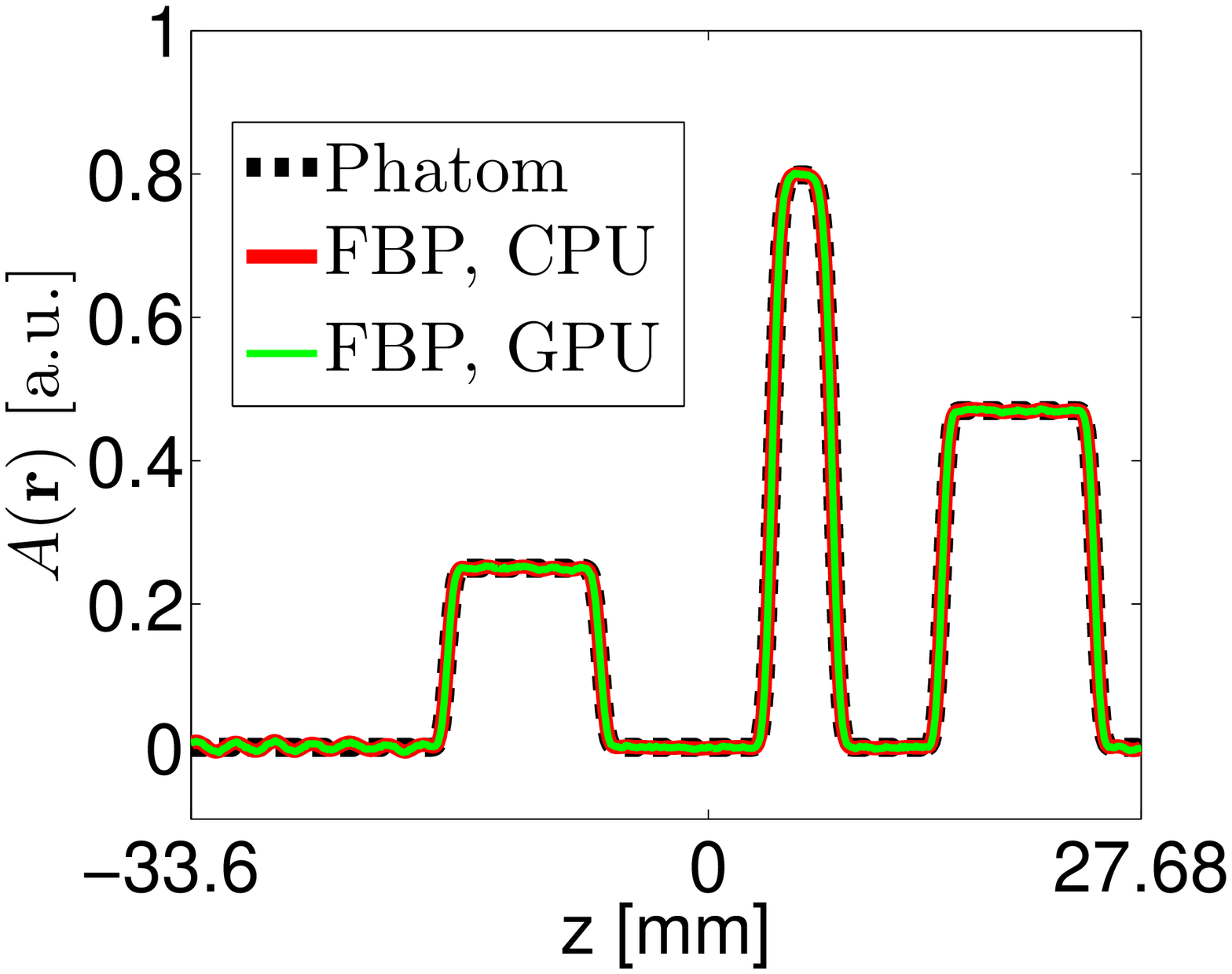}}
\subfigure[]{\includegraphics[width=8cm]{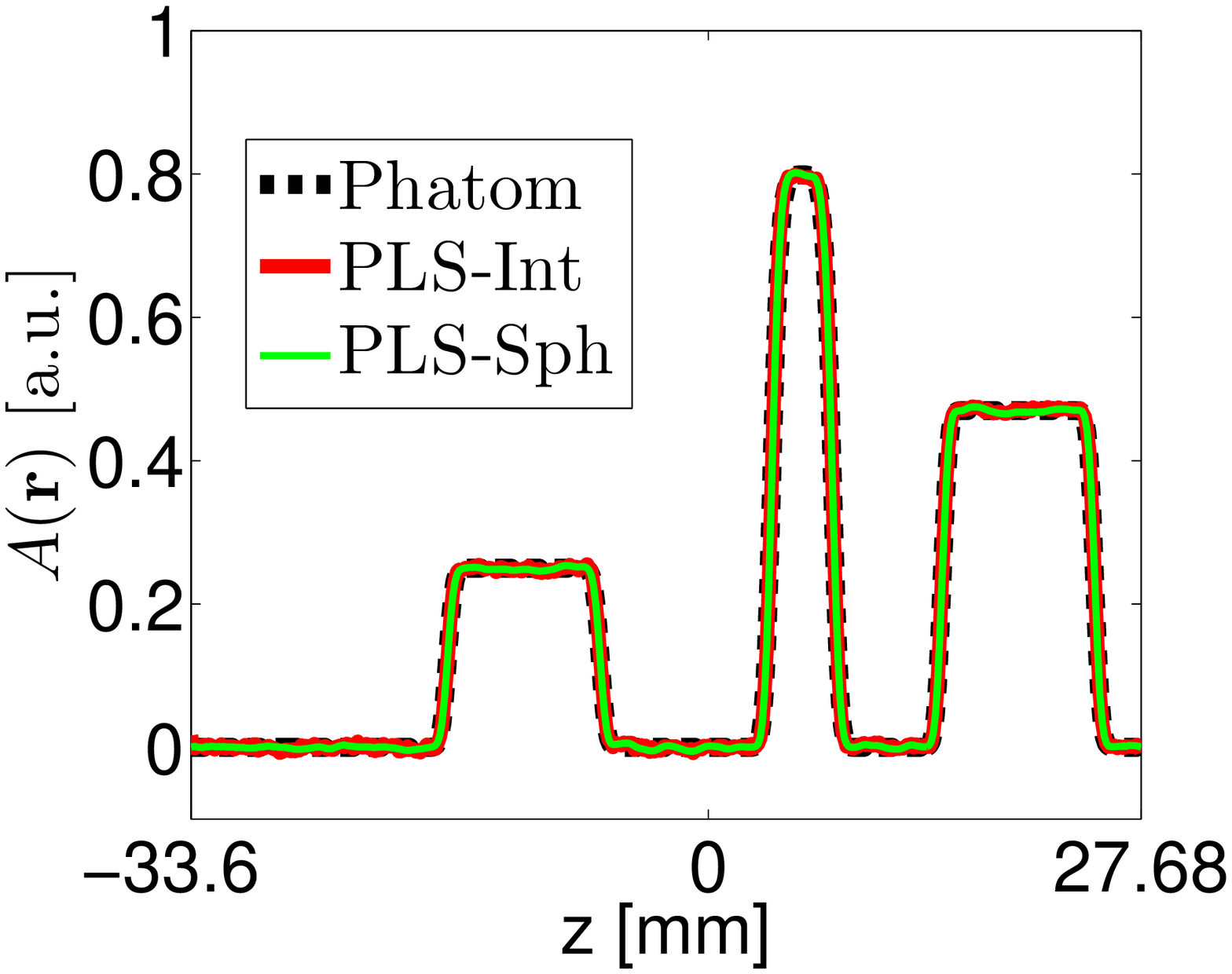}}
\caption{\label{fig:profile}
}
\end{figure}

\begin{figure}[ht]
\centering
\includegraphics[width=10cm]{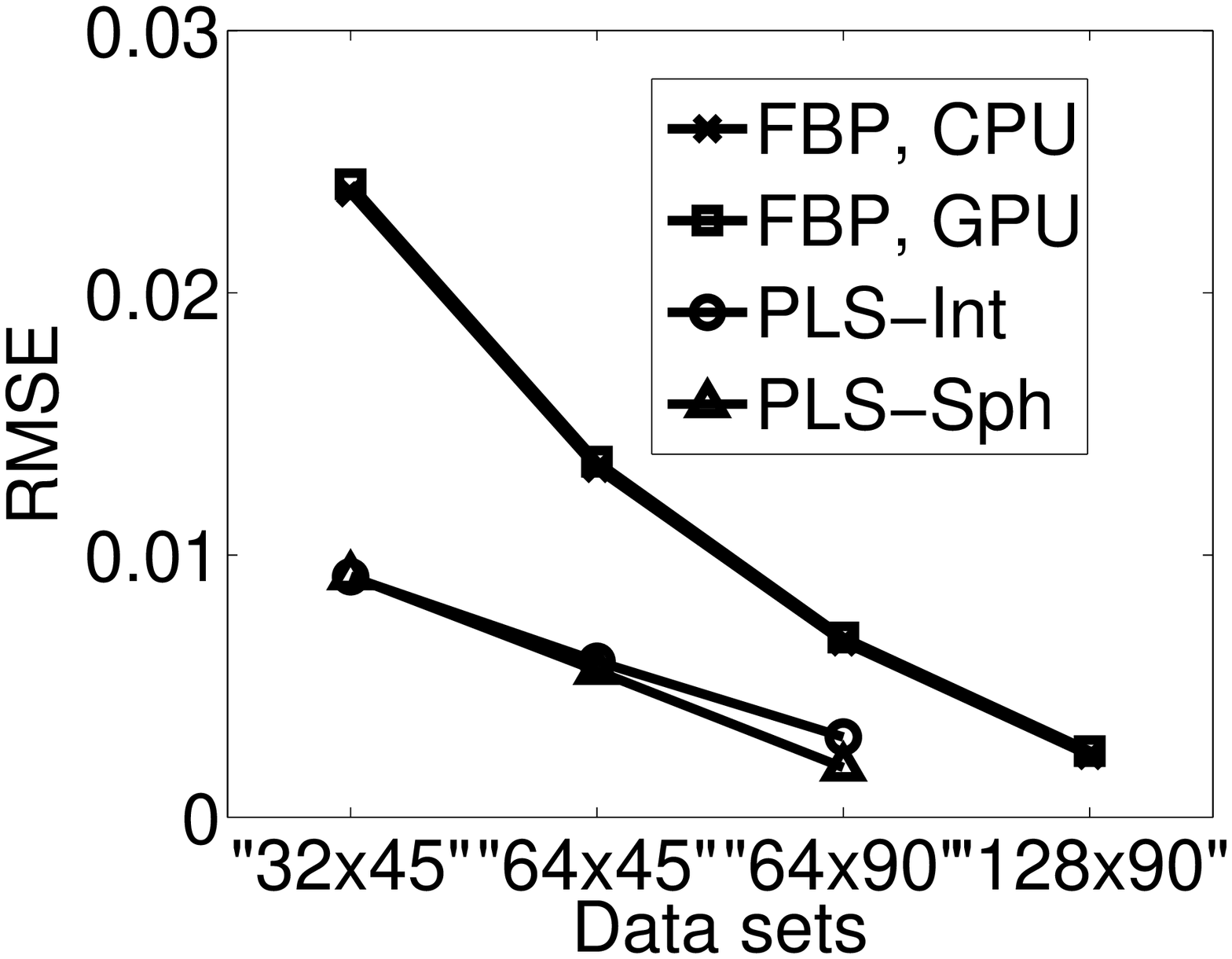}
\caption{\label{fig:error}
}
\end{figure}

Images reconstructed by use of the CPU- and GPU-based implementations of the FBP algorithm are almost identical. 
From the ``$128\times90$''-data, in which case, transducers were densely distributed over the measurement surface, 
both implementations reconstructed accurate images, as shown Fig.\  \ref{fig:fbp}-(b) and -(c).
The profiles along the three arrows in Fig.\  \ref{fig:fbp} are plotted 
in Fig. \ref{fig:profile}-(a), suggesting a nearly exact reconstruction.  
As expected, when the amount of measurement data are reduced, the reconstructed images 
contain more artifacts as shown in Fig.\ \ref{fig:fbp2}. 
However, the images reconstructed by use of GPU- and CPU-based implementations remain 
indistinguishable. 
The plots of the RMSE versus the amount of measurement data employed 
in Fig.\ \ref{fig:error} overlap, 
also suggesting the single-precision floating-point format employed by the 
GPU-based implementation has little impact on the computational accuracy.

\begin{figure}[ht]
\centering
\subfigure[]{\includegraphics[width=5cm]{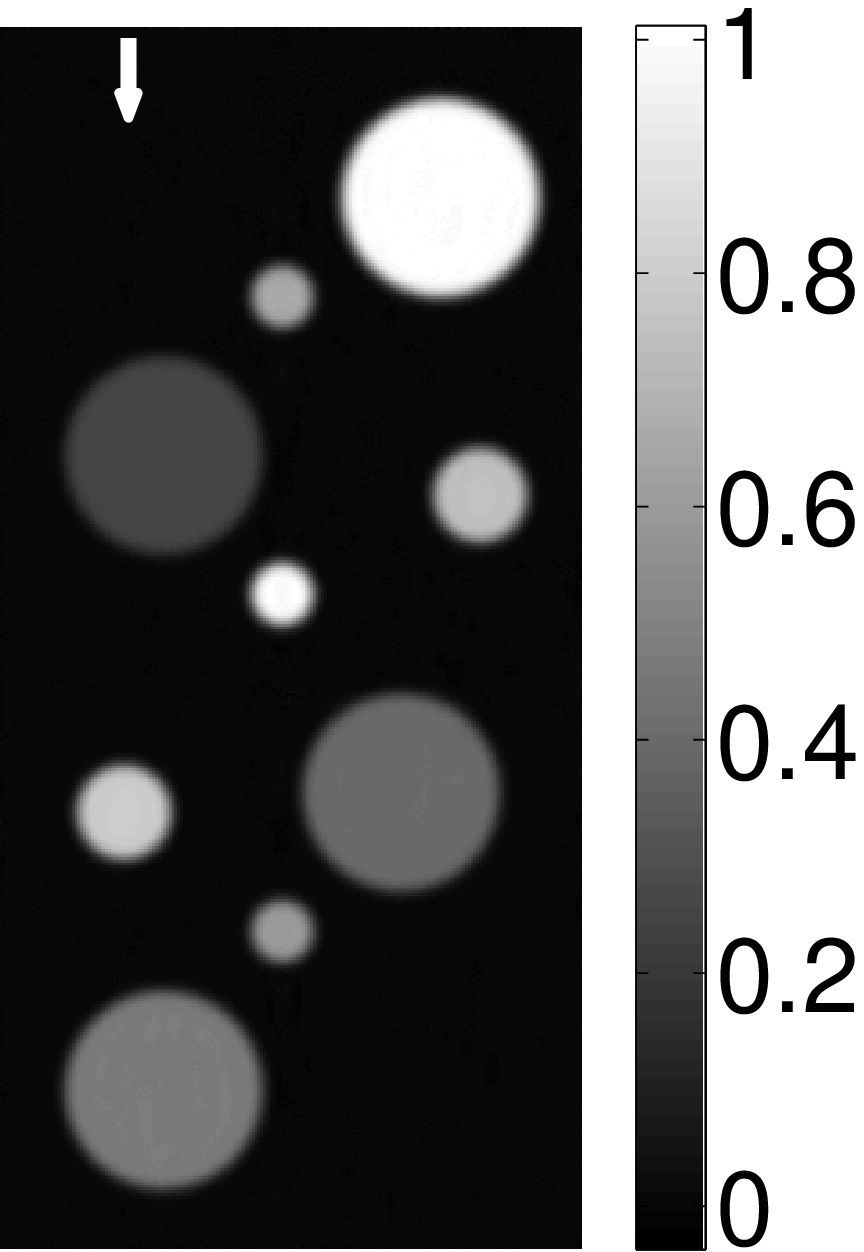}}
\subfigure[]{\includegraphics[width=5cm]{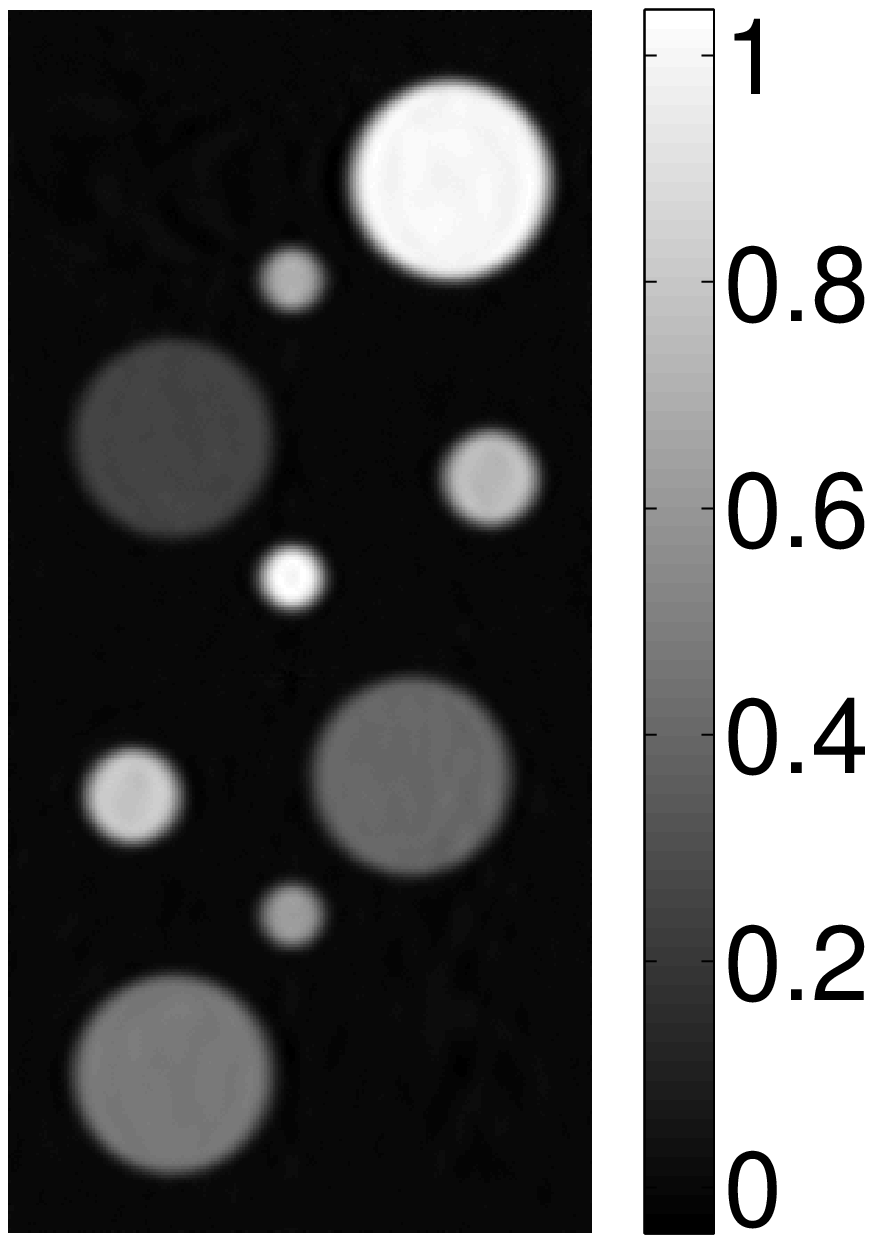}}
\subfigure[]{\includegraphics[width=5cm]{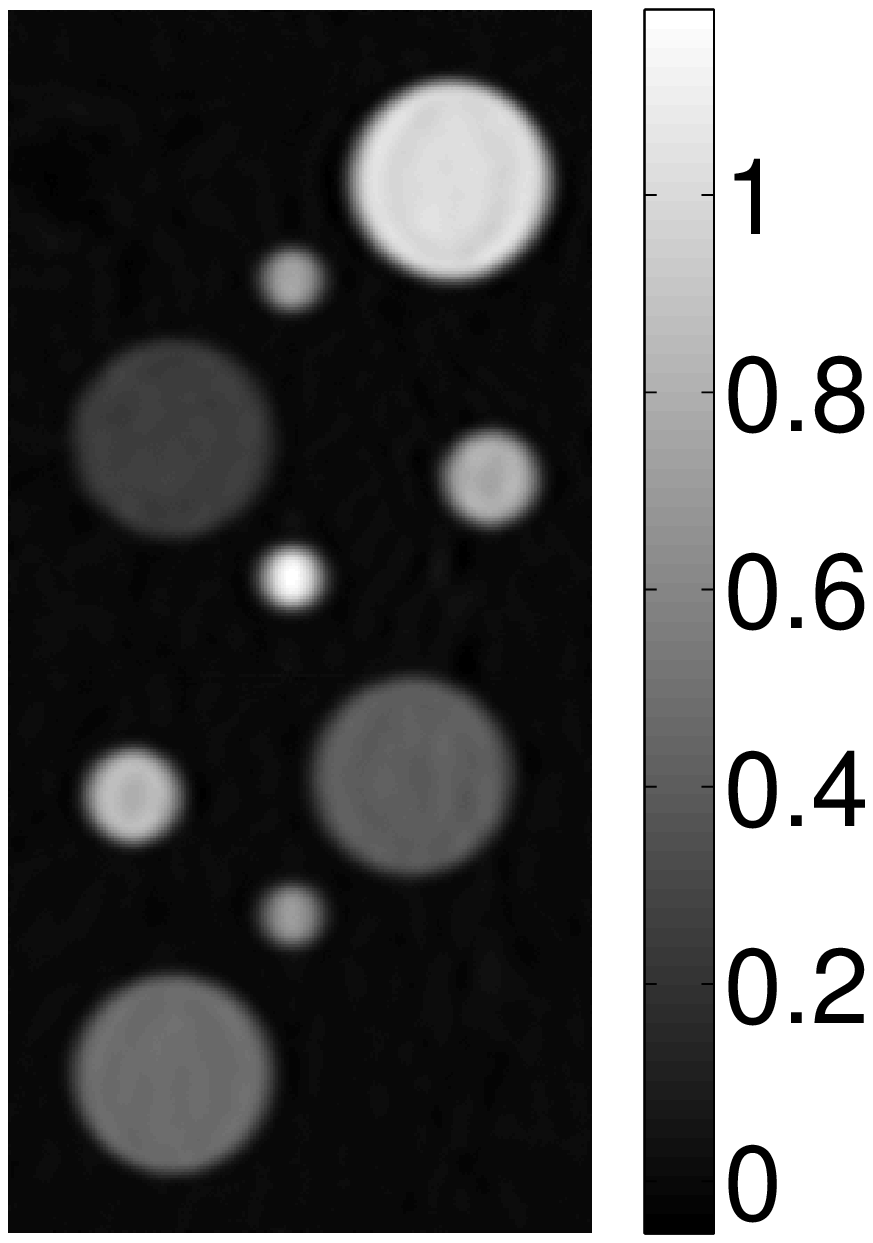}}\\
\subfigure[]{\includegraphics[width=5cm]{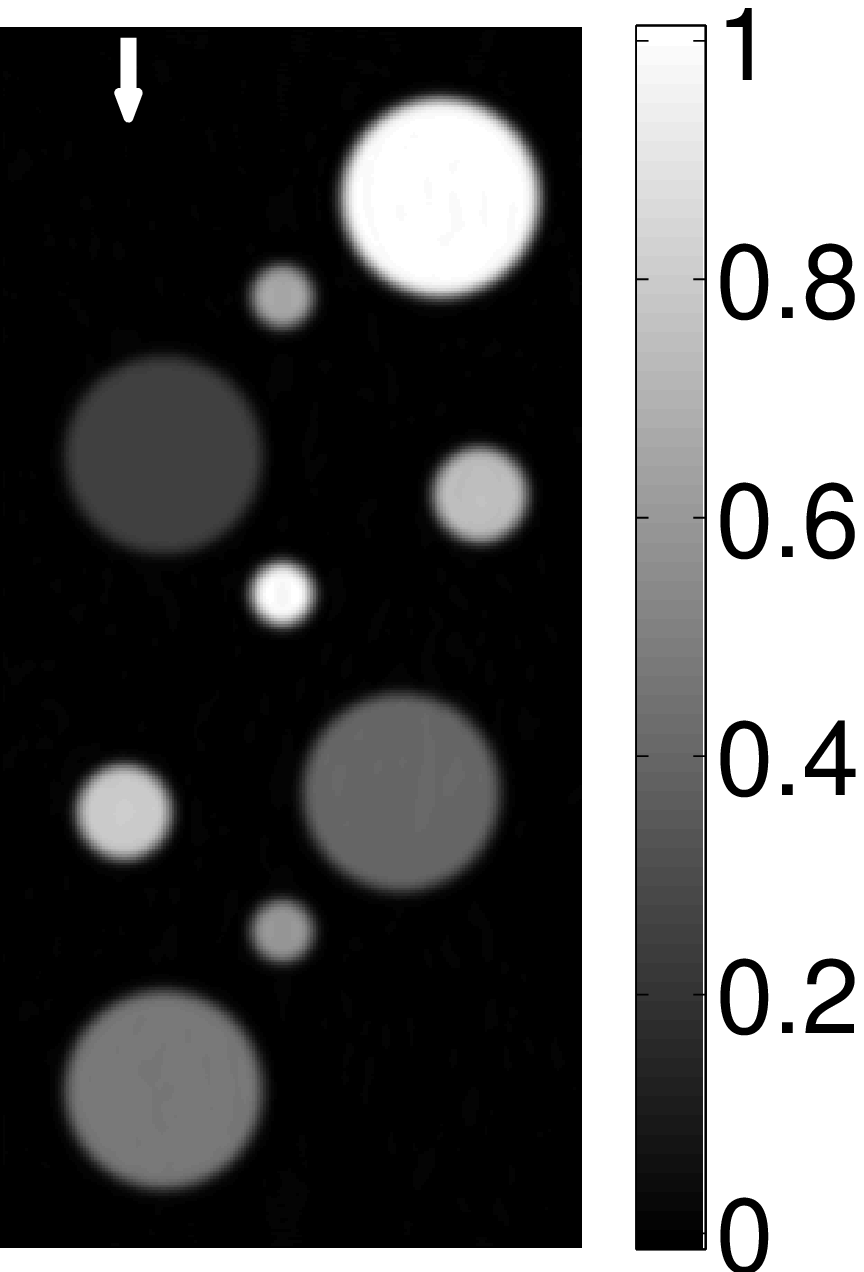}}
\subfigure[]{\includegraphics[width=5cm]{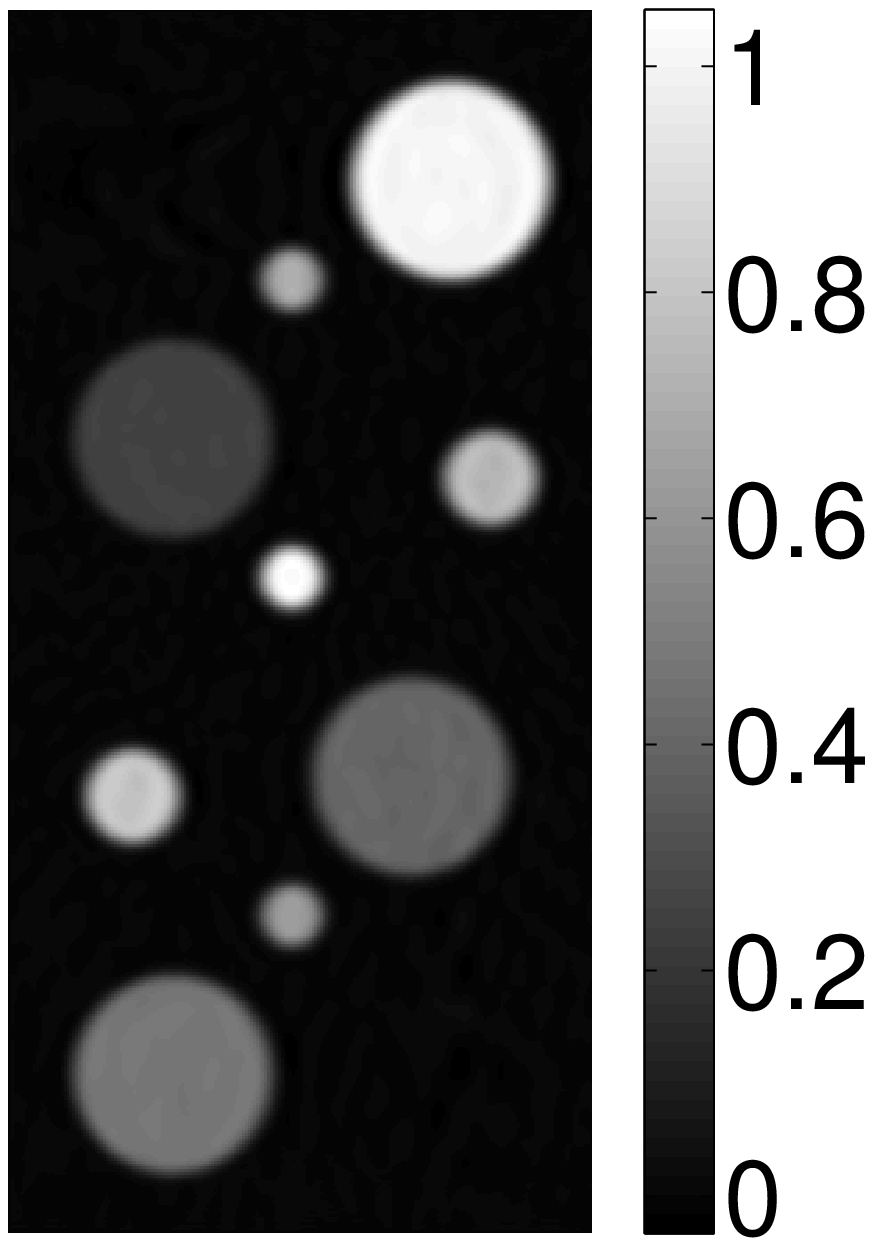}}
\subfigure[]{\includegraphics[width=5cm]{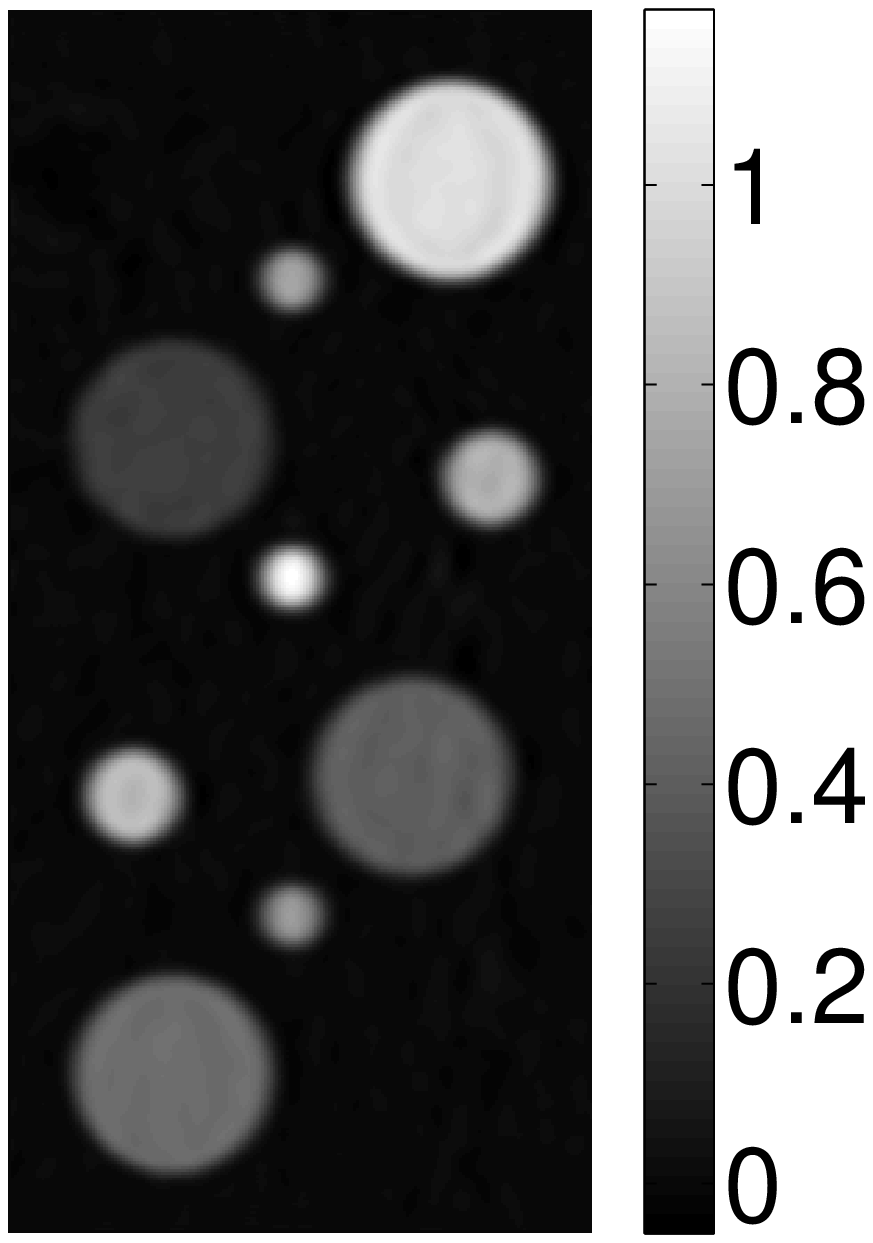}}
\caption{\label{fig:pls}
}
\end{figure}

The GPU-based implementations of the PLS-Int and PLS-Sph algorithms both reconstructed 
accurate images as displayed in Fig.\ \ref{fig:pls}. 
As expected, the images reconstructed by use of both iterative algorithms contain fewer artifacts 
than do those reconstructed by use of the FBP algorithm from the same amount of data. 
Unlike the images reconstructed by use of the FBP algorithm from the ``$64\times90$''-data (Fig.\ \ref{fig:fbp2}-(a) or -(d)), the images reconstructed by use of both iterative algorithms (Fig.\ \ref{fig:pls}-(a) and -(d)) appear to be identical to the numerical phantom. 
The profiles along the two arrows in Fig.\ \ref{fig:pls}-(a) and -(d) are plotted in Fig. \ref{fig:profile}-(b), further confirming the computational accuracy of iterative image reconstruction algorithms. 
The plots of the RMSE versus the amount of measurement data employed in Fig.\ \ref{fig:error} 
suggest the iterative image reconstruction algorithms in general outperform 
the FBP algorithm from the same amount of data. 

\subsection{Experimental results}
\begin{figure}[h]
\begin{center}
\subfigure[]{\includegraphics[width=4cm,trim=6cm 1.5cm 6cm 4cm, clip]{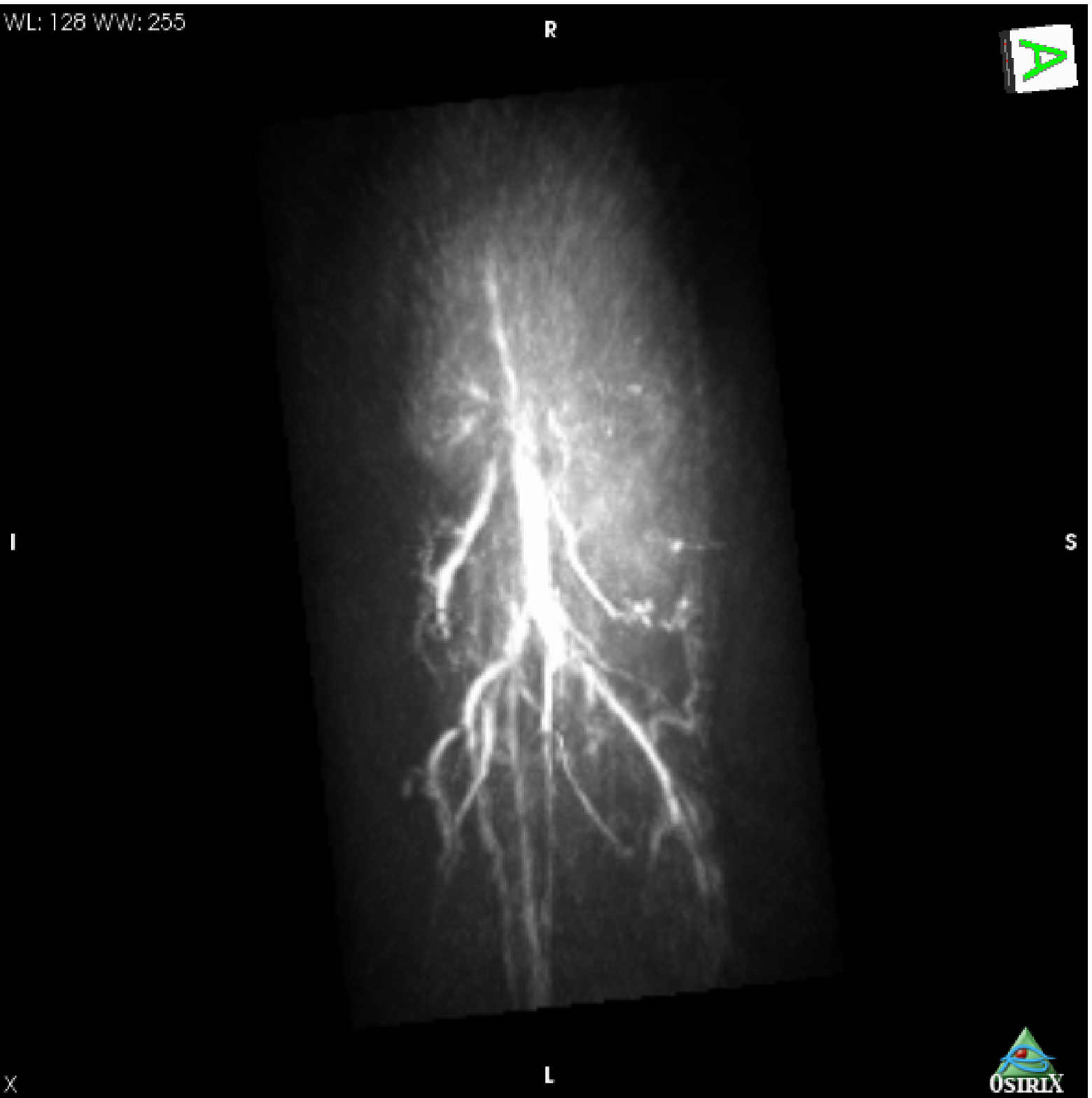}}
\hskip .5cm
\subfigure[]{\includegraphics[width=4cm,trim=6cm 1.5cm 6cm 4cm, clip]{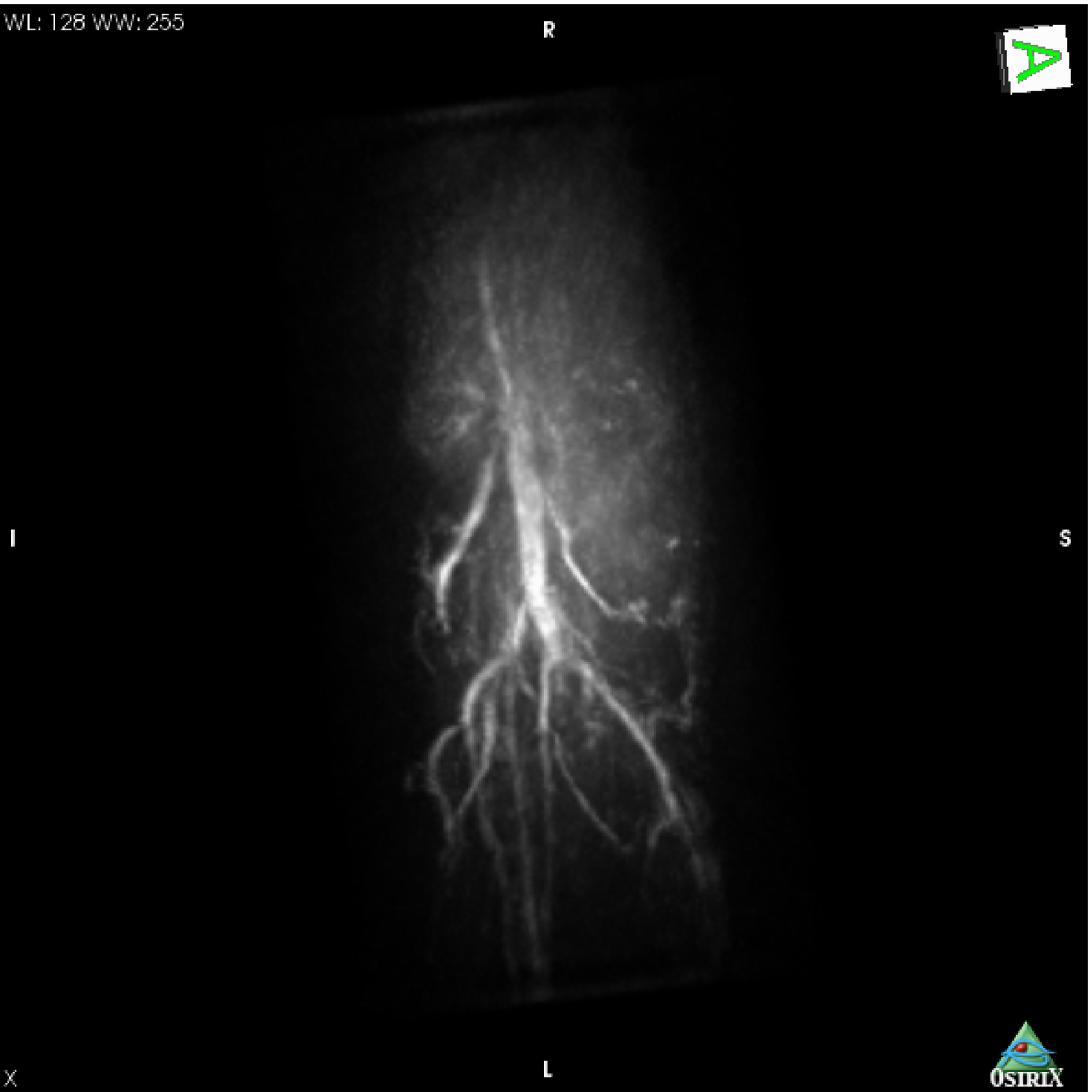}}
\hskip .5cm
\subfigure[]{\includegraphics[width=4cm,trim=6cm 1.5cm 6cm 4cm, clip]{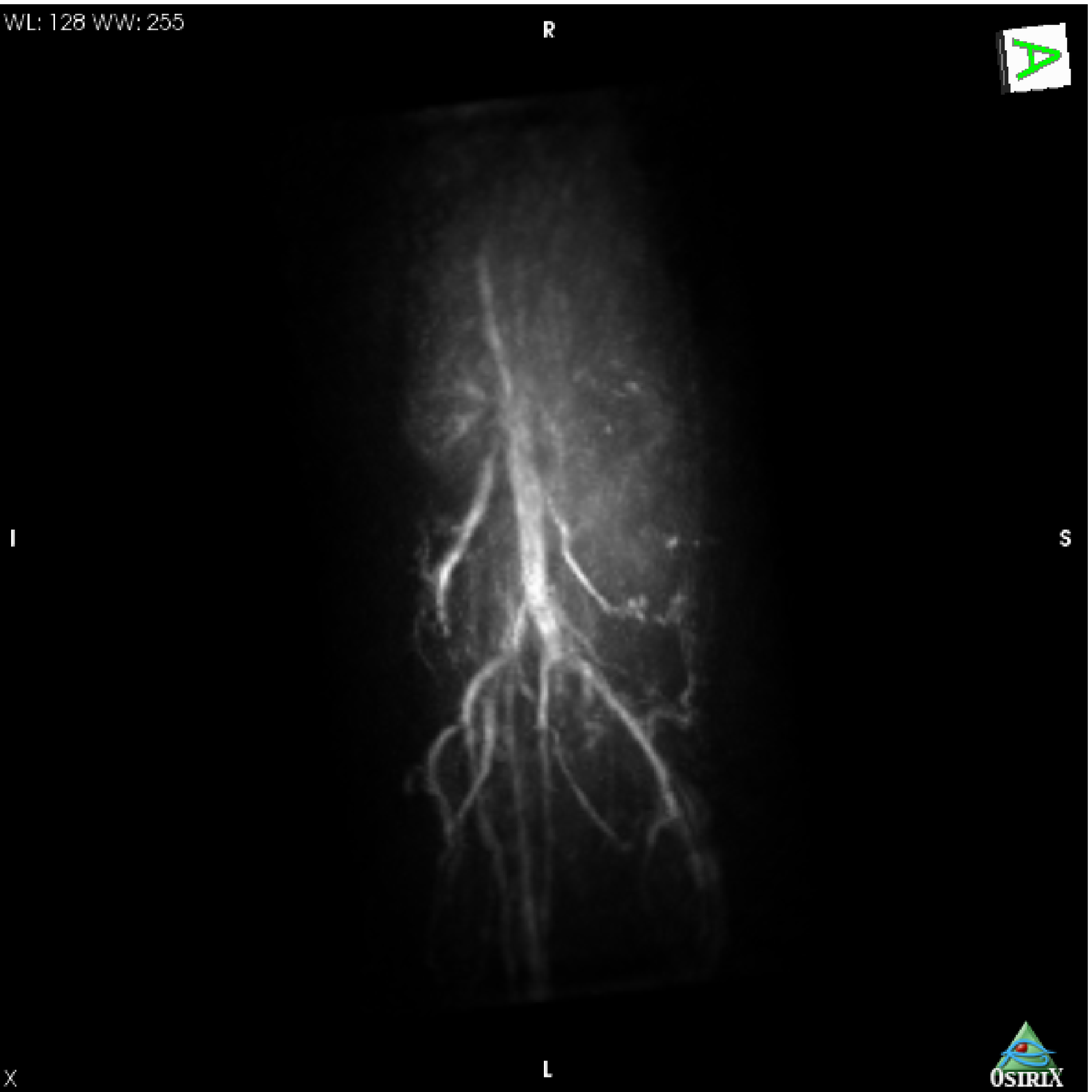}}\\
\subfigure[]{\includegraphics[width=4cm, trim=6cm 1.5cm 6cm 4cm, clip]{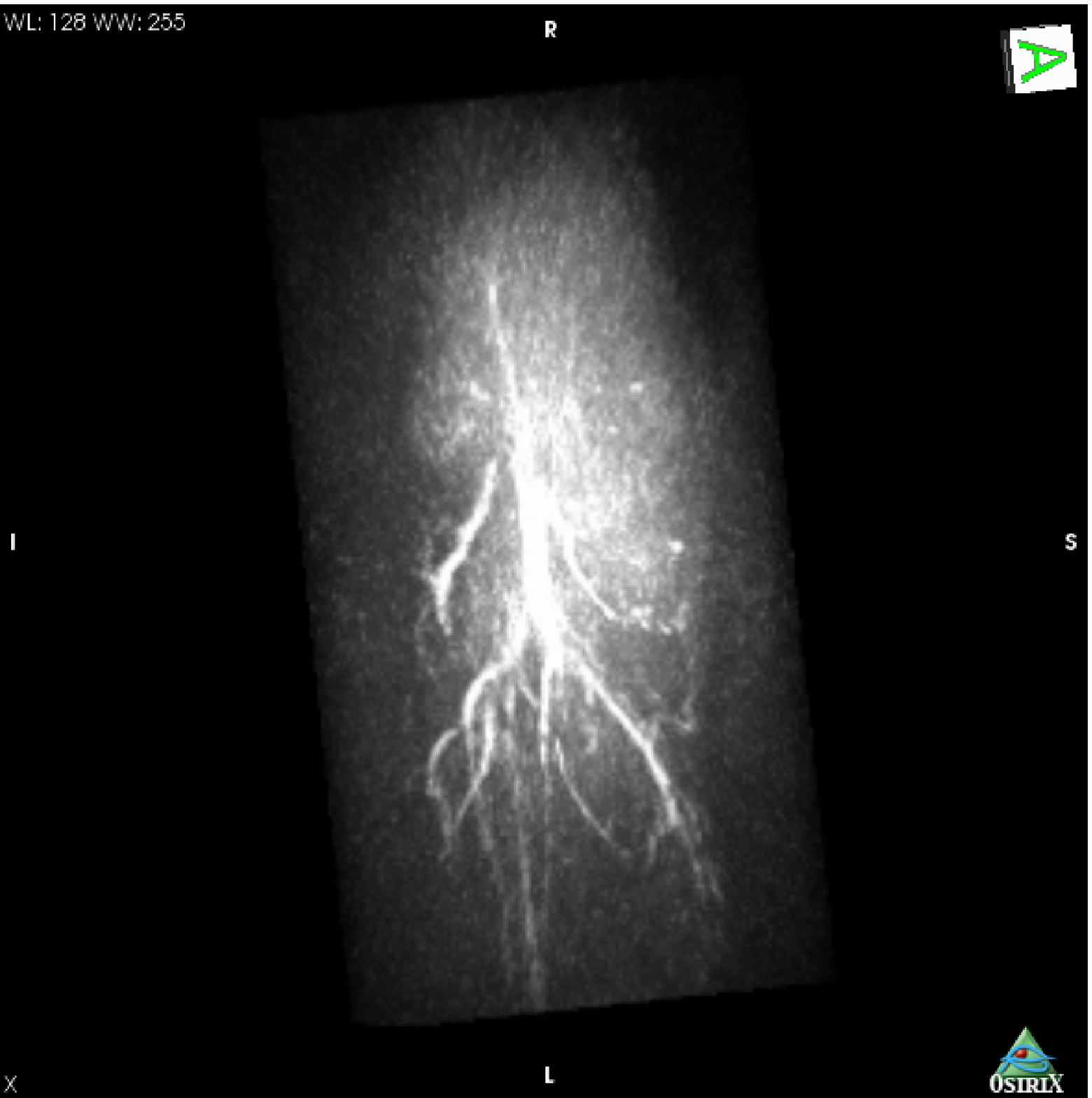}}
\hskip .5cm
\subfigure[]{\includegraphics[width=4cm, trim=6cm 1.5cm 6cm 4cm, clip]{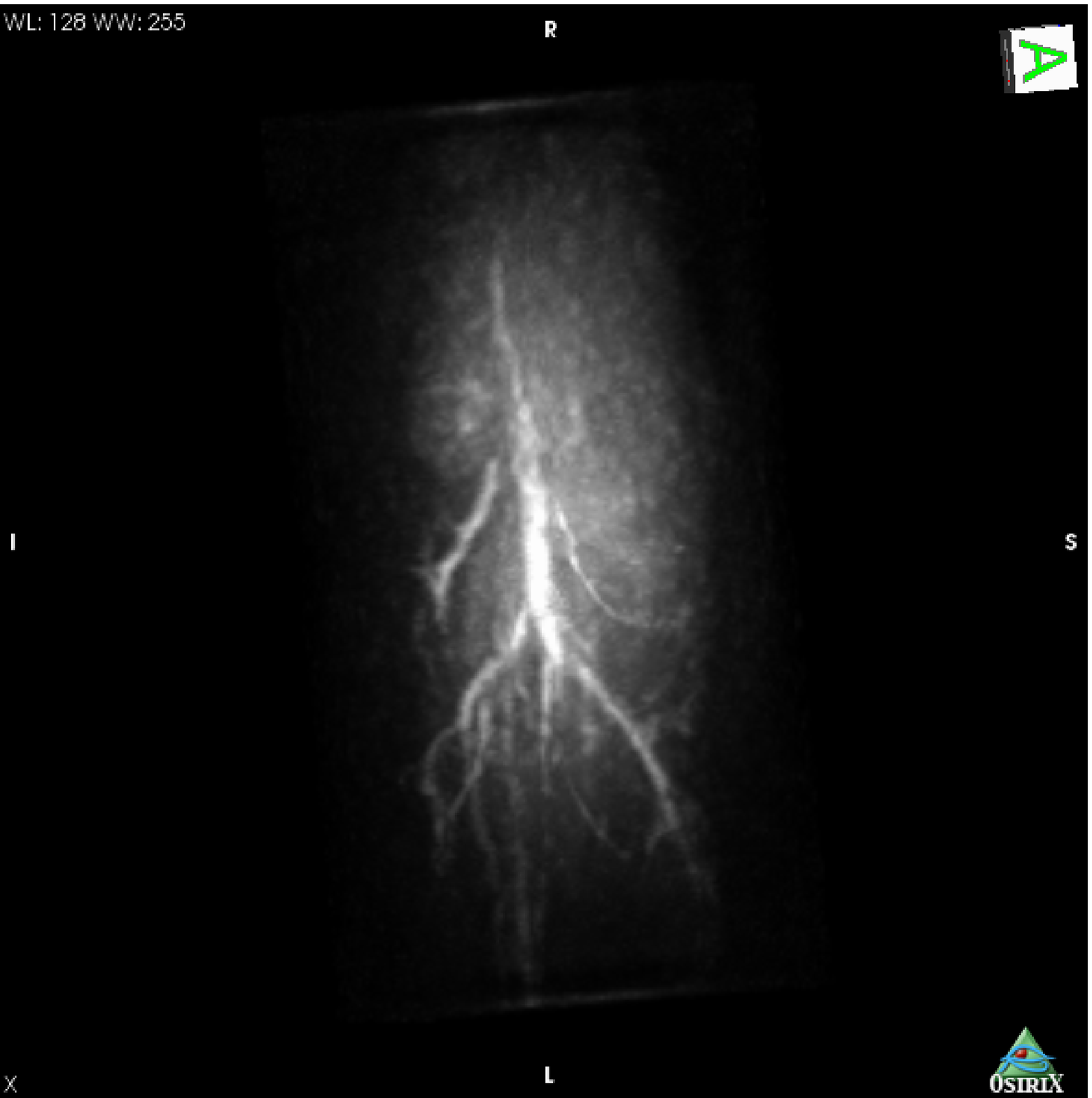}}
\hskip .5cm
\subfigure[]{\includegraphics[width=4cm, trim=6cm 1.5cm 6cm 4cm, clip]{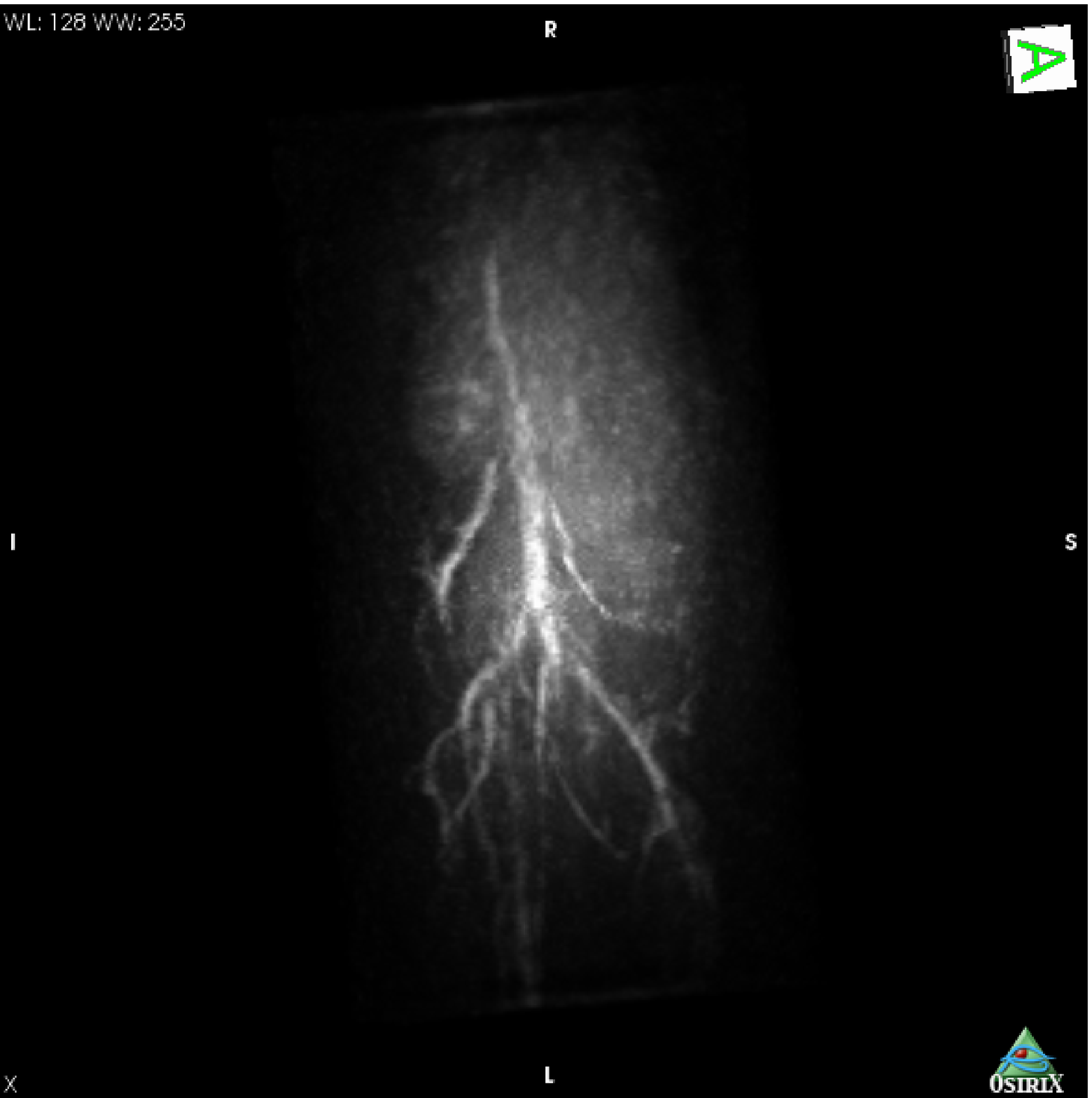}}
\end{center}
\caption{
\label{fig:mouseV1803D}
}
\end{figure}

The maximum intensity projection (MIP) of the 3D mouse images reconstructed by use of 
the GPU-based implementations reveal the mouse body vasculature as shown in Fig.\ \ref{fig:mouseV1803D}. 
Images reconstructed by use of both the PLS-Int and the PLS-Sph algorithms appear to have cleaner 
background than do the images reconstructed by use of the FBP algorithm from the same amount of data. 
All images reconstructed by iterative algorithms were obtained by $20$-iterations 
starting with uniform zeros as the initial guess. 
The PLS-Int algorithm took approximately a half day and $2$ days to process the ``quarter data''
and the ``full data'' respectively. 
The PLS-Sph algorithm took approximately one day and $4$ days to process the ``quarter data''
and the ``full data'' respectively.
Alternatively, if the CPU-based implementations were utilized, 
the PLS-Int algorithm would take an estimated $68$ days and $277$ days to process the ``quarter data''
and the ``full data'' respectively.
The PLS-Sph algorithm would take an estimated $275$ days and $1,100$ days to process the ``quarter data''
and the ``full data'' respectively.

\section{Discussion and conclusion}
\label{sect:summary}
In this study, we developed and investigated GPU-based implementations of the FBP algorithm 
and two pairs of projection/backprojection operators for 3D OAT. 
Our implementation of the FBP algorithm improved the computational efficiency 
over $1,000$ times compared to the CPU-based implementation. 
\iffalse
More importantly, our implementations of the projection/backprojection 
operators demonstrate the feasibility of  
3D iterative image reconstruction in practice. 
\fi
This work complements our earlier studies 
that demonstrated the feasibility of 3D iterative image reconstruction in practice
\cite{KunSPIETV:2012,Kun:PMB}.

Our current implementations of the iterative image reconstruction algorithms still require 
several days to process the densely sampled data set, which, however, can be further improved. 
Firstly, the amount of measurement data required for accurate image reconstruction 
can be further reduced by developing advanced image reconstruction methods \cite{OATTV09:Provost,GuoCS:2010,Meng:12,Kun:PMB}. 
Secondly, the number of iterations required can be reduced by developing fast-converging optimization algorithms \cite{FISTATIP:2009,Kun:PMB,Boyd:2011AMMD}. 

The proposed parallelization strategies by use of GPUs are of general interest. 
The implementation of the FBP algorithm \cite{Finch:02} can be adapted to  
other analytic image reconstruction algorithms, including those described in 
Refs.\ \onlinecite{Xu:2002,Xu:2005bp,Kunyansky:07,Finch:07,Schulze:2DInvPro}.  
We demonstrated the feasibility of PLS algorithm that utilized the proposed 
GPU-based implementations of the projection/backprojection operators. 
By use of these implementations, many advanced image reconstruction algorithms may also be
feasible in practice \cite{Kun:PMB}. 
Though we described our parallelization strategies for the projection/backprojection operators 
that utilized two discrete-to-discrete imaging models, 
% namely the interpolation-based and the 
% spherical-voxel-based imaging models, 
these strategies can also be applied to other D-D imaging models 
\cite{GPaltauf:2002,JCarson:2008,Ntziachristos:QPAT2012,Bu:2012}. 
Therefore, the proposed algorithms will facilitate the further investigation and application of 
advanced image reconstruction algorithms in 3D OAT.

\section*{acknowledgments}
This research was supported in part by NIH award EB010049 and CA167446.

\section*{appendix}

\noindent{\it Derivation of Equation\ \eqref{eqn:dis_srt}}

The integrated data function $g(\mathbf r^s, t )$ in Eqn.\ \eqref{eqn:srt}, 
evaluated at the $q$-th transducer and the $k$-th time instance,
can be expressed as:  
\begin{equation}\label{eqn:srtglobal}
  g(\mathbf r^s_q, t)\Big|_{t=k\Delta_t} 
 = \int_{|\mathbf r^s_q-\mathbf r|=kc_0\Delta_t}\!\!d \mathbf r\,A(\mathbf r), 
\end{equation}
where $\mathbf r^s_q$ denotes the location of the $q$-th point-like transducer. 
% is expressed in the global spherical coordinate system as 
% $(R^s, \theta^s_q, \phi^s_q)$, and $\mathbf r$ is expressed in the equivalent Cartesian coordinate system as $(x, y, z)$. 
We defined a local coordinate system, distinguished by a superscript `tr', centered at the $q$-th transducer with the $z^{\rm tr}$-axis pointing to the origin of the global coordinate system as shown in Fig.\ \ref{fig:geo}-(b). 
Assuming the object function $A(\mathbf r)$ is compactly supported in a sphere of radius $R$, 
the integral surface is symmetric about the $z^{\rm tr}$-axis. 
Thus, the orientations of the $x^{\rm tr}$- and $y^{\rm tr}$-axes can 
be arbitrary within the $z^{\rm tr}=0$ plane. 
Representing the right-hand side of Eqn.\ \eqref{eqn:srtglobal} in the local spherical coordinate system, 
one obtains  
\begin{equation}\label{eqn:srtlocal}
  g(\mathbf r^s_q,t)\Big|_{t=k\Delta_t} = (kc_0\Delta_t)^2\int_0^{\theta^{\rm tr}_{\rm max}}\!\!
   d\theta^{\rm tr}\,\sin\theta^{\rm tr}
   \int_0^{2\pi}\!\!d\phi^{\rm tr}\,
   A(kc_0\Delta_t,\theta^{\rm tr},\phi^{\rm tr}),
\end{equation}
where $\theta^{\rm tr}_{\rm max}$ is half of the apex angle of the cone that corresponds to the intersectional spherical cap as shown in Fig.\ \ref{fig:geo}-(b). 
The polar angle $\theta^{\rm tr}$ and the azimuth angle $\phi^{\rm tr}$ were discretized with intervals $\Delta_{\theta^{\rm tr}}$ and  $\Delta_{\phi^{\rm tr}}$ that satisfied 
\begin{equation}
  kc_0\Delta_t\Delta_{\theta^{\rm tr}} =  kc_0\Delta_t\sin\theta^{\rm tr}\Delta_{\phi^{\rm tr}}
   = \Delta_s. 
\end{equation}
Therefore, Eqn.\ \eqref{eqn:srtlocal} can be approximated by 
\begin{equation}\label{eqn:srtdislocal}
  g(\mathbf r^s_q,t)\Big|_{t=k\Delta_t}\approx \Delta_s^2 \sum_{i=0}^{N_i-1}
      \sum_{j=0}^{N_j-1}A(kc_0\Delta_t,\theta^{\rm tr}_i,\phi^{\rm tr}_j), 
\end{equation}
where 
$N_i = \lfloor \theta^{\rm tr}_{\rm max}/\Delta_{\theta^{\rm tr}} \rfloor$, 
$N_j = \lfloor 2\pi/\Delta_{\phi^{\rm tr}} \rfloor$, 
$\theta^{\rm tr}_i = i\Delta_{\theta^{\rm tr}}$,
and 
$\phi^{\rm tr}_j = j\Delta_{\phi^{\rm tr}}$.
We denoted by $\mathbf r_{k,i,j}$ the location in the global coordinate system corresponding to 
the location vector $(kc_0\Delta_t,\theta^{\rm tr}_i,\phi^{\rm tr}_j)$ in the local coordinate 
system in Eqn.\ \eqref{eqn:srtdislocal}. 
On substitution from the finite-dimensional representation Eqn. \eqref{eqn:dis_obj} 
into Eqn.\ \eqref{eqn:srtdislocal} with $\boldsymbol\alpha$ and $\boldsymbol\psi_n(\mathbf r)$ 
defined by Eqns.\ \eqref{eqn:coeffint} and \eqref{eqn:expfunI}, respectively, we obtained:
\begin{equation}
  g(\mathbf r^s_q,t)\Big|_{t=k\Delta_t}\approx \Delta_s^2 
      \sum_{n=0}^{N-1} \big[\boldsymbol\alpha_{\rm int}\big]_n
      \sum_{i=0}^{N_i-1}
      \sum_{j=0}^{N_j-1}\psi_n^{\rm int}(\mathbf r_{k,i,j})
      \equiv \big[\mathbf g\big]_{qK+k}. 
\end{equation}

\newpage
% \bibliography{reflect}

\begin{thebibliography}{60}%
\makeatletter
\providecommand \@ifxundefined [1]{%
 \@ifx{#1\undefined}
}%
\providecommand \@ifnum [1]{%
 \ifnum #1\expandafter \@firstoftwo
 \else \expandafter \@secondoftwo
 \fi
}%
\providecommand \@ifx [1]{%
 \ifx #1\expandafter \@firstoftwo
 \else \expandafter \@secondoftwo
 \fi
}%
\providecommand \natexlab [1]{#1}%
\providecommand \enquote  [1]{``#1''}%
\providecommand \bibnamefont  [1]{#1}%
\providecommand \bibfnamefont [1]{#1}%
\providecommand \citenamefont [1]{#1}%
\providecommand \href@noop [0]{\@secondoftwo}%
\providecommand \href [0]{\begingroup \@sanitize@url \@href}%
\providecommand \@href[1]{\@@startlink{#1}\@@href}%
\providecommand \@@href[1]{\endgroup#1\@@endlink}%
\providecommand \@sanitize@url [0]{\catcode `\\12\catcode `\$12\catcode
  `\&12\catcode `\#12\catcode `\^12\catcode `\_12\catcode `\%12\relax}%
\providecommand \@@startlink[1]{}%
\providecommand \@@endlink[0]{}%
\providecommand \url  [0]{\begingroup\@sanitize@url \@url }%
\providecommand \@url [1]{\endgroup\@href {#1}{\urlprefix }}%
\providecommand \urlprefix  [0]{URL }%
\providecommand \Eprint [0]{\href }%
\providecommand \doibase [0]{http://dx.doi.org/}%
\providecommand \selectlanguage [0]{\@gobble}%
\providecommand \bibinfo  [0]{\@secondoftwo}%
\providecommand \bibfield  [0]{\@secondoftwo}%
\providecommand \translation [1]{[#1]}%
\providecommand \BibitemOpen [0]{}%
\providecommand \bibitemStop [0]{}%
\providecommand \bibitemNoStop [0]{.\EOS\space}%
\providecommand \EOS [0]{\spacefactor3000\relax}%
\providecommand \BibitemShut  [1]{\csname bibitem#1\endcsname}%
\let\auto@bib@innerbib\@empty
\bibitem [{\citenamefont {$\mathrm {Oraevsky}$}\ and\ \citenamefont
  {Karabutov}(2003)}]{OraBook}%
  \BibitemOpen
  \bibfield  {author} {\bibinfo {author} {\bibfnamefont {A.~A.}\ \bibnamefont
  {$\mathrm {Oraevsky}$}}\ and\ \bibinfo {author} {\bibfnamefont {A.~A.}\
  \bibnamefont {Karabutov}},\ }\bibfield  {title} {\enquote {\bibinfo {title}
  {Optoacoustic tomography},}\ }in\ \href@noop {} {\emph {\bibinfo {booktitle}
  {Biomedical Photonics Handbook}}},\ \bibinfo {editor} {edited by\ \bibinfo
  {editor} {\bibfnamefont {T.}~\bibnamefont {Vo-Dinh}}}\ (\bibinfo  {publisher}
  {CRC Press LLC},\ \bibinfo {year} {2003})\ Chap.~\bibinfo {chapter}
  {34}\BibitemShut {NoStop}%
\bibitem [{\citenamefont {Wang}(2008)}]{WangTutorial}%
  \BibitemOpen
  \bibfield  {author} {\bibinfo {author} {\bibfnamefont {L.~V.}\ \bibnamefont
  {Wang}},\ }\bibfield  {title} {\enquote {\bibinfo {title} {Tutorial on
  photoacoustic microscopy and computed tomography},}\ }\href@noop {}
  {\bibfield  {journal} {\bibinfo  {journal} {IEEE Journal of Selected Topics
  in Quantum Electronics}\ }\textbf {\bibinfo {volume} {14}},\ \bibinfo {pages}
  {171--179} (\bibinfo {year} {2008})}\BibitemShut {NoStop}%
\bibitem [{\citenamefont {Kruger}, \citenamefont {Reinecke},\ and\
  \citenamefont {Kruger}(1999)}]{KrugerMP1999}%
  \BibitemOpen
  \bibfield  {author} {\bibinfo {author} {\bibfnamefont {R.}~\bibnamefont
  {Kruger}}, \bibinfo {author} {\bibfnamefont {D.}~\bibnamefont {Reinecke}}, \
  and\ \bibinfo {author} {\bibfnamefont {G.}~\bibnamefont {Kruger}},\
  }\bibfield  {title} {\enquote {\bibinfo {title} {Thermoacoustic computed
  tomography- technical considerations},}\ }\href@noop {} {\bibfield  {journal}
  {\bibinfo  {journal} {Medical Physics}\ }\textbf {\bibinfo {volume} {26}},\
  \bibinfo {pages} {1832--1837} (\bibinfo {year} {1999})}\BibitemShut {NoStop}%
\bibitem [{\citenamefont {Cox}\ \emph {et~al.}(2006)\citenamefont {Cox},
  \citenamefont {Arridge}, \citenamefont {K\"{o}stli},\ and\ \citenamefont
  {Beard}}]{CoxOSA2006}%
  \BibitemOpen
  \bibfield  {author} {\bibinfo {author} {\bibfnamefont {B.~T.}\ \bibnamefont
  {Cox}}, \bibinfo {author} {\bibfnamefont {S.~R.}\ \bibnamefont {Arridge}},
  \bibinfo {author} {\bibfnamefont {K.~P.}\ \bibnamefont {K\"{o}stli}}, \ and\
  \bibinfo {author} {\bibfnamefont {P.~C.}\ \bibnamefont {Beard}},\ }\bibfield
  {title} {\enquote {\bibinfo {title} {Two-dimensional quantitative
  photoacoustic image reconstruction of absorption distributions in scattering
  media by use of a simple iterative method},}\ }\href {\doibase
  10.1364/AO.45.001866} {\bibfield  {journal} {\bibinfo  {journal} {Appl.
  Opt.}\ }\textbf {\bibinfo {volume} {45}},\ \bibinfo {pages} {1866--1875}
  (\bibinfo {year} {2006})}\BibitemShut {NoStop}%
\bibitem [{\citenamefont {Kunyansky}(2007)}]{Kunyansky:07}%
  \BibitemOpen
  \bibfield  {author} {\bibinfo {author} {\bibfnamefont {L.~A.}\ \bibnamefont
  {Kunyansky}},\ }\bibfield  {title} {\enquote {\bibinfo {title} {Explicit
  inversion formulae for the spherical mean {R}adon transform},}\ }\href@noop
  {} {\bibfield  {journal} {\bibinfo  {journal} {Inverse Problems}\ }\textbf
  {\bibinfo {volume} {23}},\ \bibinfo {pages} {373--383} (\bibinfo {year}
  {2007})}\BibitemShut {NoStop}%
\bibitem [{\citenamefont {Finch}, \citenamefont {Patch},\ and\ \citenamefont
  {Rakesh}(2004)}]{Finch:02}%
  \BibitemOpen
  \bibfield  {author} {\bibinfo {author} {\bibfnamefont {D.}~\bibnamefont
  {Finch}}, \bibinfo {author} {\bibfnamefont {S.}~\bibnamefont {Patch}}, \ and\
  \bibinfo {author} {\bibnamefont {Rakesh}},\ }\bibfield  {title} {\enquote
  {\bibinfo {title} {Determining a function from its mean values over a family
  of spheres},}\ }\href@noop {} {\bibfield  {journal} {\bibinfo  {journal}
  {SIAM Journal of Mathematical Analysis}\ }\textbf {\bibinfo {volume} {35}},\
  \bibinfo {pages} {1213--1240} (\bibinfo {year} {2004})}\BibitemShut {NoStop}%
\bibitem [{\citenamefont {Xu}\ and\ \citenamefont {Wang}(2005)}]{Xu:2005bp}%
  \BibitemOpen
  \bibfield  {author} {\bibinfo {author} {\bibfnamefont {M.}~\bibnamefont
  {Xu}}\ and\ \bibinfo {author} {\bibfnamefont {L.~V.}\ \bibnamefont {Wang}},\
  }\bibfield  {title} {\enquote {\bibinfo {title} {Universal back-projection
  algorithm for photoacoustic computed tomography},}\ }\href@noop {} {\bibfield
   {journal} {\bibinfo  {journal} {Physical Review E}\ }\textbf {\bibinfo
  {volume} {71}} (\bibinfo {year} {2005})}\BibitemShut {NoStop}%
\bibitem [{\citenamefont {{Xu}}, \citenamefont {Feng},\ and\ \citenamefont
  {Wang}(2002)}]{Xu:planar}%
  \BibitemOpen
  \bibfield  {author} {\bibinfo {author} {\bibfnamefont {Y.}~\bibnamefont
  {{Xu}}}, \bibinfo {author} {\bibfnamefont {D.}~\bibnamefont {Feng}}, \ and\
  \bibinfo {author} {\bibfnamefont {L.~V.}\ \bibnamefont {Wang}},\ }\bibfield
  {title} {\enquote {\bibinfo {title} {Exact frequency-domain reconstruction
  for thermoacoustic tomography: {{\rm I.}} {{\rm P}}lanar geometry},}\
  }\href@noop {} {\bibfield  {journal} {\bibinfo  {journal} {IEEE Transactions
  on Medical Imaging}\ }\textbf {\bibinfo {volume} {21}},\ \bibinfo {pages}
  {823--828} (\bibinfo {year} {2002})}\BibitemShut {NoStop}%
\bibitem [{\citenamefont {Paltauf}\ \emph {et~al.}(2002)\citenamefont
  {Paltauf}, \citenamefont {Viator}, \citenamefont {Prahl},\ and\ \citenamefont
  {Jacques}}]{GPaltauf:2002}%
  \BibitemOpen
  \bibfield  {author} {\bibinfo {author} {\bibfnamefont {G.}~\bibnamefont
  {Paltauf}}, \bibinfo {author} {\bibfnamefont {J.~A.}\ \bibnamefont {Viator}},
  \bibinfo {author} {\bibfnamefont {S.~A.}\ \bibnamefont {Prahl}}, \ and\
  \bibinfo {author} {\bibfnamefont {S.~L.}\ \bibnamefont {Jacques}},\
  }\bibfield  {title} {\enquote {\bibinfo {title} {Iterative reconstruction
  algorithm for optoacoustic imaging},}\ }\href {\doibase 10.1121/1.1501898}
  {\bibfield  {journal} {\bibinfo  {journal} {The Journal of the Acoustical
  Society of America}\ }\textbf {\bibinfo {volume} {112}},\ \bibinfo {pages}
  {1536--1544} (\bibinfo {year} {2002})}\BibitemShut {NoStop}%
\bibitem [{\citenamefont {Yuan}\ and\ \citenamefont
  {Jiang}(2007)}]{HJiang:2007}%
  \BibitemOpen
  \bibfield  {author} {\bibinfo {author} {\bibfnamefont {Z.}~\bibnamefont
  {Yuan}}\ and\ \bibinfo {author} {\bibfnamefont {H.}~\bibnamefont {Jiang}},\
  }\bibfield  {title} {\enquote {\bibinfo {title} {Three-dimensional
  finite-element-based photoacoustic tomography: Reconstruction algorithm and
  simulations},}\ }\href {\doibase 10.1118/1.2409234} {\bibfield  {journal}
  {\bibinfo  {journal} {Medical Physics}\ }\textbf {\bibinfo {volume} {34}},\
  \bibinfo {pages} {538--546} (\bibinfo {year} {2007})}\BibitemShut {NoStop}%
\bibitem [{\citenamefont {Ephrat}\ \emph
  {et~al.}(2008{\natexlab{a}})\citenamefont {Ephrat}, \citenamefont
  {Keenliside}, \citenamefont {Seabrook}, \citenamefont {Prato},\ and\
  \citenamefont {Carson}}]{JCarson:2008}%
  \BibitemOpen
  \bibfield  {author} {\bibinfo {author} {\bibfnamefont {P.}~\bibnamefont
  {Ephrat}}, \bibinfo {author} {\bibfnamefont {L.}~\bibnamefont {Keenliside}},
  \bibinfo {author} {\bibfnamefont {A.}~\bibnamefont {Seabrook}}, \bibinfo
  {author} {\bibfnamefont {F.~S.}\ \bibnamefont {Prato}}, \ and\ \bibinfo
  {author} {\bibfnamefont {J.~J.~L.}\ \bibnamefont {Carson}},\ }\bibfield
  {title} {\enquote {\bibinfo {title} {Three-dimensional photoacoustic imaging
  by sparse-array detection and iterative image reconstruction},}\ }\href
  {\doibase 10.1117/1.2992131} {\bibfield  {journal} {\bibinfo  {journal}
  {Journal of Biomedical Optics}\ }\textbf {\bibinfo {volume} {13}},\ \bibinfo
  {eid} {054052} (\bibinfo {year} {2008}{\natexlab{a}})}\BibitemShut {NoStop}%
\bibitem [{\citenamefont {Zhang}\ \emph {et~al.}(2009)\citenamefont {Zhang},
  \citenamefont {Anastasio}, \citenamefont {La~Riviere},\ and\ \citenamefont
  {Wang}}]{Jin:2009}%
  \BibitemOpen
  \bibfield  {author} {\bibinfo {author} {\bibfnamefont {J.}~\bibnamefont
  {Zhang}}, \bibinfo {author} {\bibfnamefont {M.}~\bibnamefont {Anastasio}},
  \bibinfo {author} {\bibfnamefont {P.}~\bibnamefont {La~Riviere}}, \ and\
  \bibinfo {author} {\bibfnamefont {L.}~\bibnamefont {Wang}},\ }\bibfield
  {title} {\enquote {\bibinfo {title} {Effects of different imaging models on
  least-squares image reconstruction accuracy in photoacoustic tomography},}\
  }\href {\doibase 10.1109/TMI.2009.2024082} {\bibfield  {journal} {\bibinfo
  {journal} {Medical Imaging, IEEE Transactions on}\ }\textbf {\bibinfo
  {volume} {28}},\ \bibinfo {pages} {1781 --1790} (\bibinfo {year}
  {2009})}\BibitemShut {NoStop}%
\bibitem [{\citenamefont {Provost}\ and\ \citenamefont
  {Lesage}(2009)}]{OATTV09:Provost}%
  \BibitemOpen
  \bibfield  {author} {\bibinfo {author} {\bibfnamefont {J.}~\bibnamefont
  {Provost}}\ and\ \bibinfo {author} {\bibfnamefont {F.}~\bibnamefont
  {Lesage}},\ }\bibfield  {title} {\enquote {\bibinfo {title} {The application
  of compressed sensing for photo-acoustic tomography},}\ }\href {\doibase
  10.1109/TMI.2008.2007825} {\bibfield  {journal} {\bibinfo  {journal} {Medical
  Imaging, IEEE Transactions on}\ }\textbf {\bibinfo {volume} {28}},\ \bibinfo
  {pages} {585 --594} (\bibinfo {year} {2009})}\BibitemShut {NoStop}%
\bibitem [{\citenamefont {Wang}\ \emph {et~al.}(2011)\citenamefont {Wang},
  \citenamefont {Ermilov}, \citenamefont {Su}, \citenamefont {Brecht},
  \citenamefont {Oraevsky},\ and\ \citenamefont {Anastasio}}]{TMI:transmodel}%
  \BibitemOpen
  \bibfield  {author} {\bibinfo {author} {\bibfnamefont {K.}~\bibnamefont
  {Wang}}, \bibinfo {author} {\bibfnamefont {S.~A.}\ \bibnamefont {Ermilov}},
  \bibinfo {author} {\bibfnamefont {R.}~\bibnamefont {Su}}, \bibinfo {author}
  {\bibfnamefont {H.-P.}\ \bibnamefont {Brecht}}, \bibinfo {author}
  {\bibfnamefont {A.~A.}\ \bibnamefont {Oraevsky}}, \ and\ \bibinfo {author}
  {\bibfnamefont {M.~A.}\ \bibnamefont {Anastasio}},\ }\bibfield  {title}
  {\enquote {\bibinfo {title} {An imaging model incorporating ultrasonic
  transducer properties for three-dimensional optoacoustic tomography},}\
  }\href {\doibase 10.1109/TMI.2010.2072514} {\bibfield  {journal} {\bibinfo
  {journal} {Medical Imaging, IEEE Transactions on}\ }\textbf {\bibinfo
  {volume} {30}},\ \bibinfo {pages} {203 --214} (\bibinfo {year}
  {2011})}\BibitemShut {NoStop}%
\bibitem [{\citenamefont {Guo}\ \emph {et~al.}(2010)\citenamefont {Guo},
  \citenamefont {Li}, \citenamefont {Song},\ and\ \citenamefont
  {Wang}}]{GuoCS:2010}%
  \BibitemOpen
  \bibfield  {author} {\bibinfo {author} {\bibfnamefont {Z.}~\bibnamefont
  {Guo}}, \bibinfo {author} {\bibfnamefont {C.}~\bibnamefont {Li}}, \bibinfo
  {author} {\bibfnamefont {L.}~\bibnamefont {Song}}, \ and\ \bibinfo {author}
  {\bibfnamefont {L.~V.}\ \bibnamefont {Wang}},\ }\bibfield  {title} {\enquote
  {\bibinfo {title} {Compressed sensing in photoacoustic tomography in vivo},}\
  }\href {\doibase 10.1117/1.3381187} {\bibfield  {journal} {\bibinfo
  {journal} {Journal of Biomedical Optics}\ }\textbf {\bibinfo {volume} {15}},\
  \bibinfo {eid} {021311} (\bibinfo {year} {2010})}\BibitemShut {NoStop}%
\bibitem [{\citenamefont {Huang}, \citenamefont {Oraevsky},\ and\ \citenamefont
  {Anastasio}(2010)}]{Huang:SPIEBoundEnhance}%
  \BibitemOpen
  \bibfield  {author} {\bibinfo {author} {\bibfnamefont {C.}~\bibnamefont
  {Huang}}, \bibinfo {author} {\bibfnamefont {A.~A.}\ \bibnamefont {Oraevsky}},
  \ and\ \bibinfo {author} {\bibfnamefont {M.~A.}\ \bibnamefont {Anastasio}},\
  }\bibfield  {title} {\enquote {\bibinfo {title} {Investigation of
  limited-view image reconstruction in optoacoustic tomography employing a
  priori structural information},}\ \ }(\bibinfo  {publisher} {SPIE},\ \bibinfo
  {year} {2010})\ p.\ \bibinfo {pages} {780004}\BibitemShut {NoStop}%
\bibitem [{\citenamefont {Xu}, \citenamefont {Li},\ and\ \citenamefont
  {Wang}(2010)}]{xu:water}%
  \BibitemOpen
  \bibfield  {author} {\bibinfo {author} {\bibfnamefont {Z.}~\bibnamefont
  {Xu}}, \bibinfo {author} {\bibfnamefont {C.}~\bibnamefont {Li}}, \ and\
  \bibinfo {author} {\bibfnamefont {L.~V.}\ \bibnamefont {Wang}},\ }\bibfield
  {title} {\enquote {\bibinfo {title} {Photoacoustic tomography of water in
  phantoms and tissue},}\ }\href {\doibase 10.1117/1.3443793} {\bibfield
  {journal} {\bibinfo  {journal} {Journal of Biomedical Optics}\ }\textbf
  {\bibinfo {volume} {15}},\ \bibinfo {eid} {036019} (\bibinfo {year}
  {2010})}\BibitemShut {NoStop}%
\bibitem [{\citenamefont {Xu}, \citenamefont {Zhu},\ and\ \citenamefont
  {Wang}(2011)}]{xu:edema}%
  \BibitemOpen
  \bibfield  {author} {\bibinfo {author} {\bibfnamefont {Z.}~\bibnamefont
  {Xu}}, \bibinfo {author} {\bibfnamefont {Q.}~\bibnamefont {Zhu}}, \ and\
  \bibinfo {author} {\bibfnamefont {L.~V.}\ \bibnamefont {Wang}},\ }\bibfield
  {title} {\enquote {\bibinfo {title} {In vivo photoacoustic tomography of
  mouse cerebral edema induced by cold injury},}\ }\href {\doibase
  10.1117/1.3584847} {\bibfield  {journal} {\bibinfo  {journal} {Journal of
  Biomedical Optics}\ }\textbf {\bibinfo {volume} {16}},\ \bibinfo {eid}
  {066020} (\bibinfo {year} {2011})}\BibitemShut {NoStop}%
\bibitem [{\citenamefont {Buehler}\ \emph {et~al.}(2011)\citenamefont
  {Buehler}, \citenamefont {Rosenthal}, \citenamefont {Jetzfellner},
  \citenamefont {Dima}, \citenamefont {Razansky},\ and\ \citenamefont
  {Ntziachristos}}]{Ntziachristos:2011}%
  \BibitemOpen
  \bibfield  {author} {\bibinfo {author} {\bibfnamefont {A.}~\bibnamefont
  {Buehler}}, \bibinfo {author} {\bibfnamefont {A.}~\bibnamefont {Rosenthal}},
  \bibinfo {author} {\bibfnamefont {T.}~\bibnamefont {Jetzfellner}}, \bibinfo
  {author} {\bibfnamefont {A.}~\bibnamefont {Dima}}, \bibinfo {author}
  {\bibfnamefont {D.}~\bibnamefont {Razansky}}, \ and\ \bibinfo {author}
  {\bibfnamefont {V.}~\bibnamefont {Ntziachristos}},\ }\bibfield  {title}
  {\enquote {\bibinfo {title} {Model-based optoacoustic inversions with
  incomplete projection data},}\ }\href {\doibase 10.1118/1.3556916} {\bibfield
   {journal} {\bibinfo  {journal} {Medical Physics}\ }\textbf {\bibinfo
  {volume} {38}},\ \bibinfo {pages} {1694--1704} (\bibinfo {year}
  {2011})}\BibitemShut {NoStop}%
\bibitem [{\citenamefont {Bu}\ \emph {et~al.}(2012)\citenamefont {Bu},
  \citenamefont {Liu}, \citenamefont {Shiina}, \citenamefont {Kondo},
  \citenamefont {Yamakawa}, \citenamefont {Fukutani}, \citenamefont {Someda},\
  and\ \citenamefont {Asao}}]{Bu:2012}%
  \BibitemOpen
  \bibfield  {author} {\bibinfo {author} {\bibfnamefont {S.}~\bibnamefont
  {Bu}}, \bibinfo {author} {\bibfnamefont {Z.}~\bibnamefont {Liu}}, \bibinfo
  {author} {\bibfnamefont {T.}~\bibnamefont {Shiina}}, \bibinfo {author}
  {\bibfnamefont {K.}~\bibnamefont {Kondo}}, \bibinfo {author} {\bibfnamefont
  {M.}~\bibnamefont {Yamakawa}}, \bibinfo {author} {\bibfnamefont
  {K.}~\bibnamefont {Fukutani}}, \bibinfo {author} {\bibfnamefont
  {Y.}~\bibnamefont {Someda}}, \ and\ \bibinfo {author} {\bibfnamefont
  {Y.}~\bibnamefont {Asao}},\ }\bibfield  {title} {\enquote {\bibinfo {title}
  {Model-based reconstruction integrated with fluence compensation for
  photoacoustic tomography},}\ }\href {\doibase 10.1109/TBME.2012.2187649}
  {\bibfield  {journal} {\bibinfo  {journal} {Biomedical Engineering, IEEE
  Transactions on}\ }\textbf {\bibinfo {volume} {59}},\ \bibinfo {pages} {1354
  --1363} (\bibinfo {year} {2012})}\BibitemShut {NoStop}%
\bibitem [{\citenamefont {Wang}\ \emph
  {et~al.}(2012{\natexlab{a}})\citenamefont {Wang}, \citenamefont {Su},
  \citenamefont {Oraevsky},\ and\ \citenamefont {Anastasio}}]{KunSPIETV:2012}%
  \BibitemOpen
  \bibfield  {author} {\bibinfo {author} {\bibfnamefont {K.}~\bibnamefont
  {Wang}}, \bibinfo {author} {\bibfnamefont {R.}~\bibnamefont {Su}}, \bibinfo
  {author} {\bibfnamefont {A.~A.}\ \bibnamefont {Oraevsky}}, \ and\ \bibinfo
  {author} {\bibfnamefont {M.~A.}\ \bibnamefont {Anastasio}},\ }\bibfield
  {title} {\enquote {\bibinfo {title} {Investigation of iterative image
  reconstruction in optoacoustic tomography},}\ \ }(\bibinfo  {publisher}
  {SPIE},\ \bibinfo {year} {2012})\ p.\ \bibinfo {pages} {82231Y}\BibitemShut
  {NoStop}%
\bibitem [{\citenamefont {Wang}\ \emph
  {et~al.}(2012{\natexlab{b}})\citenamefont {Wang}, \citenamefont {Su},
  \citenamefont {Oraevsky},\ and\ \citenamefont {Anastasio}}]{Kun:PMB}%
  \BibitemOpen
  \bibfield  {author} {\bibinfo {author} {\bibfnamefont {K.}~\bibnamefont
  {Wang}}, \bibinfo {author} {\bibfnamefont {R.}~\bibnamefont {Su}}, \bibinfo
  {author} {\bibfnamefont {A.~A.}\ \bibnamefont {Oraevsky}}, \ and\ \bibinfo
  {author} {\bibfnamefont {M.~A.}\ \bibnamefont {Anastasio}},\ }\bibfield
  {title} {\enquote {\bibinfo {title} {Investigation of iterative image
  reconstruction in three-dimensional optoacoustic tomography},}\ }\href
  {http://stacks.iop.org/0031-9155/57/i=17/a=5399} {\bibfield  {journal}
  {\bibinfo  {journal} {Physics in Medicine and Biology}\ }\textbf {\bibinfo
  {volume} {57}},\ \bibinfo {pages} {5399} (\bibinfo {year}
  {2012}{\natexlab{b}})}\BibitemShut {NoStop}%
\bibitem [{\citenamefont {Huang}\ \emph
  {et~al.}(2012{\natexlab{a}})\citenamefont {Huang}, \citenamefont {Nie},
  \citenamefont {Schoonover}, \citenamefont {Guo}, \citenamefont {Schirra},
  \citenamefont {Anastasio},\ and\ \citenamefont {Wang}}]{HuangchaoJBObrain}%
  \BibitemOpen
  \bibfield  {author} {\bibinfo {author} {\bibfnamefont {C.}~\bibnamefont
  {Huang}}, \bibinfo {author} {\bibfnamefont {L.}~\bibnamefont {Nie}}, \bibinfo
  {author} {\bibfnamefont {R.~W.}\ \bibnamefont {Schoonover}}, \bibinfo
  {author} {\bibfnamefont {Z.}~\bibnamefont {Guo}}, \bibinfo {author}
  {\bibfnamefont {C.~O.}\ \bibnamefont {Schirra}}, \bibinfo {author}
  {\bibfnamefont {M.~A.}\ \bibnamefont {Anastasio}}, \ and\ \bibinfo {author}
  {\bibfnamefont {L.~V.}\ \bibnamefont {Wang}},\ }\bibfield  {title} {\enquote
  {\bibinfo {title} {Aberration correction for transcranial photoacoustic
  tomography of primates employing adjunct image data},}\ }\href {\doibase
  10.1117/1.JBO.17.6.066016} {\bibfield  {journal} {\bibinfo  {journal}
  {Journal of Biomedical Optics}\ }\textbf {\bibinfo {volume} {17}},\ \bibinfo
  {eid} {066016} (\bibinfo {year} {2012}{\natexlab{a}})}\BibitemShut {NoStop}%
\bibitem [{\citenamefont {Dean-Ben}\ \emph {et~al.}(2012)\citenamefont
  {Dean-Ben}, \citenamefont {Buehler}, \citenamefont {Ntziachristos},\ and\
  \citenamefont {Razansky}}]{German:3D12}%
  \BibitemOpen
  \bibfield  {author} {\bibinfo {author} {\bibfnamefont {X.}~\bibnamefont
  {Dean-Ben}}, \bibinfo {author} {\bibfnamefont {A.}~\bibnamefont {Buehler}},
  \bibinfo {author} {\bibfnamefont {V.}~\bibnamefont {Ntziachristos}}, \ and\
  \bibinfo {author} {\bibfnamefont {D.}~\bibnamefont {Razansky}},\ }\bibfield
  {title} {\enquote {\bibinfo {title} {Accurate model-based reconstruction
  algorithm for three-dimensional optoacoustic tomography},}\ }\href {\doibase
  10.1109/TMI.2012.2208471} {\bibfield  {journal} {\bibinfo  {journal} {Medical
  Imaging, IEEE Transactions on}\ }\textbf {\bibinfo {volume} {31}},\ \bibinfo
  {pages} {1922 --1928} (\bibinfo {year} {2012})}\BibitemShut {NoStop}%
\bibitem [{\citenamefont {Huang}\ \emph
  {et~al.}(2012{\natexlab{b}})\citenamefont {Huang}, \citenamefont {Nie},
  \citenamefont {Schoonover}, \citenamefont {Wang},\ and\ \citenamefont
  {Anastasio}}]{HuangchaoJBOatten}%
  \BibitemOpen
  \bibfield  {author} {\bibinfo {author} {\bibfnamefont {C.}~\bibnamefont
  {Huang}}, \bibinfo {author} {\bibfnamefont {L.}~\bibnamefont {Nie}}, \bibinfo
  {author} {\bibfnamefont {R.~W.}\ \bibnamefont {Schoonover}}, \bibinfo
  {author} {\bibfnamefont {L.~V.}\ \bibnamefont {Wang}}, \ and\ \bibinfo
  {author} {\bibfnamefont {M.~A.}\ \bibnamefont {Anastasio}},\ }\bibfield
  {title} {\enquote {\bibinfo {title} {Photoacoustic computed tomography
  correcting for heterogeneity and attenuation},}\ }\href {\doibase
  10.1117/1.JBO.17.6.061211} {\bibfield  {journal} {\bibinfo  {journal}
  {Journal of Biomedical Optics}\ }\textbf {\bibinfo {volume} {17}},\ \bibinfo
  {eid} {061211} (\bibinfo {year} {2012}{\natexlab{b}})}\BibitemShut {NoStop}%
\bibitem [{\citenamefont {Wang}\ \emph
  {et~al.}(2012{\natexlab{c}})\citenamefont {Wang}, \citenamefont {Xiang},
  \citenamefont {Jiang}, \citenamefont {Yang}, \citenamefont {Zhang},
  \citenamefont {Carney},\ and\ \citenamefont {Jiang}}]{Wang:12}%
  \BibitemOpen
  \bibfield  {author} {\bibinfo {author} {\bibfnamefont {B.}~\bibnamefont
  {Wang}}, \bibinfo {author} {\bibfnamefont {L.}~\bibnamefont {Xiang}},
  \bibinfo {author} {\bibfnamefont {M.~S.}\ \bibnamefont {Jiang}}, \bibinfo
  {author} {\bibfnamefont {J.}~\bibnamefont {Yang}}, \bibinfo {author}
  {\bibfnamefont {Q.}~\bibnamefont {Zhang}}, \bibinfo {author} {\bibfnamefont
  {P.~R.}\ \bibnamefont {Carney}}, \ and\ \bibinfo {author} {\bibfnamefont
  {H.}~\bibnamefont {Jiang}},\ }\bibfield  {title} {\enquote {\bibinfo {title}
  {Photoacoustic tomography system for noninvasive real-time three-dimensional
  imaging of epilepsy},}\ }\href {\doibase 10.1364/BOE.3.001427} {\bibfield
  {journal} {\bibinfo  {journal} {Biomed. Opt. Express}\ }\textbf {\bibinfo
  {volume} {3}},\ \bibinfo {pages} {1427--1432} (\bibinfo {year}
  {2012}{\natexlab{c}})}\BibitemShut {NoStop}%
\bibitem [{\citenamefont {Buehler}\ \emph {et~al.}(2012)\citenamefont
  {Buehler}, \citenamefont {De\'{a}n-Ben}, \citenamefont {Claussen},
  \citenamefont {Ntziachristos},\ and\ \citenamefont {Razansky}}]{Buehler:12}%
  \BibitemOpen
  \bibfield  {author} {\bibinfo {author} {\bibfnamefont {A.}~\bibnamefont
  {Buehler}}, \bibinfo {author} {\bibfnamefont {X.~L.}\ \bibnamefont
  {De\'{a}n-Ben}}, \bibinfo {author} {\bibfnamefont {J.}~\bibnamefont
  {Claussen}}, \bibinfo {author} {\bibfnamefont {V.}~\bibnamefont
  {Ntziachristos}}, \ and\ \bibinfo {author} {\bibfnamefont {D.}~\bibnamefont
  {Razansky}},\ }\bibfield  {title} {\enquote {\bibinfo {title}
  {Three-dimensional optoacoustic tomography at video rate},}\ }\href {\doibase
  10.1364/OE.20.022712} {\bibfield  {journal} {\bibinfo  {journal} {Opt.
  Express}\ }\textbf {\bibinfo {volume} {20}},\ \bibinfo {pages} {22712--22719}
  (\bibinfo {year} {2012})}\BibitemShut {NoStop}%
\bibitem [{\citenamefont {Lindholm}\ \emph {et~al.}(2008)\citenamefont
  {Lindholm}, \citenamefont {Nickolls}, \citenamefont {Oberman},\ and\
  \citenamefont {Montrym}}]{NVIDIA:09arch}%
  \BibitemOpen
  \bibfield  {author} {\bibinfo {author} {\bibfnamefont {E.}~\bibnamefont
  {Lindholm}}, \bibinfo {author} {\bibfnamefont {J.}~\bibnamefont {Nickolls}},
  \bibinfo {author} {\bibfnamefont {S.}~\bibnamefont {Oberman}}, \ and\
  \bibinfo {author} {\bibfnamefont {J.}~\bibnamefont {Montrym}},\ }\bibfield
  {title} {\enquote {\bibinfo {title} {Nvidia tesla: A unified graphics and
  computing architecture},}\ }\href {\doibase 10.1109/MM.2008.31} {\bibfield
  {journal} {\bibinfo  {journal} {Micro, IEEE}\ }\textbf {\bibinfo {volume}
  {28}},\ \bibinfo {pages} {39 --55} (\bibinfo {year} {2008})}\BibitemShut
  {NoStop}%
\bibitem [{\citenamefont {{NVIDIA}}(2008)}]{CUDAGuide2.0}%
  \BibitemOpen
  \bibfield  {author} {\bibinfo {author} {\bibnamefont {{NVIDIA}}},\
  }\href@noop {} {\emph {\bibinfo {title} {{NVIDIA CUDA Programming Guide
  2.0}}}}\ (\bibinfo {year} {2008})\BibitemShut {NoStop}%
\bibitem [{\citenamefont {Zhao}, \citenamefont {Hu},\ and\ \citenamefont
  {Zhang}(2009)}]{Zhao09:GPU}%
  \BibitemOpen
  \bibfield  {author} {\bibinfo {author} {\bibfnamefont {X.}~\bibnamefont
  {Zhao}}, \bibinfo {author} {\bibfnamefont {J.-J.}\ \bibnamefont {Hu}}, \ and\
  \bibinfo {author} {\bibfnamefont {P.}~\bibnamefont {Zhang}},\ }\bibfield
  {title} {\enquote {\bibinfo {title} {{GPU}-based 3{D} cone-beam {CT} image
  reconstruction for large data volume},}\ }\href {\doibase
  10.1155/2009/149079} {\bibfield  {journal} {\bibinfo  {journal} {Journal of
  Biomedical Imaging}\ }\textbf {\bibinfo {volume} {2009}},\ \bibinfo {pages}
  {8:1--8:8} (\bibinfo {year} {2009})}\BibitemShut {NoStop}%
\bibitem [{\citenamefont {Okitsu}, \citenamefont {Ino},\ and\ \citenamefont
  {Hagihara}(2010)}]{Okitsu10:GPU}%
  \BibitemOpen
  \bibfield  {author} {\bibinfo {author} {\bibfnamefont {Y.}~\bibnamefont
  {Okitsu}}, \bibinfo {author} {\bibfnamefont {F.}~\bibnamefont {Ino}}, \ and\
  \bibinfo {author} {\bibfnamefont {K.}~\bibnamefont {Hagihara}},\ }\bibfield
  {title} {\enquote {\bibinfo {title} {High-performance cone beam
  reconstruction using {CUDA} compatible {GPU}s},}\ }\href {\doibase
  10.1016/j.parco.2010.01.004} {\bibfield  {journal} {\bibinfo  {journal}
  {Parallel Computing}\ }\textbf {\bibinfo {volume} {36}},\ \bibinfo {pages}
  {129 -- 141} (\bibinfo {year} {2010})}\BibitemShut {NoStop}%
\bibitem [{\citenamefont {Chou}\ \emph {et~al.}(2011)\citenamefont {Chou},
  \citenamefont {Chuo}, \citenamefont {Hung},\ and\ \citenamefont
  {Wang}}]{ChouMP:GPU}%
  \BibitemOpen
  \bibfield  {author} {\bibinfo {author} {\bibfnamefont {C.-Y.}\ \bibnamefont
  {Chou}}, \bibinfo {author} {\bibfnamefont {Y.-Y.}\ \bibnamefont {Chuo}},
  \bibinfo {author} {\bibfnamefont {Y.}~\bibnamefont {Hung}}, \ and\ \bibinfo
  {author} {\bibfnamefont {W.}~\bibnamefont {Wang}},\ }\bibfield  {title}
  {\enquote {\bibinfo {title} {A fast forward projection using multithreads for
  multirays on {GPU}s in medical image reconstruction},}\ }\href {\doibase
  10.1118/1.3591994} {\bibfield  {journal} {\bibinfo  {journal} {Medical
  Physics}\ }\textbf {\bibinfo {volume} {38}},\ \bibinfo {pages} {4052--4065}
  (\bibinfo {year} {2011})}\BibitemShut {NoStop}%
\bibitem [{\citenamefont {Stone}\ \emph {et~al.}(2008)\citenamefont {Stone},
  \citenamefont {Haldar}, \citenamefont {Tsao}, \citenamefont {m.W. Hwu},
  \citenamefont {Sutton},\ and\ \citenamefont {Liang}}]{MRIGPU:2008}%
  \BibitemOpen
  \bibfield  {author} {\bibinfo {author} {\bibfnamefont {S.}~\bibnamefont
  {Stone}}, \bibinfo {author} {\bibfnamefont {J.}~\bibnamefont {Haldar}},
  \bibinfo {author} {\bibfnamefont {S.}~\bibnamefont {Tsao}}, \bibinfo {author}
  {\bibfnamefont {W.}~\bibnamefont {m.W. Hwu}}, \bibinfo {author}
  {\bibfnamefont {B.}~\bibnamefont {Sutton}}, \ and\ \bibinfo {author}
  {\bibfnamefont {Z.-P.}\ \bibnamefont {Liang}},\ }\bibfield  {title} {\enquote
  {\bibinfo {title} {Accelerating advanced {MRI} reconstructions on {GPU}s},}\
  }\href {\doibase 10.1016/j.jpdc.2008.05.013} {\bibfield  {journal} {\bibinfo
  {journal} {Journal of Parallel and Distributed Computing}\ }\textbf {\bibinfo
  {volume} {68}},\ \bibinfo {pages} {1307 -- 1318} (\bibinfo {year}
  {2008})}\BibitemShut {NoStop}%
\bibitem [{\citenamefont {Treeby}\ and\ \citenamefont
  {Cox}(2010)}]{kwave:toolbox}%
  \BibitemOpen
  \bibfield  {author} {\bibinfo {author} {\bibfnamefont {B.~E.}\ \bibnamefont
  {Treeby}}\ and\ \bibinfo {author} {\bibfnamefont {B.~T.}\ \bibnamefont
  {Cox}},\ }\bibfield  {title} {\enquote {\bibinfo {title} {k-wave: Matlab
  toolbox for the simulation and reconstruction of photoacoustic wave
  fields},}\ }\href {\doibase 10.1117/1.3360308} {\bibfield  {journal}
  {\bibinfo  {journal} {Journal of Biomedical Optics}\ }\textbf {\bibinfo
  {volume} {15}},\ \bibinfo {eid} {021314} (\bibinfo {year}
  {2010})}\BibitemShut {NoStop}%
\bibitem [{\citenamefont {Barrett}\ and\ \citenamefont
  {Myers}(2004)}]{BarrettBook}%
  \BibitemOpen
  \bibfield  {author} {\bibinfo {author} {\bibfnamefont {H.}~\bibnamefont
  {Barrett}}\ and\ \bibinfo {author} {\bibfnamefont {K.}~\bibnamefont
  {Myers}},\ }\href@noop {} {\emph {\bibinfo {title} {Foundations of Image
  Science}}}\ (\bibinfo  {publisher} {Wiley Series in Pure and Applied
  Optics},\ \bibinfo {year} {2004})\BibitemShut {NoStop}%
\bibitem [{\citenamefont {Wang}\ and\ \citenamefont
  {Wu}(2007)}]{LVbook:optics}%
  \BibitemOpen
  \bibfield  {author} {\bibinfo {author} {\bibfnamefont {L.~V.}\ \bibnamefont
  {Wang}}\ and\ \bibinfo {author} {\bibfnamefont {H.-I.}\ \bibnamefont {Wu}},\
  }\href@noop {} {\emph {\bibinfo {title} {Biomedical Optics, Principles and
  Imaging}}}\ (\bibinfo  {publisher} {Wiley},\ \bibinfo {address} {Hoboken,
  N.J.},\ \bibinfo {year} {2007})\BibitemShut {NoStop}%
\bibitem [{\citenamefont {Xu}\ and\ \citenamefont {Wang}(2006)}]{XuReview}%
  \BibitemOpen
  \bibfield  {author} {\bibinfo {author} {\bibfnamefont {M.}~\bibnamefont
  {Xu}}\ and\ \bibinfo {author} {\bibfnamefont {L.~V.}\ \bibnamefont {Wang}},\
  }\bibfield  {title} {\enquote {\bibinfo {title} {Photoacoustic imaging in
  biomedicine},}\ }\href@noop {} {\bibfield  {journal} {\bibinfo  {journal}
  {Review of Scientific Instruments}\ }\textbf {\bibinfo {volume} {77}}
  (\bibinfo {year} {2006})}\BibitemShut {NoStop}%
\bibitem [{\citenamefont {Wang}\ and\ \citenamefont
  {Anastasio}(2011)}]{MarkBookChapter}%
  \BibitemOpen
  \bibfield  {author} {\bibinfo {author} {\bibfnamefont {K.}~\bibnamefont
  {Wang}}\ and\ \bibinfo {author} {\bibfnamefont {M.~A.}\ \bibnamefont
  {Anastasio}},\ }\bibfield  {title} {\enquote {\bibinfo {title} {Photoacoustic
  and thermoacoustic tomography: image formation principles},}\ }in\ \href@noop
  {} {\emph {\bibinfo {booktitle} {Handbook of Mathematical Methods in
  Imaging}}},\ \bibinfo {editor} {edited by\ \bibinfo {editor} {\bibfnamefont
  {O.}~\bibnamefont {Scherzer}}}\ (\bibinfo  {publisher} {Springer},\ \bibinfo
  {year} {2011})\ Chap.~\bibinfo {chapter} {18}\BibitemShut {NoStop}%
\bibitem [{\citenamefont {Rosenthal}, \citenamefont {Ntziachristos},\ and\
  \citenamefont {Razansky}(2011)}]{Ntziachristos:CaliTrans11}%
  \BibitemOpen
  \bibfield  {author} {\bibinfo {author} {\bibfnamefont {A.}~\bibnamefont
  {Rosenthal}}, \bibinfo {author} {\bibfnamefont {V.}~\bibnamefont
  {Ntziachristos}}, \ and\ \bibinfo {author} {\bibfnamefont {D.}~\bibnamefont
  {Razansky}},\ }\bibfield  {title} {\enquote {\bibinfo {title} {Optoacoustic
  methods for frequency calibration of ultrasonic sensors},}\ }\href {\doibase
  10.1109/TUFFC.2011.1809} {\bibfield  {journal} {\bibinfo  {journal}
  {Ultrasonics, Ferroelectrics and Frequency Control, IEEE Transactions on}\
  }\textbf {\bibinfo {volume} {58}},\ \bibinfo {pages} {316 --326} (\bibinfo
  {year} {2011})}\BibitemShut {NoStop}%
\bibitem [{\citenamefont {Kak}\ and\ \citenamefont {Slaney}(1988)}]{KakBook}%
  \BibitemOpen
  \bibfield  {author} {\bibinfo {author} {\bibfnamefont {A.~C.}\ \bibnamefont
  {Kak}}\ and\ \bibinfo {author} {\bibfnamefont {M.}~\bibnamefont {Slaney}},\
  }\href@noop {} {\emph {\bibinfo {title} {Principles of Computerized
  Tomographic Imaging}}}\ (\bibinfo  {publisher} {IEEE Press},\ \bibinfo {year}
  {1988})\BibitemShut {NoStop}%
\bibitem [{\citenamefont {Anastasio}\ \emph
  {et~al.}(2005{\natexlab{a}})\citenamefont {Anastasio}, \citenamefont {Zhang},
  \citenamefont {Pan}, \citenamefont {Zou}, \citenamefont {Keng},\ and\
  \citenamefont {Wang}}]{AnastasioTATHT}%
  \BibitemOpen
  \bibfield  {author} {\bibinfo {author} {\bibfnamefont {{\rm
  M.A}.}~\bibnamefont {Anastasio}}, \bibinfo {author} {\bibfnamefont
  {J.}~\bibnamefont {Zhang}}, \bibinfo {author} {\bibfnamefont
  {X.}~\bibnamefont {Pan}}, \bibinfo {author} {\bibfnamefont {Y.}~\bibnamefont
  {Zou}}, \bibinfo {author} {\bibfnamefont {G.}~\bibnamefont {Keng}}, \ and\
  \bibinfo {author} {\bibfnamefont {{\rm L.V}.}~\bibnamefont {Wang}},\
  }\bibfield  {title} {\enquote {\bibinfo {title} {Half-time image
  reconstruction in thermoacoustic tomography},}\ }\href@noop {} {\bibfield
  {journal} {\bibinfo  {journal} {IEEE Transactions on Medical Imaging}\
  }\textbf {\bibinfo {volume} {24}},\ \bibinfo {pages} {199--210} (\bibinfo
  {year} {2005}{\natexlab{a}})}\BibitemShut {NoStop}%
\bibitem [{\citenamefont {Anastasio}\ \emph
  {et~al.}(2005{\natexlab{b}})\citenamefont {Anastasio}, \citenamefont {Zhang},
  \citenamefont {Sidky}, \citenamefont {Zou}, \citenamefont {Xia},\ and\
  \citenamefont {Pan}}]{AnastasioTATHTfeasible}%
  \BibitemOpen
  \bibfield  {author} {\bibinfo {author} {\bibfnamefont {M.}~\bibnamefont
  {Anastasio}}, \bibinfo {author} {\bibfnamefont {J.}~\bibnamefont {Zhang}},
  \bibinfo {author} {\bibfnamefont {E.}~\bibnamefont {Sidky}}, \bibinfo
  {author} {\bibfnamefont {Y.}~\bibnamefont {Zou}}, \bibinfo {author}
  {\bibfnamefont {D.}~\bibnamefont {Xia}}, \ and\ \bibinfo {author}
  {\bibfnamefont {X.}~\bibnamefont {Pan}},\ }\bibfield  {title} {\enquote
  {\bibinfo {title} {Feasibility of half-data image reconstruction in 3-{D}
  reflectivity tomography with a spherical aperture},}\ }\href {\doibase
  10.1109/TMI.2005.852055} {\bibfield  {journal} {\bibinfo  {journal} {Medical
  Imaging, IEEE Transactions on}\ }\textbf {\bibinfo {volume} {24}},\ \bibinfo
  {pages} {1100 --1112} (\bibinfo {year} {2005}{\natexlab{b}})}\BibitemShut
  {NoStop}%
\bibitem [{\citenamefont {Claerbout}(1992)}]{ClaerboutBook}%
  \BibitemOpen
  \bibfield  {author} {\bibinfo {author} {\bibfnamefont {J.~F.}\ \bibnamefont
  {Claerbout}},\ }\href@noop {} {\emph {\bibinfo {title} {{Earth Sounding
  Analysis: Processing Versus Inversion}}}}\ (\bibinfo  {publisher} {Blackwell
  Scientific Publications},\ \bibinfo {address} {Cambridge, MA},\ \bibinfo
  {year} {1992})\BibitemShut {NoStop}%
\bibitem [{\citenamefont {Khokhlova}\ \emph {et~al.}(2007)\citenamefont
  {Khokhlova}, \citenamefont {Pelivanov}, \citenamefont {Kozhushko},
  \citenamefont {Zharinov}, \citenamefont {Solomatin},\ and\ \citenamefont
  {Karabutov}}]{Khokhlova:07}%
  \BibitemOpen
  \bibfield  {author} {\bibinfo {author} {\bibfnamefont {T.~D.}\ \bibnamefont
  {Khokhlova}}, \bibinfo {author} {\bibfnamefont {I.~M.}\ \bibnamefont
  {Pelivanov}}, \bibinfo {author} {\bibfnamefont {V.~V.}\ \bibnamefont
  {Kozhushko}}, \bibinfo {author} {\bibfnamefont {A.~N.}\ \bibnamefont
  {Zharinov}}, \bibinfo {author} {\bibfnamefont {V.~S.}\ \bibnamefont
  {Solomatin}}, \ and\ \bibinfo {author} {\bibfnamefont {A.~A.}\ \bibnamefont
  {Karabutov}},\ }\bibfield  {title} {\enquote {\bibinfo {title} {Optoacoustic
  imaging of absorbing objects in a turbid medium: ultimate sensitivity and
  application to breast cancer diagnostics},}\ }\href@noop {} {\bibfield
  {journal} {\bibinfo  {journal} {Appl. Opt.}\ }\textbf {\bibinfo {volume}
  {46}},\ \bibinfo {pages} {262--272} (\bibinfo {year} {2007})}\BibitemShut
  {NoStop}%
\bibitem [{\citenamefont {Ephrat}\ \emph
  {et~al.}(2008{\natexlab{b}})\citenamefont {Ephrat}, \citenamefont
  {Keenliside}, \citenamefont {Seabrook}, \citenamefont {Prato},\ and\
  \citenamefont {Carson}}]{JCarson:2009}%
  \BibitemOpen
  \bibfield  {author} {\bibinfo {author} {\bibfnamefont {P.}~\bibnamefont
  {Ephrat}}, \bibinfo {author} {\bibfnamefont {L.}~\bibnamefont {Keenliside}},
  \bibinfo {author} {\bibfnamefont {A.}~\bibnamefont {Seabrook}}, \bibinfo
  {author} {\bibfnamefont {F.~S.}\ \bibnamefont {Prato}}, \ and\ \bibinfo
  {author} {\bibfnamefont {J.~J.~L.}\ \bibnamefont {Carson}},\ }\bibfield
  {title} {\enquote {\bibinfo {title} {Three-dimensional photoacoustic imaging
  by sparse-array detection and iterative image reconstruction},}\ }\href
  {\doibase 10.1117/1.2992131} {\bibfield  {journal} {\bibinfo  {journal}
  {Journal of Biomedical Optics}\ }\textbf {\bibinfo {volume} {13}},\ \bibinfo
  {eid} {054052} (\bibinfo {year} {2008}{\natexlab{b}})}\BibitemShut {NoStop}%
\bibitem [{\citenamefont {$\mathrm {Wernick}$ M.~N.}\ and\ \citenamefont
  {Aarsvold}(2004)}]{Wernickbookchap21}%
  \BibitemOpen
  \bibfield  {author} {\bibinfo {author} {\bibnamefont {$\mathrm {Wernick}$
  M.~N.}}\ and\ \bibinfo {author} {\bibfnamefont {J.~N.}\ \bibnamefont
  {Aarsvold}},\ }\href@noop {} {\emph {\bibinfo {title} {Emission Tomography,
  the Fundamentals of PET and SPECT}}}\ (\bibinfo  {publisher} {Elsevier
  Academic Press},\ \bibinfo {address} {San Diego, California},\ \bibinfo
  {year} {2004})\BibitemShut {NoStop}%
\bibitem [{\citenamefont {Zeng}\ and\ \citenamefont
  {Gullberg}(2000)}]{Zeng:unmatch}%
  \BibitemOpen
  \bibfield  {author} {\bibinfo {author} {\bibfnamefont {G.}~\bibnamefont
  {Zeng}}\ and\ \bibinfo {author} {\bibfnamefont {G.}~\bibnamefont
  {Gullberg}},\ }\bibfield  {title} {\enquote {\bibinfo {title} {Unmatched
  projector/backprojector pairs in an iterative reconstruction algorithm},}\
  }\href {\doibase 10.1109/42.870265} {\bibfield  {journal} {\bibinfo
  {journal} {Medical Imaging, IEEE Transactions on}\ }\textbf {\bibinfo
  {volume} {19}},\ \bibinfo {pages} {548 --555} (\bibinfo {year}
  {2000})}\BibitemShut {NoStop}%
\bibitem [{\citenamefont {Morton}\ and\ \citenamefont
  {Mayers}(2005)}]{MortonBook:2005}%
  \BibitemOpen
  \bibfield  {author} {\bibinfo {author} {\bibfnamefont {K.~W.}\ \bibnamefont
  {Morton}}\ and\ \bibinfo {author} {\bibfnamefont {D.~F.}\ \bibnamefont
  {Mayers}},\ }\href@noop {} {\emph {\bibinfo {title} {Numerical Solution of
  Partial Differential Equations: An Introduction}}}\ (\bibinfo  {publisher}
  {Cambridge University Press},\ \bibinfo {address} {New York, NY, USA},\
  \bibinfo {year} {2005})\BibitemShut {NoStop}%
\bibitem [{\citenamefont {Brecht}\ \emph {et~al.}(2009)\citenamefont {Brecht},
  \citenamefont {Su}, \citenamefont {Fronheiser}, \citenamefont {Ermilov},
  \citenamefont {Conjusteau},\ and\ \citenamefont {Oraevsky}}]{petermice:jbo}%
  \BibitemOpen
  \bibfield  {author} {\bibinfo {author} {\bibfnamefont {H.-P.}\ \bibnamefont
  {Brecht}}, \bibinfo {author} {\bibfnamefont {R.}~\bibnamefont {Su}}, \bibinfo
  {author} {\bibfnamefont {M.}~\bibnamefont {Fronheiser}}, \bibinfo {author}
  {\bibfnamefont {S.~A.}\ \bibnamefont {Ermilov}}, \bibinfo {author}
  {\bibfnamefont {A.}~\bibnamefont {Conjusteau}}, \ and\ \bibinfo {author}
  {\bibfnamefont {A.~A.}\ \bibnamefont {Oraevsky}},\ }\bibfield  {title}
  {\enquote {\bibinfo {title} {Whole-body three-dimensional optoacoustic
  tomography system for small animals},}\ }\href {\doibase 10.1117/1.3259361}
  {\bibfield  {journal} {\bibinfo  {journal} {Journal of Biomedical Optics}\
  }\textbf {\bibinfo {volume} {14}},\ \bibinfo {eid} {064007} (\bibinfo {year}
  {2009})}\BibitemShut {NoStop}%
\bibitem [{\citenamefont {Anastasio}\ \emph {et~al.}(2007)\citenamefont
  {Anastasio}, \citenamefont {Zhang}, \citenamefont {Modgil},\ and\
  \citenamefont {{La Riviere}}}]{FourierShell:07}%
  \BibitemOpen
  \bibfield  {author} {\bibinfo {author} {\bibfnamefont {{\rm M.
  A}.}~\bibnamefont {Anastasio}}, \bibinfo {author} {\bibfnamefont
  {J.}~\bibnamefont {Zhang}}, \bibinfo {author} {\bibfnamefont
  {D.}~\bibnamefont {Modgil}}, \ and\ \bibinfo {author} {\bibfnamefont
  {P.}~\bibnamefont {{La Riviere}}},\ }\bibfield  {title} {\enquote {\bibinfo
  {title} {Application of inverse source concepts to photoacoustic
  tomography},}\ }\href@noop {} {\bibfield  {journal} {\bibinfo  {journal}
  {Inverse Problems}\ }\textbf {\bibinfo {volume} {23}},\ \bibinfo {pages}
  {S21--S35} (\bibinfo {year} {2007})}\BibitemShut {NoStop}%
\bibitem [{\citenamefont {{Fessler}}(1994)}]{Fessler:94}%
  \BibitemOpen
  \bibfield  {author} {\bibinfo {author} {\bibfnamefont {J.~A.}\ \bibnamefont
  {{Fessler}}},\ }\bibfield  {title} {\enquote {\bibinfo {title} {Penalized
  weighted least-squares reconstruction for positron emission tomography},}\
  }\href@noop {} {\bibfield  {journal} {\bibinfo  {journal} {IEEE Transactions
  on Medical Imaging}\ }\textbf {\bibinfo {volume} {13}},\ \bibinfo {pages}
  {290--300} (\bibinfo {year} {1994})}\BibitemShut {NoStop}%
\bibitem [{\citenamefont {Shewchuk}(1994)}]{Shewchuk:1994}%
  \BibitemOpen
  \bibfield  {author} {\bibinfo {author} {\bibfnamefont {J.~R.}\ \bibnamefont
  {Shewchuk}},\ }\href@noop {} {\enquote {\bibinfo {title} {An introduction to
  the conjugate gradient method without the agonizing pain},}\ }\bibinfo {type}
  {Tech. Rep.}\ (\bibinfo {address} {Pittsburgh, PA, USA},\ \bibinfo {year}
  {1994})\BibitemShut {NoStop}%
\bibitem [{\citenamefont {Fessler}\ and\ \citenamefont
  {Booth}(1999)}]{Fessler:1999}%
  \BibitemOpen
  \bibfield  {author} {\bibinfo {author} {\bibfnamefont {J.}~\bibnamefont
  {Fessler}}\ and\ \bibinfo {author} {\bibfnamefont {S.}~\bibnamefont
  {Booth}},\ }\bibfield  {title} {\enquote {\bibinfo {title}
  {Conjugate-gradient preconditioning methods for shift-variant {PET} image
  reconstruction},}\ }\href {\doibase 10.1109/83.760336} {\bibfield  {journal}
  {\bibinfo  {journal} {Image Processing, IEEE Transactions on}\ }\textbf
  {\bibinfo {volume} {8}},\ \bibinfo {pages} {688 --699} (\bibinfo {year}
  {1999})}\BibitemShut {NoStop}%
\bibitem [{\citenamefont {Meng}\ \emph {et~al.}(2012)\citenamefont {Meng},
  \citenamefont {Wang}, \citenamefont {Ying}, \citenamefont {Liang},\ and\
  \citenamefont {Song}}]{Meng:12}%
  \BibitemOpen
  \bibfield  {author} {\bibinfo {author} {\bibfnamefont {J.}~\bibnamefont
  {Meng}}, \bibinfo {author} {\bibfnamefont {L.~V.}\ \bibnamefont {Wang}},
  \bibinfo {author} {\bibfnamefont {L.}~\bibnamefont {Ying}}, \bibinfo {author}
  {\bibfnamefont {D.}~\bibnamefont {Liang}}, \ and\ \bibinfo {author}
  {\bibfnamefont {L.}~\bibnamefont {Song}},\ }\bibfield  {title} {\enquote
  {\bibinfo {title} {Compressed-sensing photoacoustic computed tomography in
  vivo with partially known support},}\ }\href {\doibase 10.1364/OE.20.016510}
  {\bibfield  {journal} {\bibinfo  {journal} {Opt. Express}\ }\textbf {\bibinfo
  {volume} {20}},\ \bibinfo {pages} {16510--16523} (\bibinfo {year}
  {2012})}\BibitemShut {NoStop}%
\bibitem [{\citenamefont {Beck}\ and\ \citenamefont
  {Teboulle}(2009)}]{FISTATIP:2009}%
  \BibitemOpen
  \bibfield  {author} {\bibinfo {author} {\bibfnamefont {A.}~\bibnamefont
  {Beck}}\ and\ \bibinfo {author} {\bibfnamefont {M.}~\bibnamefont
  {Teboulle}},\ }\bibfield  {title} {\enquote {\bibinfo {title} {Fast
  gradient-based algorithms for constrained total variation image denoising and
  deblurring problems},}\ }\href {\doibase 10.1109/TIP.2009.2028250} {\bibfield
   {journal} {\bibinfo  {journal} {Image Processing, IEEE Transactions on}\
  }\textbf {\bibinfo {volume} {18}},\ \bibinfo {pages} {2419 --2434} (\bibinfo
  {year} {2009})}\BibitemShut {NoStop}%
\bibitem [{\citenamefont {Boyd}\ \emph {et~al.}(2011)\citenamefont {Boyd},
  \citenamefont {Parikh}, \citenamefont {Chu}, \citenamefont {Peleato},\ and\
  \citenamefont {Eckstein}}]{Boyd:2011AMMD}%
  \BibitemOpen
  \bibfield  {author} {\bibinfo {author} {\bibfnamefont {S.}~\bibnamefont
  {Boyd}}, \bibinfo {author} {\bibfnamefont {N.}~\bibnamefont {Parikh}},
  \bibinfo {author} {\bibfnamefont {E.}~\bibnamefont {Chu}}, \bibinfo {author}
  {\bibfnamefont {B.}~\bibnamefont {Peleato}}, \ and\ \bibinfo {author}
  {\bibfnamefont {J.}~\bibnamefont {Eckstein}},\ }\bibfield  {title} {\enquote
  {\bibinfo {title} {Distributed optimization and statistical learning via the
  alternating direction method of multipliers},}\ }\href {\doibase
  10.1561/2200000016} {\bibfield  {journal} {\bibinfo  {journal} {Found. Trends
  Mach. Learn.}\ }\textbf {\bibinfo {volume} {3}},\ \bibinfo {pages} {1--122}
  (\bibinfo {year} {2011})}\BibitemShut {NoStop}%
\bibitem [{\citenamefont {Xu}\ and\ \citenamefont {Wang}(2002)}]{Xu:2002}%
  \BibitemOpen
  \bibfield  {author} {\bibinfo {author} {\bibfnamefont {M.}~\bibnamefont
  {Xu}}\ and\ \bibinfo {author} {\bibfnamefont {{\rm L.V}.}~\bibnamefont
  {Wang}},\ }\bibfield  {title} {\enquote {\bibinfo {title} {Time-domain
  reconstruction for thermoacoustic tomography in a spherical geometry},}\
  }\href@noop {} {\bibfield  {journal} {\bibinfo  {journal} {IEEE Transactions
  on Medical Imaging}\ }\textbf {\bibinfo {volume} {21}},\ \bibinfo {pages}
  {814--822} (\bibinfo {year} {2002})}\BibitemShut {NoStop}%
\bibitem [{\citenamefont {Finch}, \citenamefont {Haltmeier},\ and\
  \citenamefont {Rakesh}(2007)}]{Finch:07}%
  \BibitemOpen
  \bibfield  {author} {\bibinfo {author} {\bibfnamefont {D.}~\bibnamefont
  {Finch}}, \bibinfo {author} {\bibfnamefont {M.}~\bibnamefont {Haltmeier}}, \
  and\ \bibinfo {author} {\bibnamefont {Rakesh}},\ }\bibfield  {title}
  {\enquote {\bibinfo {title} {Inversion of spherical means and the wave
  equation in even dimensions},}\ }\href {\doibase 10.1137/070682137}
  {\bibfield  {journal} {\bibinfo  {journal} {SIAM Journal on Applied
  Mathematics}\ }\textbf {\bibinfo {volume} {68}},\ \bibinfo {pages} {392--412}
  (\bibinfo {year} {2007})}\BibitemShut {NoStop}%
\bibitem [{\citenamefont {Elbau}, \citenamefont {Scherzer},\ and\ \citenamefont
  {Schulze}(2012)}]{Schulze:2DInvPro}%
  \BibitemOpen
  \bibfield  {author} {\bibinfo {author} {\bibfnamefont {P.}~\bibnamefont
  {Elbau}}, \bibinfo {author} {\bibfnamefont {O.}~\bibnamefont {Scherzer}}, \
  and\ \bibinfo {author} {\bibfnamefont {R.}~\bibnamefont {Schulze}},\
  }\bibfield  {title} {\enquote {\bibinfo {title} {Reconstruction formulas for
  photoacoustic sectional imaging},}\ }\href@noop {} {\bibfield  {journal}
  {\bibinfo  {journal} {Inverse Problems}\ }\textbf {\bibinfo {volume} {28}},\
  \bibinfo {pages} {045004} (\bibinfo {year} {2012})}\BibitemShut {NoStop}%
\bibitem [{\citenamefont {Rosenthal}, \citenamefont {Razansky},\ and\
  \citenamefont {Ntziachristos}(2010)}]{Ntziachristos:QPAT2012}%
  \BibitemOpen
  \bibfield  {author} {\bibinfo {author} {\bibfnamefont {A.}~\bibnamefont
  {Rosenthal}}, \bibinfo {author} {\bibfnamefont {D.}~\bibnamefont {Razansky}},
  \ and\ \bibinfo {author} {\bibfnamefont {V.}~\bibnamefont {Ntziachristos}},\
  }\bibfield  {title} {\enquote {\bibinfo {title} {Fast semi-analytical
  model-based acoustic inversion for quantitative optoacoustic tomography},}\
  }\href {\doibase 10.1109/TMI.2010.2044584} {\bibfield  {journal} {\bibinfo
  {journal} {Medical Imaging, IEEE Transactions on}\ }\textbf {\bibinfo
  {volume} {29}},\ \bibinfo {pages} {1275 --1285} (\bibinfo {year}
  {2010})}\BibitemShut {NoStop}%
\end{thebibliography}
%

\newpage

\section*{Figure Captions}
\noindent 1. (a) Schematic of the 3D OAT scanning geometry.
(b) Schematic of the local coordinate system for the implementation of interpolation-based D-D imaging model.  
\vskip .3cm

\noindent 2. Slices corresponding to the plane $y=0$ of
(a) the phantom
and the images reconstructed by use of
(b) the CPU-based and (c) the GPU-based implementations of the FBP algorithm
from the ``$128\times90$''-data. 
\vskip .3cm

\noindent 3.  Slices corresponding to the plane $y=0$ of the images reconstructed by use of the FBP algorithm
with
(a) the CPU-based implementation from the ``$64\times90$''-data,
(b) the CPU-based implementation from the ``$64\times45$''-data,
(c) the CPU-based implementation from the ``$32\times45$''-data,
(d) the GPU-based implementation from the ``$64\times90$''-data,
(e) the GPU-based implementation from the ``$64\times45$''-data,
and
(f) the GPU-based implementation from the ``$32\times45$''-data.
\vskip .3cm

\noindent 4. Slices corresponding to the plane $y=0$ of the images reconstructed by use of the GPU-based 
implementations of 
(a) the PLS-Int algorithm from the ``$64\times90$''-data,
(b) the PLS-Int algorithm from the ``$64\times45$''-data,
(c) the PLS-Int algorithm from the ``$32\times45$''-data,
(d) the PLS-Sph algorithm from the ``$64\times90$''-data,
(e) the PLS-Sph algorithm from the ``$64\times45$''-data,
and
(f) the PLS-Sph algorithm from the ``$32\times45$''-data.
\vskip .3cm

\noindent 5. Profiles along the line $(x,y)=(-6.58,0)$-mm of the images reconstructed by use of 
(a) the CPU- and GPU-based implementations of the FBP algorithm from the
``$128\times90$''-data,
and 
(b) the GPU-based implementations of the PLS-Int and the PLS-Sph algorithms from the 
``$64\times90$''-data.
\vskip .3cm

\noindent 6. Plots of the RMSE against the amount of data by use of the FBP, the PLS-Int
and the PLS-Sph algorithms.
\vskip .3cm

\noindent 7. MIP renderings of the 3D images of the mouse body reconstructed
by use of the GPU-based implementations of
(a) the FBP algorithm from the ``full data'',
(b) the PLS-Int algorithm from the ``full data'' with $\mu=1.0\times 10^4$,
(c) the PLS-Sph algorithm from the ``full data'' with $\mu=1.0\times 10^4$,
(d) the FBP algorithm from the ``quarter data'',
(e) the PLS-Int algorithm from the ``quarter data'' with $\mu=1.0\times 10^3$,
and
(f) the PLS-Sph algorithm from the ``quarter data'' with $\mu=1.0\times 10^3$.
The grayscale window is [0,12.0].

\newpage

\clearpage

\centering

\begin{algorithm}[H]
\caption{ \label{alg:fbp}
Implementation of the FBP algorithm (on host)
}
\algsetup{indent=2em}
\begin{algorithmic}[1]
\REQUIRE $\mathbf p$
\ENSURE $\hat{\boldsymbol\alpha}_{\rm fbp}$
\STATE $w = -C_pR^s\Delta_{\theta^s}\Delta_{\phi^s}/(\pi\beta c_0^3\Delta_t)$
       \COMMENT{Precalculate the common coefficient}
\STATE ${\rm T}\!\_\mathbf p  \leftarrow \mathbf p$
    \COMMENT{Bound data to texture memory}
\STATE K\_fbp$\,\, \langle\langle\langle\quad ( N_y, N_x ),\, (N_z, 1, 1)
                 \quad\rangle\rangle\rangle$
           ($\omega$,
            ${\rm D}\!\_\hat{\boldsymbol\alpha}_{\rm fbp}$)
\STATE $\hat{\boldsymbol\alpha}_{\rm fbp} \leftarrow {\rm D}\!\_{\hat{\boldsymbol\alpha}_{\rm fbp}}$
    \COMMENT{Copy data from global memory to host}
\end{algorithmic}
\end{algorithm}

\begin{algorithm}[H]
\caption{\label{alg:K_fbp}
Implementation of kernel K\_fbp 
     $\,\, \langle\langle\langle\quad  ( N_y, N_x ),\, (N_z, 1, 1) 
                 \quad\rangle\rangle\rangle$
}
\algsetup{indent=2em}
\begin{algorithmic}[1]
\REQUIRE $\omega$, 
         ${\rm T}\!\_\mathbf p$,
         ${\rm D}\!\_{\hat{\boldsymbol\alpha}_{\rm fbp}}$
\ENSURE ${\rm D}\!\_{\hat{\boldsymbol\alpha}_{\rm fbp}}$
\STATE $x = ({\rm blockIdx.y})\Delta_s + x_{\rm min}$;
$y = ({\rm blockIdx.x})\Delta_s + y_{\rm min}$;
$z = ({\rm threadIdx.x})\Delta_s + z_{\rm min}$
\STATE $\Sigma = 0$ 
\FOR{$n_r=0$ \TO $N_r-1$}
\STATE $\theta^s = n_r\Delta_{\theta^s} + \theta^s_{\rm min}$; $z^s = R^s\cos\theta^s$; $r^s = R^s\sin\theta^s$; 
$w'=w\sin\theta^s$
\FOR{$n_v=0$ \TO $N_v-1$}
\STATE $\phi^s = n_v\Delta_{\phi^s} + \phi^s_{\rm min}$; 
$x^s = r^s\cos\phi^s$; $y^s = r^s\sin\phi^s$
\STATE $\bar t = ((x-x^s)^2+(y-y^s)^2+(z-z^s)^2)^{1/2}$
\STATE $t_n = (\bar t/c_0-t_{\rm min})/\Delta_t$; $n_t = {\rm floor}(t_n)$
\STATE $\Sigma \pa \omega'\Big\{
      \big[ (n_t\Delta_t+t_{\rm min})/(t_n\Delta_t+t_{\rm min}) - 1.5\big] {\rm T}\!\_\mathbf p[n_r][n_v][n_t] 
     +\big[ 1.5 - (n_t\Delta_t+t_{\rm min})/(t_n\Delta_t+t_{\rm min})\big] {\rm T}\!\_\mathbf p[n_r][n_v][n_t+1]
           \Big\}$
  \COMMENT{Fetch data from texture memory}
\ENDFOR
\ENDFOR
\STATE ${\rm D}\!\_{\hat {\boldsymbol\alpha}_{\rm fbp}}
       [{\rm blockIdx.y}][{\rm blockIdx.x}][{\rm threadIdx.x}]=\Sigma$
\end{algorithmic}
\end{algorithm}
\clearpage

\begin{algorithm}[H]
\caption{\label{alg:intH}
Implementation of $\mathbf g = \mathbf G \boldsymbol \alpha_{\rm int}$ (on host)
}
\algsetup{indent=2em}
\begin{algorithmic}[1]
\REQUIRE $\boldsymbol\alpha_{\rm int}$
\ENSURE $\mathbf g$
\STATE ${\rm T}\!\_\boldsymbol\alpha_{\rm int} \leftarrow \boldsymbol\alpha_{\rm int}$
    \COMMENT{Bound data to texture memory}
\STATE K\_srt$\,\, \langle\langle\langle\quad (N_v, N_t),\, (N_r, 1, 1)
                 \quad\rangle\rangle\rangle$
           (${\rm D}\!\_\mathbf g$)
\STATE $\mathbf g \leftarrow {\rm D}\!\_\mathbf g$
    \COMMENT{Copy data from global memory to host}
% \FOR{$n_r=0$ \TO $N_r-1$}
% \FOR{$n_v=0$ \TO $N_v-1$}
% \FOR{$n_t=1$ \TO $N_t-2$}
% \STATE $\mathbf p_{\rm int}[n_r][n_v][n_t] = \beta\big\{\mathbf g_{\rm int}[n_r][n_v][n_t+1]/(n_t+1+t_{\rm min}/\Delta_t) - \mathbf g_{\rm int}[n_r][n_v][n_t-1]/(n_t-1+t_{\rm min}/\Delta_t)\big\}/(8\pi C_p\Delta_t^2)$
% \ENDFOR
% \ENDFOR
% \ENDFOR
\end{algorithmic}
\end{algorithm}

\begin{algorithm}[H]
\caption{\label{alg:K_intH}
Implementation of kernel K\_srt 
     $\,\, \langle\langle\langle\quad (N_v, N_t),\, (N_r, 1, 1)
                 \quad\rangle\rangle\rangle$
}
\algsetup{indent=2em}
\begin{algorithmic}[1]
\REQUIRE ${\rm D}\!\_\mathbf g$, ${\rm T}\!\_\boldsymbol \alpha_{\rm int}$
\ENSURE ${\rm D}\!\_\mathbf g$
\STATE $\bar t = ({\rm blockIdx.y})c_0\Delta_t+c_0t_{\rm min}$;
\quad $\theta^s = ({\rm threadIdx.x})\Delta_{\theta^s} +\theta^s_{\rm min}$;
\quad $\phi^s = ({\rm blockIdx.x})\Delta_{\phi^s}+\phi^s_{\rm min}$
\STATE $\theta'_{\rm max}=\arccos\big(((R^s)^2 + \bar t^2 - R^2)/(2\bar tR^s)\big)$
\STATE $\Sigma = 0$;
\quad $\theta' = \theta'_{\rm max}$
\WHILE{$\theta'>0$}
\STATE $z' = \bar t \cos\theta'$;
\quad $r' = \bar t \sin\theta'$;
\quad $\phi' = 0$
\WHILE{$\phi' < 2\pi$}
\STATE $x' = r' \cos\phi'$;\quad $y' = r' \sin\phi'$
\STATE $x = -x'\sin\theta'-(z'-R^s)\cos\theta'$;
\quad $y = y'$;
\quad $z = x'\cos\theta' - (z'-R^s)\sin\theta'$
\COMMENT{Convert to global coordinate system}
\STATE $x_n = (x-x_{\rm min})/\Delta_s$; 
\quad $y_n = (y-y_{\rm min})/\Delta_s$; 
\quad $z_n = (z-z_{\rm min})/\Delta_s$
\STATE $\Sigma \pa {\rm tex3D}(x_n,y_n,z_n)$
  \COMMENT{Tri-linear interpolation}
\STATE $\phi' = \phi' + \Delta_s/r'$
\ENDWHILE
\STATE $\theta' = \theta' - \Delta_s/\bar t$
\ENDWHILE
\STATE ${\rm D}\!\_\mathbf g
      [{\rm threadIdx.x}][{\rm blockIdx.x}][{\rm blockIdx.y}]=\Sigma \Delta_s^2$
\end{algorithmic}
\end{algorithm}
\clearpage

\begin{algorithm}[H]
\caption{\label{alg:intHT}
Implementation of $\boldsymbol\alpha'_{\rm int} = \mathbf G^\dagger \mathbf g'$ (on host)
}
\algsetup{indent=2em}
\begin{algorithmic}[1]
\REQUIRE $\mathbf g'$ 
\ENSURE $\boldsymbol\alpha'_{\rm int}$
% \FOR{$n_r=0$ \TO $N_r-1$}
% \FOR{$n_v=0$ \TO $N_v-1$}
% \FOR{$n_t=1$ \TO $N_t-2$}
% \STATE $\mathbf g_{\rm int}[n_r][n_v][n_t] = \beta\big\{\mathbf p_{\rm int}[n_r][n_v][n_t-1] - \mathbf p_{\rm int}[n_r][n_v][n_t+1]\big\}/\big\{(n_t+t_{\rm min}/\Delta_t)8\pi C_p\Delta_t^2\big\}$
% \ENDFOR
% \ENDFOR
% \ENDFOR
\STATE ${\rm T}\!\_\mathbf g' \leftarrow \mathbf g'$
    \COMMENT{Bound data to texture memory}
\STATE K\_srtT$\,\, \langle\langle\langle\quad (N_v, N_t),\, (N_r, 1, 1)
                 \quad\rangle\rangle\rangle$
           (${\rm D}\!\_\boldsymbol \alpha'_{\rm int}$)
\STATE $\boldsymbol\alpha'_{\rm int} \leftarrow {\rm D}\!\_\boldsymbol \alpha'_{\rm int}$
    \COMMENT{Copy data from global memory to host}
\end{algorithmic}
\end{algorithm}

\begin{algorithm}[H]
\caption{\label{alg:K_intHT}
Implementation of kernel K\_srtT 
     $\,\, \langle\langle\langle\quad (N_v, N_t),\, (N_r, 1, 1)
                 \quad\rangle\rangle\rangle$
}
\algsetup{indent=2em}
\begin{algorithmic}[1]
\REQUIRE ${\rm D}\!\_\boldsymbol\alpha'_{\rm int}$, 
         ${\rm T}\!\_\mathbf g'_{\rm int}$
\ENSURE ${\rm D}\!\_\boldsymbol\alpha'_{\rm int}$
\STATE $\bar t = ({\rm blockIdx.y})c_0\Delta_t+c_0t_{\rm min}$;
\quad $\theta^s = ({\rm threadIdx.x})\Delta_{\theta^s} +\theta^s_{\rm min}$;
\quad $\phi^s = ({\rm blockIdx.x})\Delta_{\phi^s}+\phi^s_{\rm min}$
\STATE $\theta'_{\rm max}=\arccos\big(((R^s)^2 + \bar t^2 - R^2)/(2\bar tR^s)\big)$;
\quad $\theta' = \theta'_{\rm max}$
\WHILE{$\theta'>0$}
\STATE $z' = \bar t \cos\theta'$;
\quad $r' = \bar t \sin\theta'$;
\quad $\phi' = 0$
\WHILE{$\phi' < 2\pi$}
\STATE $x' = r' \cos\phi'$;\quad $y' = r' \sin\phi'$
\STATE $x = -x'\sin\theta-(z'-R')\cos\theta$;
\quad $y = y'$;
\quad $z = x'\cos\theta - (z'-R')\sin\theta$
\COMMENT{Convert to global coordinate system}
\STATE $x_n = (x-x_{\rm min})/\Delta_s$; 
\quad $y_n = (y-y_{\rm min})/\Delta_s$; 
\quad $z_n = (z-z_{\rm min})/\Delta_s$
\STATE $n_x = {\rm floor}(x_n)$; 
\quad $n_y = {\rm floor}(y_n)$; 
\quad $n_z = {\rm floor}(z_n)$
\STATE ${\rm D\_}\boldsymbol\alpha'_{\rm int}[n_z][n_y][n_x] \pa \Delta_s^2(n_x+1-x_n)(n_y+1-y_n)(n_z+1-z_n) {\rm T\_}\mathbf g'_{\rm int}[\rm {threadIdx.x}][{\rm blockIdx.x}][{\rm blockIdx.y}]$
\COMMENT{Add weights to one of the eight neighboring nodes by use of `atomicAdd'; Repeat this operation for all other seven neighboring nodes}
\STATE $\phi' = \phi' + \Delta_s/r'$
\ENDWHILE
\STATE $\theta' = \theta' - \Delta_s/\bar t$
\ENDWHILE
\end{algorithmic}
\end{algorithm}
\clearpage

\begin{algorithm}[H]
\caption{\label{alg:sphH}
Implementation of $\tilde{\mathbf u}_{\rm sph} = \mathbf H_{\rm sph} \boldsymbol\alpha_{\rm sph}$
(on host)
}
\algsetup{indent=2em}
\begin{algorithmic}[1]
\REQUIRE $\boldsymbol \alpha_{\rm sph}$, $\tilde {\mathbf p}_0$
\ENSURE $\tilde {\mathbf u}_{\rm sph}$
\FOR {$n_{\rm pth}=0$ \TO $N_{\rm pth}-1$}
\STATE parm\_fwdarg[$n_{\rm pth}$].$n_{\rm pth}$ = $n_{\rm pth}$
\STATE parm\_fwdarg[$n_{\rm pth}$].$\tilde p_0$ = \&$\tilde {\mathbf p}_0[0]$
\STATE parm\_fwdarg[$n_{\rm pth}$].$\alpha_{\rm pth}$=\&$\boldsymbol\alpha_{\rm sph}[n_{\rm pth}N_xN_yN_z/N_{\rm pth}]$
\STATE parm\_fwdarg[$n_{\rm pth}$].$\tilde u'_{\rm pth}$=$\&\tilde{\mathbf u}'_{\rm sph}[ n_{\rm pth}N_rN_vN_f]$ 
\COMMENT{Pass addresses of arrays to each pthread}
\STATE pthread\_create(\&pthreads[$n_{\rm pth}$], NULL, fwd\_pthread, (void *)(parm\_fwdarg+$n_{\rm pth}$))
\COMMENT{Call function fwd\_pthread}
\ENDFOR
\FOR {$n_{\rm pth}=0$ \TO $N_{\rm pth}-1$}
\FOR {$n=0$ \TO $N_rN_vN_f$}
\STATE $\tilde{\mathbf u}_{\rm sph}[n] \,+\!\!= \tilde{\mathbf u}'_{\rm sph}[n+n_{\rm pth}N_rN_vN_f]$
\ENDFOR
\ENDFOR
\end{algorithmic}
\end{algorithm}
\clearpage

\begin{algorithm}[H]
\caption{\label{alg:fwd_pthread}
Implementation of function fwd\_pthread (on host)}
\algsetup{indent=2em}
\begin{algorithmic}[1]
\REQUIRE $n_{\rm pth}$, 
         $\tilde {\mathbf p}_0$, 
         $\boldsymbol \alpha_{\rm pth}$, 
         $\tilde {\mathbf u}'_{\rm pth}$
\ENSURE $\tilde {\mathbf u}'_{\rm pth}$
\STATE ${\rm C\!\_\tilde{\mathbf p}}_0 \leftarrow \tilde{\mathbf p}_0$
    \COMMENT{Copy from host to constant memory}
\FOR {$n_x=0$ \TO $N_x/N_{\rm pth}-1$}
  \STATE $x=(n_x + n_{\rm pth}N_x/N_{\rm pth})\Delta_s + x_{\rm min}$
  \FOR {$n_y=0$ \TO $N_y-1$}
    \STATE $y=n_y \Delta_s + y_{\rm min}$
    \STATE ${\rm C\!\_\boldsymbol{\alpha}_{\rm pth}} 
              \leftarrow \boldsymbol \alpha_{\rm pth}[n_x][n_y][:]$ 
    \COMMENT{Copy from host to constant memory}
    \STATE K\_fwdsph$\,\, \langle\langle\langle\quad (N_v,N_r),(N_f,1,1)
                 \quad\rangle\rangle\rangle$
           ($x,y,$ D\!\_$\tilde{\mathbf u}'_{\rm pth}$)
  \ENDFOR
\ENDFOR
  \STATE $\tilde{\mathbf u}'_{\rm pth} \leftarrow {\rm D}\!\_\tilde{\mathbf u}'_{\rm pth}$
    \COMMENT{Copy from global memory to host}
\end{algorithmic}
\end{algorithm}
\clearpage

\begin{algorithm}[H]
\caption{\label{alg:Kfwdsph}
Implementation of Kernel K\_fwdsph$\,\, \langle\langle\langle\quad (N_v,N_r),(N_f,1,1)
                 \quad\rangle\rangle\rangle$
}
\algsetup{indent=2em}
\begin{algorithmic}[1]
\REQUIRE $x$, $y$, 
         D\!\_$\tilde{\mathbf u}'_{\rm pth}$, 
         C\!\_$\boldsymbol\alpha_{\rm pth}$, 
         C\!\_$\tilde{\mathbf p}_0$
\ENSURE D\!\_$\tilde{\mathbf u}'_{\rm pth}$ 
\STATE  $\theta^s = ({\rm blockIdx.y})\Delta_{\theta^s} +\theta^s_{\rm min}$;
\quad $\phi^s = ({\rm blockIdx.x})\Delta_{\phi^s}+\phi^s_{\rm min}$;
\quad $f=({\rm threadIdx.x})\Delta_f+f_{\rm min}$
\STATE $z^s = R^s\cos\theta^s$;
\quad $x^s = R^s\sin\theta^s\cos\phi^s$;
\quad $y^s = R^s\sin\theta^s\sin\phi^s$
\COMMENT{Calculate locations of transducers}
\STATE $\Sigma^r = 0$; \quad $\Sigma^i = 0$ 
\COMMENT{Initiate the partial summation including the real and imaginary parts}
\FOR {$n_z=0$ \TO $N_z-1$}
  \STATE $z=n_z \Delta_s + z_{\rm min}$
  \STATE $d=\Big((x-x^s)^2+(y-y^s)^2+(z-z^s)^2\Big)^{1/2}$  
  \STATE $\tilde h^r = \cos(2\pi fd/c_0)/(2\pi d)$;
\quad $\tilde h^i = -\sin(2\pi fd/c_0)/(2\pi d)$ 
\COMMENT{Calculate SIR; Example here assumes point-like transducers}
  \STATE $\Sigma^r \,+\!\!= {\rm C}\!\_\boldsymbol\alpha_{\rm pth}[n_z]\Big(
          \tilde{h}^r {\rm C}\!\_\tilde{\mathbf p}_0[{\rm threadIdx.x}].r
         -\tilde{h}^i {\rm C}\!\_\tilde{\mathbf p}_0[{\rm threadIdx.x}].i
         \Big)$
  \STATE $\Sigma^i \,+\!\!= {\rm C}\!\_\boldsymbol\alpha_{\rm pth}[n_z]\Big(
          \tilde{h}^r {\rm C}\!\_\tilde{\mathbf p}_0[{\rm threadIdx.x}].i
         +\tilde{h}^i {\rm C}\!\_\tilde{\mathbf p}_0[{\rm threadIdx.x}].r
         \Big)$
\ENDFOR
  \STATE ${\rm D}\!\_\tilde{\mathbf u}'_{\rm pth}[{\rm blockIdx.y}][{\rm blockIdx.x}][{\rm threadIdx.x}].r\,+\!\!= \Sigma^r$
\quad
  \STATE ${\rm D}\!\_\tilde{\mathbf u}'_{\rm pth}[{\rm blockIdx.y}][{\rm blockIdx.x}][{\rm threadIdx.x}].i\,+\!\!= \Sigma^i$
\end{algorithmic}
\end{algorithm}
\clearpage

\begin{algorithm}[H]
\caption{\label{alg:sphHT}
Implementation of $\boldsymbol \alpha'_{\rm sph} = \mathbf H_{\rm sph}^\dagger \tilde{\mathbf u}$ 
(on host)
}
\algsetup{indent=2em}
\begin{algorithmic}[1]
\REQUIRE $\tilde {\mathbf u}$, $\tilde {\mathbf p}_0$
\ENSURE $\boldsymbol \alpha'_{\rm sph}$
\FOR {$n_{\rm pth}=0$ \TO $N_{\rm pth}-1$}
\STATE parm\_bwdarg[$n_{\rm pth}$].$n_{\rm pth}$ = $n_{\rm pth}$
\STATE parm\_bwdarg[$n_{\rm pth}$].$\tilde p_0$ = \&$\tilde {\mathbf p}_0[0]$
\STATE parm\_bwdarg[$n_{\rm pth}$].$\tilde u_{\rm pth}$=\&$\tilde{\mathbf u}[n_{\rm pth}N_rN_vN_f/N_{\rm pth}]$
\STATE parm\_bwdarg[$n_{\rm pth}$].$\alpha''_{\rm pth}$=$\&\boldsymbol \alpha''_{\rm sph}[ n_{\rm pth}N_xN_yN_z]$ 
\COMMENT{Pass addresses of arrays to each pthread}
\STATE pthread\_create(\&pthreads[$n_{\rm pth}$], NULL, bwd\_pthread, (void *)(parm\_bwdarg+$n_{\rm pth}$))
\COMMENT{Call function bwd\_pthread}
\ENDFOR
\FOR {$n_{\rm pth}=0$ \TO $N_{\rm pth}-1$}
\FOR {$n=0$ \TO $N_xN_yN_z$}
\STATE $\boldsymbol\alpha'_{\rm sph}[n]\,+\!\!= \boldsymbol\alpha''_{\rm sph}[n+n_{\rm pth}N_xN_yN_z]$
\ENDFOR
\ENDFOR
\end{algorithmic}
\end{algorithm}
\clearpage

\begin{algorithm}[H]
\caption{\label{alg:bwd_pthread}
Implementation of function bwd\_pthread (on host)}
\algsetup{indent=2em}
\begin{algorithmic}[1]
\REQUIRE $n_{\rm pth}$, 
         $\tilde {\mathbf p}_0$, 
         $\tilde {\mathbf u}_{\rm pth}$,
         $\boldsymbol \alpha''_{\rm pth}$ 
\ENSURE $\boldsymbol \alpha''_{\rm pth}$
\STATE ${\rm C\!\_\tilde{\mathbf p}}_0 \leftarrow \tilde{\mathbf p}_0$
    \COMMENT{Copy from host to constant memory}
\FOR {$n_r=0$ \TO $N_r/N_{\rm pth}-1$}
  \STATE{$\theta^s=(n_r+n_{\rm pth}N_r/N_{\rm pth})\Delta_{\theta^s}+\theta^s_{\rm min}$}; 
  \quad $z^s = R^s\cos\theta^s$;
  \quad $r^s = R^s\sin\theta^s$
  \FOR {$n_v=0$ \TO $N_v-1$}
    \STATE $\phi^s = n_v\Delta_{\phi^s} + \phi^s_{\rm min}$;
    \quad $x^s = r^s\cos\phi^s$;
    \quad $y^s = r^s\sin\phi^s$
    \STATE ${\rm C\!\_\tilde{\mathbf u}_{\rm pth}} 
              \leftarrow \tilde{\mathbf u}_{\rm pth}[n_r][n_v][:]$ 
    \COMMENT{Copy from host to constant memory}
    \STATE K\_bwdsph$\,\, \langle\langle\langle\quad (N_y,N_x),(N_z,1,1)
                 \quad\rangle\rangle\rangle$
           ($x^s, y^s, z^s$ D\!\_$\boldsymbol\alpha''_{\rm pth}$)
  \ENDFOR
\ENDFOR
  \STATE $\boldsymbol\alpha''_{\rm pth} \leftarrow {\rm D}\!\_\boldsymbol\alpha''_{\rm pth}$
    \COMMENT{Copy from global memory to host}
\end{algorithmic}
\end{algorithm}
\clearpage

\begin{algorithm}[H]
\caption{\label{alg:Kbwdsph}
Implementation of Kernel K\_bwdsph$\,\, \langle\langle\langle\quad (N_y,N_x),(N_z,1,1)
                 \quad\rangle\rangle\rangle$
}
\algsetup{indent=2em}
\begin{algorithmic}[1]
\REQUIRE $x^s$, 
         $y^s$, 
         $z^s$, 
         D\!\_$\boldsymbol\alpha''_{\rm pth}$, 
         C\!\_$\tilde{\mathbf u}_{\rm pth}$, 
         C\!\_$\tilde{\mathbf p}_0$
\ENSURE D\!\_$\boldsymbol\alpha''_{\rm pth}$
\STATE  $x = ({\rm blockIdx.y})\Delta_s + x_{\rm min}$;
\quad $y = ({\rm blockIdx.x})\Delta_s + y_{\rm min}$;
\quad $z=({\rm threadIdx.x})\Delta_s+z_{\rm min}$
\STATE $d=\Big((x-x^s)^2+(y-y^s)^2+(z-z^s)^2\Big)^{1/2}$;
\quad $\Sigma = 0$
\COMMENT{Initiate the partial summation}
\FOR {$n_f=0$ \TO $N_f-1$}
 \STATE{$f = n_f\Delta_f+f_{\rm min}$} 
  \STATE $\tilde h^r = \cos(2\pi fd/c_0)/(2\pi d)$;
\quad $\tilde h^i = -\sin(2\pi fd/c_0)/(2\pi d)$ 
 \COMMENT{Calculate SIR; Example here assumes point-like transducers}
  \STATE $\Sigma \,+\!\!= {\rm C}\!\_\tilde{\mathbf u}_{\rm pth}[n_f].r\Big(
          \tilde{h}^r {\rm C}\!\_\tilde{\mathbf p}_0[n_f].r
         -\tilde{h}^i {\rm C}\!\_\tilde{\mathbf p}_0[n_f].i
         \Big)
         +{\rm C}\!\_\tilde{\mathbf u}_{\rm pth}[n_f].i\Big(
           \tilde{h}^i {\rm C}\!\_\tilde{\mathbf p}_0[n_f].r
          +\tilde{h}^r {\rm C}\!\_\tilde{\mathbf p}_0[n_f].i
         \Big)$
\ENDFOR
  \STATE ${\rm D}\!\_\boldsymbol\alpha''_{\rm pth}[{\rm blockIdx.y}][{\rm blockIdx.x}][{\rm threadIdx.x}]\,+\!\!= \Sigma$
\end{algorithmic}
\end{algorithm}
\clearpage

\begin{table}[ht]
\caption{
Computational times of the 3D image reconstructions by use of the 
CPU- and GPU-based implementations
}
\begin{tabular}{l |cc |cc |cc}
\hline\hline  
               & \multicolumn{2}{c|}{FBP [sec]}
               & \multicolumn{2}{c|}{PLS-Int [min/iteration]}
               & \multicolumn{2}{c}{PLS-Sph [min/iteration]}\\
Data sets   &CPU&\quad GPU \quad &\quad CPU& GPU&\quad CPU & GPU \\[0.5ex]
\hline
``$32\times45$'' & $6,189$  & $6$  & $2,448$ & $20$  & $7,961$   & $22$ \\
``$64\times45$'' & $12,975$ & $12$ & $4,896$ & $35$  & $15,923$  & $43$\\ 
``$64\times90$'' & $26,190$ & $23$ & $9,792$ & $68$  & $31,845$  & $86$\\ 
``$128\times90$''& $53,441$ & $46$ &  -      &  -    & -         & -\\ 
``quarter data''      & $12,975$ & $12$ & $4,896$ & $35$  & $19,776$  & $78$\\
``full data''         & $53,441$ & $46$ & $19,968$& $137$ & $79,177$  & $313$\\ 
\hline\hline
\end{tabular}
\label{tab:time}
\end{table}
\clearpage

\clearpage

\clearpage

\clearpage

\clearpage

\clearpage

\clearpage

\end{document}